\documentclass[a4paper,11pt,headings=big,DIV=12]{scrartcl}
\usepackage{amsmath,amssymb}
\usepackage{color,xcolor}
\definecolor{blue}{rgb}{0,0,0.5}
\usepackage[colorlinks,linkcolor=blue,urlcolor=blue,citecolor=blue]{hyperref}
\usepackage{graphicx}
\addtokomafont{disposition}{\rmfamily\boldmath}
\usepackage[utf8]{inputenc}
\usepackage{enumerate}
\usepackage{slashed}
\usepackage{cite}
\usepackage{multirow}
\usepackage{tikz}
\usetikzlibrary{matrix}

\allowdisplaybreaks
\newcommand{\ifb}{{\rm fb}^{-1}}
\newcommand{\NN}{{\cal N}}

\newcommand{\hel}{\phi}
\newcommand{\Rea}{\textrm{Re}}
\newcommand{\Ima}{\textrm{Im}}
\newcommand{\SP}[2]{\langle #1 | #2 \rangle}

\newcommand{\SPthth}[2]{\langle #1 | #2 \rangle_{\theta_K\theta_\ell}}
\newcommand{\SPthc}[3]{\langle #1 | #2 \rangle_{\theta_{#3} \hel}}
\newcommand{\SPththc}[2]{\langle #1 | #2 \rangle_{\theta_K\theta_\ell \hel}}

\newcommand{\Had}{H}

\newcommand{\vev}[1]{\langle #1 \rangle}

\newcommand{\matel}[3]{\langle #1|#2|#3\rangle}
\newcommand{\al}{\alpha}
\newcommand{\be}{\beta}
\newcommand{\ga}{\gamma}
\newcommand{\de}{\delta}
\newcommand{\la}{\lambda}

\newcommand{\GeV}{\;\mbox{GeV}}
\newcommand{\MeV}{\;\mbox{MeV}}

\newcommand{\Ks}{K}
\newcommand{\Kone}{K_1}
\newcommand{\Ktwo}{K_2}
\newcommand{\lone}{\ell_1}
\newcommand{\lonebar}{\bar{\ell}_{1}}
\newcommand{\ltwo}{\ell_{2}}
\newcommand{\ltwobar}{\bar{\ell}_{2}}
\newcommand{\IKsF}{I^{(0)}_{K^*}}
\newcommand{\IKF}{I^{(0)}_{K}}

\newcommand{\WignerD}[4]{D^{#1}_{#2, #3} \left( #4 \right)}

\newcommand{\bWignerD}[4]{\bar{D}^{#1}_{#2, #3} \left( #4 \right)}

\newcommand{\sca}[2]{#1 \cdot #2}
\newcommand{\famp}{\amp}
\newcommand{\Op}[1]{O^{(#1)}}
\newcommand{\mlone}{m_{\ell_1}}
\newcommand{\mltwo}{m_{\ell_2}}
\newcommand{\mKs}{m_{K^*}}
\newcommand{\mpi}{m_{\pi}}
\newcommand{\mK}{m_{K}}
\newcommand{\mB}{m_{B}}
\newcommand{\bonep}{\beta_1^{+}}
\newcommand{\bonem}{\beta_1^{-}}
\newcommand{\btwop}{\beta_2^{+}}
\newcommand{\btwom}{\beta_2^{-}}
\newcommand{\ml}{m_{\ell}}
\newcommand{\J}{g}

\newcommand{\gpol}{\omega}
\newcommand{\gpolt}{\epsilon}

\newcommand{\amp}{{\cal A}}
\newcommand{\cL}{\cos \thetal}

\newcommand{\thetal}{\theta_{\ell}}

\newcommand{\cK}{\cos \theta_K}

\newcommand{\Bin}{B}
\newcommand{\KallenB}{\lambda_{B}}
\newcommand{\KallenBK}{\lambda_{BK}}
\newcommand{\Kallengas}{\lambda_{\ga^*}}
\newcommand{\KallenKs}{\lambda_{K^*}}

\newcommand{\WC}[1]{ C_{#1}}

\newcommand{\HTenstzero}{H^{T_t}_0}
\newcommand{\HTenstzerobar}{\bar{H}^{T_t}_0}
\newcommand{\HTenstp}{H^{T_t}_{+}}
\newcommand{\HTenstpbar}{\bar{H}^{T_t}_{+}}
\newcommand{\HTenstm}{H^{T_t}_{-}}
\newcommand{\HTenstmbar}{\bar{H}^{T_t}_{-}}
\newcommand{\HTenszero}{H^{T}_0}
\newcommand{\HTenszerobar}{\bar{H}^{T}_0}
\newcommand{\HTensp}{H^{T}_{+}}
\newcommand{\HTenspbar}{\bar{H}^{T}_{+}}
\newcommand{\HTensm}{H^{T}_{-}}
\newcommand{\HTensmbar}{\bar{H}^{T}_{-}}
\newcommand{\HTime}{ H^{t}}

\newcommand{\HSca}{H^{S}}

\newcommand{\HVK}{h^V}
\newcommand{\HAK}{h^A}
\newcommand{\HSK}{h^S}
\newcommand{\HPK}{h^P}
\newcommand{\HTensK}{h^T}
\newcommand{\HTenstK}{h^{T_t}}

\newcommand{\HSKbar}{\bar{h}^S}
\newcommand{\HPKbar}{\bar{h}^P}
\newcommand{\HTensKbar}{\bar{h}^T}
\newcommand{\HTenstKbar}{\bar{h}^{T_t}}

\newcommand{\HPS}{ H^{P}}
\newcommand{\HPSbar}{ \bar{H}^{P}}
\newcommand{\HS}{H^{S}}
\newcommand{\HSbar}{\bar{H}^{S}}

\newcommand{\HVzero}{H^{V}_{0}}
\newcommand{\HVzerobar}{\bar{H}^{V}_{0}}
\newcommand{\HVm}{H^{V}_{-}}
\newcommand{\HVmbar}{\bar{H}^{V}_{-}}
\newcommand{\HVp}{H^{V}_{+}}

\newcommand{\HAzero}{H^{A}_{0}}
\newcommand{\HAzerobar}{\bar{H}^{A}_{0}}
\newcommand{\HAm}{H^{A}_{-}}
\newcommand{\HAmbar}{\bar{H}^{A}_{-}}
\newcommand{\HAp}{H^{A}_{+}}
\newcommand{\HApbar}{\bar{H}^{A}_{+}}

\newcommand{\GKstar}[3]{G^{#1, #2}_{#3}}
\newcommand{\GKstart}[3]{\tilde{G}^{#1, #2}_{#3}}
\newcommand{\GKstaro}[3]{\mathbb{G}^{#1, #2}_{#3}}

\newcommand{\GK}[1]{G^{(#1)}}
\newcommand{\GKh}[1]{\hat{G}^{(#1)}}
\newcommand{\GKt}[1]{\tilde{G}^{(#1)}}
\newcommand{\GKo}[1]{\mathbb{G}^{(#1)}}
\newcommand{\Gbaryon}[3]{\mathcal{K}^{#1, #2}_{#3}}
\newcommand{\DD}[3]{\Omega^{#1,#2}_{#3}}
\newcommand{\partialphi}[4]{p^{#1, #2}_{#3,#4}( \hel )}
\newcommand{\cc}[3]{c^{#1,#2}_{#3}}
\newcommand{\MM}[3]{M^{#1,#2}_{#3}}
\newcommand{\kpartial}[2]{k^{#1}_{#2}(\theta_K)}
\newcommand{\lpartial}[2]{l^{#1}_{#2}(\thetal)}

\newcommand{\intbin}{\int_{\textrm{bin}} dq^2} 
\newcommand{\binavg}[1]{\left\langle #1 \right\rangle_{\textrm{bin}}}
\newcommand{\Nbin}{\mathcal{N}_{\textrm{bin}}'} 
\newcommand{\Nbinz}{\mathcal{N}_{\textrm{bin}}} 

\begin{document}
\pagenumbering{gobble}
\begin{flushright}
\begin{tabular}{l}
CP3-Origins-2015-017 DNRF90 \\
DIAS-2015-17
\end{tabular}
\end{flushright}
\vskip1.5cm

\begin{center}
{\Large\bfseries \boldmath
Generalised helicity formalism, higher moments and \\
the  $B \to K_{J_K}(\to K \pi)\lonebar \ltwo$  angular distributions}\\[0.8 cm]
{\Large%
James Gratrex$^a$,
Markus Hopfer$^{a,b}$,
and Roman Zwicky$^a$
\\[0.5 cm]
\small
$^a$ Higgs Centre for Theoretical Physics, School of Physics and Astronomy,\\
University of Edinburgh, Edinburgh EH9 3JZ, Scotland \\
$^b$ Institut f\"ur Physik, Karl-Franzens-Universit\"at Graz, Universit\"atsplatz 5, 8010 Graz, Austria
}
\\[0.5 cm]
\small
E-Mail:
\texttt{\href{mailto:j.gratrex@ed.ac.uk}{j.gratrex@ed.ac.uk}},
\texttt{\href{mailto:markus.hopfer@emtensor.com}{markus.hopfer@emtensor.com}},
\texttt{\href{mailto:roman.zwicky@ed.ac.uk}{roman.zwicky@ed.ac.uk}}.
\end{center}

\bigskip

\begin{abstract}\noindent
We generalise the Jacob-Wick helicity formalism, which applies to sequential decays,
to effective field theories of rare decays of the type $B \to K_{J_K}(\to K \pi)\lonebar \ltwo$.
This is achieved by reinterpreting  local interaction vertices
$\bar b \Gamma'_{\mu_1 ..\mu_n} s \bar \ell  \Gamma^{\mu_1 ..\mu_n}  \ell$ as a coherent
sum of $1 \to 2$ processes mediated by particles whose spin ranges between zero and $n$.
 We illustrate the framework
 by deriving the full angular distributions for  $\bar{B} \to \bar{K}\lone \ltwobar$ and $\bar{B} \to \bar{K}^*(\to \bar{K} \pi)\lone \ltwobar$
 for the complete dimension-six effective Hamiltonian  for non-equal lepton masses.
  Amplitudes and decay rates are expressed in terms of Wigner rotation matrices, leading naturally to the method of moments in various forms.
  We discuss how higher-spin operators and QED corrections alter the standard angular
 distribution used throughout the literature, 
 potentially leading to differences between the method of moments and the likelihood fits.
We propose to diagnose these effects by assessing higher angular moments.
 These could be relevant in investigating the nature of the current LHCb anomalies in
$R_K = {\cal B}( B \to K \mu^+\mu^-) /{\cal B}( B \to K e^+e^-)$ as well as angular observables in $B \to K^* \mu^+\mu^-$.
\end{abstract}

\newpage

\newpage
\setcounter{tocdepth}{2}
\tableofcontents
\newpage
\pagenumbering{arabic}

\section{Introduction}

Helicity amplitudes (HAs), as defined by Jacob and Wick \cite{JW1959}, 
describe $A \to BC$ ($1 \to 2$) transitions and have definite transformation properties under rotation.
The key idea is that the angular and  helicity information are equivalent to each other. 
Angular decay distributions 
follow (e.g. \cite{Richman:1984gh,Haber1994,Kutschke1996}) 
from evaluating the HAs with $B$ and $C$ in the forward direction, with the angular information encoded  in Wigner $D$ matrix functions, 
reminiscent of the Wigner-Eckart theorem.

 The intent of this paper is to generalise 
this method to decays of the type $A \to (B_1 B_2)C$  
which are schematically described by local interactions of the form
\begin{equation}
\label{eq:Hsym}
H^{\rm eff} \sim (A C)_{\mu_1 \dots \mu_n}  (B_1B_2)^{\mu_1 \dots \mu_n}   \;.
\end{equation}
We do so by rewriting the $1 \to 3$ decay as a sequence of $1 \to 2$ processes, by inserting multiple complete sets
of polarisation states between the Lorentz contractions of $AC$ and $B_1 B_2$  above.
This leads to a  reinterpretation of the decay in terms of a sum over intermediate particles of spin 
$J$, where $J$ can range from $0$ up to $n$ depending on the specific 
structure of the operators.  Symbolically we may write 
\begin{equation}
{\cal A}(A \to (B_1 B_2)C)  =  \sum_{J \geq 0}^n {\cal A}(A \to B_J (\to B_1 B_2)C) \;,
\end{equation}
with  ${\cal A}$ denoting   the amplitude.  We refer to this case as the 
$B$-particle factorisation approximation. 
At the formal level, the main work is the decomposition of the Lorentz tensors into irreducible objects under the spatial rotation group (reminiscent of the $3 +1$ decomposition of cosmological perturbation theory for example). 
 
Important examples of such  decays are given by the rare radiative decays 
$B \to K \ell^+\ell^-$ and $B \to K^{*}(\to K \pi)\ell^+\ell^-$.
Besides evaluating non-perturbative matrix elements to these decays (e.g. \cite{Ball:2004rg,Khodjamirian:2006st,BFS2001,FM02,AAGW2001,DLZ2012,LZ2013,KMPW10,BJZ2006,MXZ2008,
Ball:2004ye,Khodjamirian:2012rm,Bouchard:2013mia}), it has become clear that it is beneficial to consider general properties of the amplitudes entering the angular distributions (e.g.  \cite{HZ2013,DHJS14,HM15,DV15}).
Our work can be seen to be part of the latter category.

We evaluate the $\bar{B} \to \bar{K}^{(*)} \ell^+\ell^-$ angular decay distributions within the generalised helicity 
framework developed in this paper, providing an alternative method to traditional techniques using 
Dirac trace technology  \cite{KSSS99,Tub02}. An important consequence of the manner in which we derive the distribution is that 
it lends itself to the methods of 
moments (MoM), which use the decomposition of the distribution into orthogonal functions to obtain observables independently of each other.  
This is a complementary  method to the likelihood fit to extract the dynamical information from the decay, 
and was recently studied from an experimental viewpoint in \cite{BCDS2015}.
We discuss the impact of including higher partial waves in both the $(K\pi)$- and especially the dilepton-system.
The latter give rise to corrections, in the form of higher moments, to the standard form of the angular distribution used in the literature. 
The sources of higher dilepton partial waves are higher spin operators and electroweak corrections, both of which we discuss 
qualitatively. The two sources can be distinguished by their different behaviour in higher partial moments. 
We encourage experimental investigation of higher moments from various viewpoints. In particular, 
we discuss how higher moments can be used to diagnose the size of QED effects in $B \to K \ell^+\ell^-$ (with $\ell = e,\mu$) 
 and test leakage of $J/\Psi$-contributions into the lower dilepton-spectrum. Both 
are of importance in view of $R_K$ as well as the angular anomalies in the low
dilepton-spectrum, which have recently been reported by the  LHCb collaboration  in \cite{RKLHCb} and  \cite{LHCb-P5p,LHCbBKmumu2015} respectively.

The paper is organised as follows. In section \ref{sec:method} the methodology is introduced ending 
with a formal expression for the fourfold decay distribution in terms of rotation matrices and HAs.
Specific angular distributions for $\bar{B} \to \bar{K}^{(*)} \lone \ltwobar$,\footnote{Throughout this work, we use non-equal leptons $\ell_1 \neq \ell_2$, in order to accommodate semi-leptonic decays of the type $B \to \rho \ell \nu$ as well as potential lepton flavour violation \cite{GGL2014}, motivated by the $R_K$ measurement.} with  detailed results in appendices \ref{app:BtoKs}  (and a Mathematica notebook in the arXiv version) 
and \ref{app:BtoK},  are given in section \ref{sec:angularD}. 
The method of total and partial moments is presented in section \ref{sec:Moments}. 
Section \ref{sec:limit} contains the discussion of including higher partial waves: a qualitative assessment 
of higher spin operators and QED corrections is presented in subsections \ref{sec:higher} and \ref{sec:beyond}
respectively. The relevance of testing for higher moments is emphasised in subsection \ref{sec:aktuell}.
The paper ends with conclusions in section \ref{sec:conclusions}. Additional material, such as 
the leptonic HAs and a few brief remarks on $\Lambda_b \to \Lambda \left( \to (p,n) \pi \right) \lone \ltwobar $, 
is presented in appendices \ref{app:leptonHAs} and \ref{app:baryons} respectively.
In appendix \ref{app:Conventions} we provide the kinematic conventions for computation 
of the angular distribution by the sole use of Dirac trace technology. 

\section{Generalised Helicity Formalism for Effective Theories}
\label{sec:method}

We first review the standard helicity formalism in section~\ref{sec:hel-qual}, and qualitatively apply it to
sequential $1 \to 2$ decays in section~\ref{sec:standard}, specialising to the spin configuration relevant 
for our decays at the end. 
In section~\ref{sec:effective}  the formalism is  extended  to include decays like $B \to K_{J_K}(\to K \pi)\lonebar \ltwo$ 
described by effective field theories for $b \to s \lonebar \ltwo$ transitions. 
The framework can be straightforwardly applied to the entire zoo of semi-leptonic and rare flavour decays 
such as $B_s \to K^* \ell \nu$, $B \to D^{(*)} \ell \nu$, $D \to (\pi,\rho) \mu \mu$, $D \to (\pi,\rho) \mu \nu$, $K \to \pi \mu \mu$ etc., 
and can also be extended to include initial particles with non-zero spin.

\subsection{The basic idea of the Helicity Formalism and its Extension}
\label{sec:hel-qual}

The discussion in this section is standard and we refer the reader to \cite{Richman:1984gh,Haber1994,Kutschke1996} for  more extensive reviews, as well as the pioneering paper of Jacob and Wick \cite{JW1959}.
In a $1 \to 2$ (say $A \to B_1 B_2$) decay a particle of spin $J_A$ and helicity $M_A$ decays into 
two particles of momentum $\vec{p}_1$ and $\vec{p}_2$ with helicities  $\la_1$ and $\la_2$ 
respectively. In the centre-of-mass frame ($\vec{p}_1= - \vec{p}_2$)  
 the system can be characterised by the two helicities and the direction (i.e. the solid angles $\theta$ and $\phi$).
By inserting a complete set of two-particle angular momentum states 
the corresponding matrix element can be written 
\begin{eqnarray}
\label{eq:one}
 {\cal A}(A \to B_1 B_2)  &=& \langle \theta,\phi, \la_1,\la_2 | J_A,M_A \rangle = \sum_{j,m  }\langle \theta,\phi, \la_1,\la_2 |    j,m , \la_1, \la_2\rangle \langle j,m, \la_1, \la_2|  J_A,M_A \rangle \nonumber  \\[0.1cm]
 &=& \underbrace{ \langle \theta,\phi, \la_1,\la_2 |    J_A,M_A , \la_1, \la_2\rangle}_{ \sqrt{\frac{2J_A+1}{4\pi}}  \WignerD{J_A}{M_A}{\la_1-\la_2}{\phi,\theta,-\phi }} 
  \underbrace{\langle J_A,M_A, \la_1, \la_2|  J_A,M_A \rangle}_{ \!\!\!{\phantom{ \sqrt{\frac{1}{1}}}}   
  {\cal A}^{J_A}_{M_A,\la_1,\la_2}}
 \end{eqnarray}
  as a product of Wigner $D$-functions 
 and a HA ${\cal A}^{J_A}_{M_A,\la_1,\la_2}$. 
 The Wigner matrix is a  $(2J_A+1)$-dimensional $SO(3)$ representation in the helicity basis.
 The essence is that the distribution of the amplitude over the angles is then governed by the rotation matrix as a function of the helicities. 
 In practice one only needs to compute the HA. 
 
 The process $B \to J/\Psi(\to \ell^+ \ell^-) K^*(\to  K\pi)$  constitutes  a well-known 
  example 
 of a sequential $1 \to 2$ decay where the formalism can be applied \cite{Dighe:1998vk}. 
 The idea of this paper is to extend this formalism to the case where the $\lonebar \ltwo$-pair emerges from a local interaction 
 vertex $O_{ij} \sim \bar s \Gamma_i b \bar \ell \Gamma_j \ell$ with effective Hamiltonian $H^{\rm eff} \sim \sum_{ij} C_{ij} O_{ij}$.
 This is achieved by reinterpreting the local interaction vertex as originating from a sum of particles whose spin 
depends on the number of Lorentz contractions between the $\Gamma_{i,j}$ structures.   
Elements of this program have appeared in the literature, e.g. \cite{LW2007} 
for $B \to K_J \ell^+ \ell^-$, but we are unaware of a systematic presentation that allows 
the incorporation of a generic effective Hamiltonian as well as other decay types.

\subsubsection{Helicity formalism for $\Bin_{J_{\Bin}} \to \Ks_{J_{\Ks} }(\to \Kone \Ktwo)\ga_{J_\ga}( \to \lonebar \ltwo ) $}

\label{sec:standard}

Let us consider the following sequential decay 
\begin{equation}
\Bin_{J_{\Bin}} \to  \Ks_{J_{\Ks} }(\to \Kone \Ktwo)\ga_{J_\ga}^{\vphantom{+}}( \to \lonebar \ltwo )
\end{equation}
where $J_\Bin$, $J_\ga$ and $J_\Ks$ denote the spin of the particles $\Bin$, $\ga_J^{\vphantom{+}}$ and $K_J$. 
The notation is close to the main application of this paper but we emphasise that 
at this point the methodology is completely general.  
Assuming the decay to be a series of sequential $1 \to 2$ decays the amplitude can 
be written in terms of a product of $1 \to 2$ HAs times the corresponding 
Wigner functions 
\begin{equation}
\label{eq:helamp}
\begin{split}
& \amp(\Omega_{\Bin},\Omega_\ell ,\Omega_{\Ks}| \la_{\Bin },\la_{\Kone},\la_{\Ktwo},\la_1,\la_2 )  \sim  \\
& \quad  \sum_{\la_{\ga} \la_{\Ks}} 
 \bWignerD{J_{\Bin}}{\la_{\Bin} }{ \la_{\ga}-\la_{ \Ks } }{\Omega_{\Bin}}
 {\cal \Had}_{\la_{\ga} \la_{\Ks} }
 \bWignerD {J_{\Ks}}{\la_{\Ks} }{\la_{\Ktwo}-\la_{\Kone}}{\Omega_{\Ks}} 
{\cal \Ks}_{\la_{\Kone}, \la_{\Ktwo}}
 \bWignerD {J_\ga}{\la_{\ga}}{\la_\ell}{\Omega_\ell}{\cal L}_{\la_{\lone}
\la_{\ltwo}} \;,
\end{split}
\end{equation}
where the $\la_i$ are  helicity indices, and  
\begin{equation}
\label{eq:lal}
\la_\ell \equiv \la_1 - \la_2 \; , 
\end{equation}
 is a shorthand that we use frequently  throughout the paper.   The HAs ${\cal H}$, ${\cal K}$ and ${\cal L}$ 
 correspond to the transitions $B_{J_B}  \to K_{J_K} \ga_{J_\ga}$, $ K_{J_K} \to K_1 K_2$ and 
 $ \ga_{J_\ga} \to \bar \ell_1 \ell_2$ respectively.
 The helicities of the internal particles $\ga_J$ and $K_J$  have to be coherently summed over. 
The Wigner $D$-functions   
\begin{equation}
\WignerD{j}{m'}{m}{\Omega = (\alpha,\beta,\gamma)} 
  = \langle j m'|e^{-i\alpha J_z}e^{-i\beta J_y}e^{-i\gamma J_z}|jm\rangle
  \end{equation}
 are irreducible $SO(3)$-representations  of dimension $2j+1$. The $J_i$ are the generators of angular momentum, and the states $|j,m \rangle$ carry angular momentum $j$ and helicity $m$ and are orthonormalised  
 $\langle j,m| j',m' \rangle = \de_{jj'} \de_{mm'}$. 
 To avoid proliferation of indices we denote complex conjugation by a bar instead of the more standard asterisk.
   
\paragraph{Adaptation to  $J_B = 0$ and $K_1 = K$ and $K_2 =\pi$}

\begin{figure}
\includegraphics[scale=0.7,clip=true,trim= 0 370 0 250]{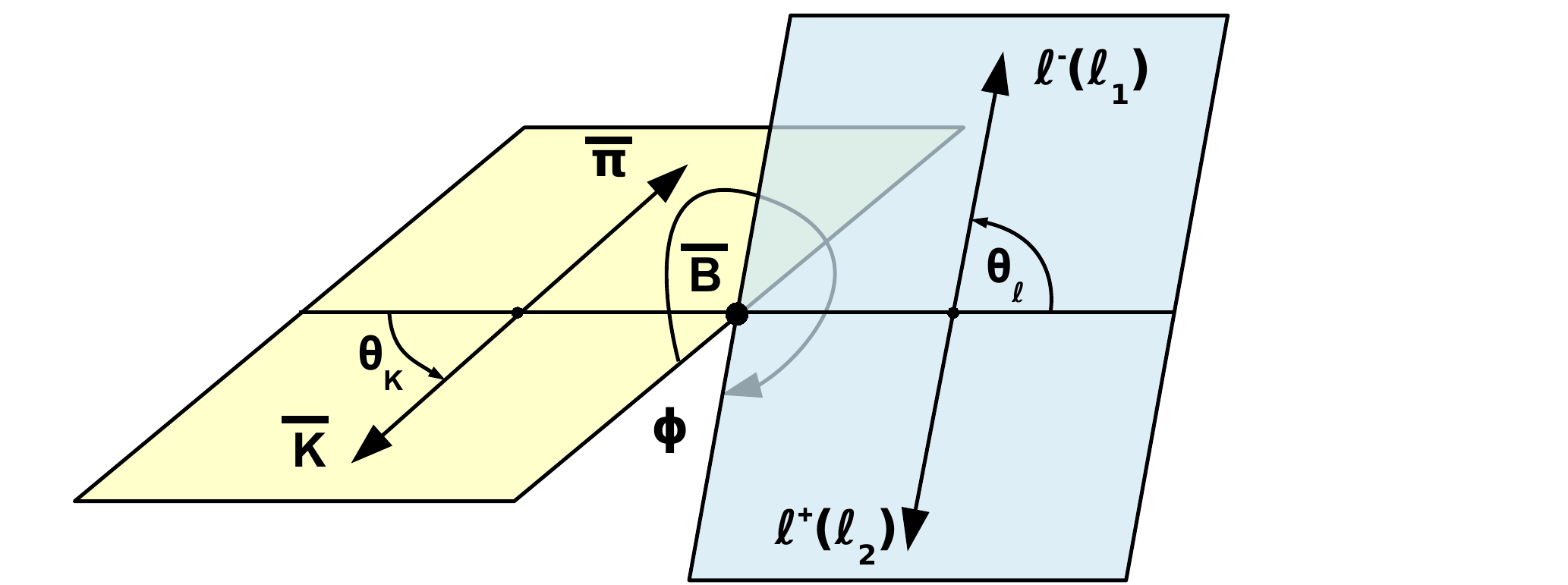} \\[0.5cm]
\includegraphics[scale=0.7,clip=true,trim= 0 370 0 250]{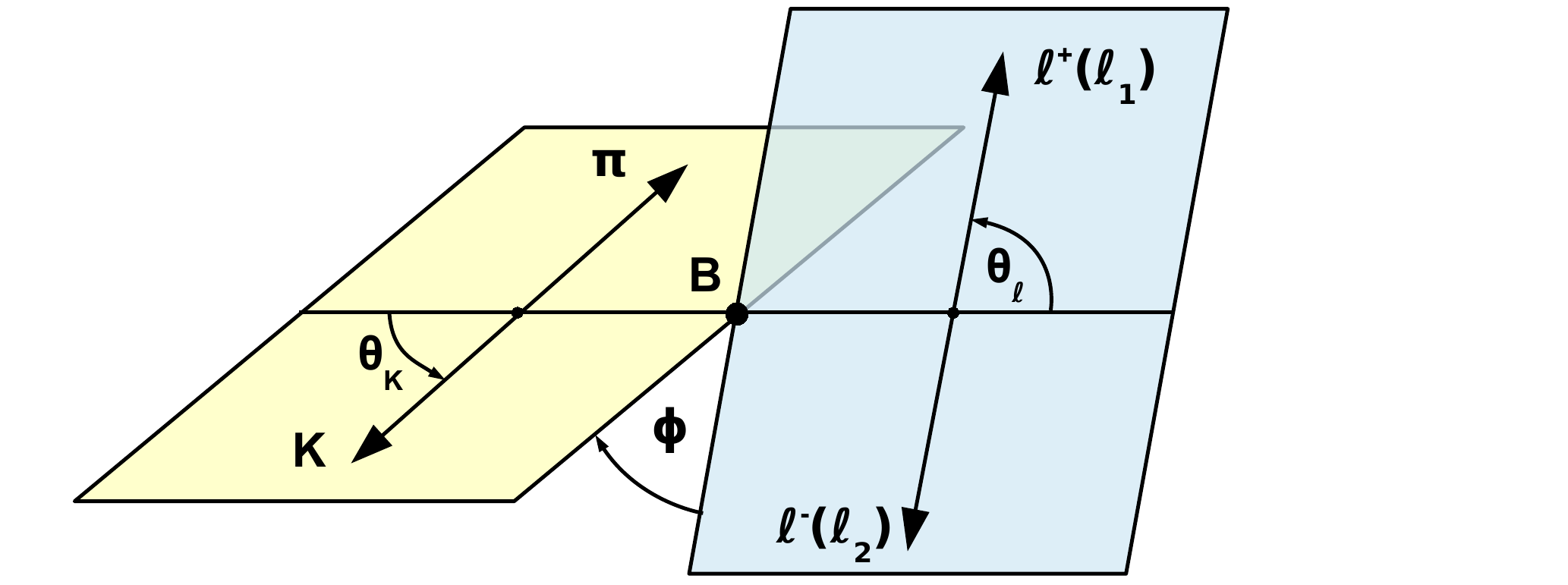} 
\caption{\small Decay geometries for $\bar B \to \bar K^*  \lone   \ltwobar $ (above) and 
$ B \to  K^*    \lonebar  \ltwo$ (below). 
In both cases  $\ell_1 =  \ell^- $, $\ell_2 = \ell^-$ denote the negatively charged lepton.
The  conventions  are the same as used by the  LHCb collaboration in \cite{LHCbKmumu2013} (cf. appendix A therein).  
Comparison to the convention used by the theory community can be found in appendix~\ref{app:ang-conventions}. 
The pictures are slightly misleading in that the angles $\theta_{\ell,K}$ are drawn 
in the restframe of the lepton-pair and the $K^*$-meson. For decays which are not self-tagging, 
such as $B_s,\bar B_s \to \phi(\to K^+ K^-) \mu^+ \mu^-$ at the LHCb, the angles $( \thetal,\theta_K,\phi)   \to (\pi - \thetal,\pi - \theta_K,2 \pi -\phi),$ and one can only measure
the sum of both decay rates (second equation\eqref{eq:sum}).}
 \label{fig:decay_planes}
 \end{figure}

In order to ease the notation slightly we move straight to the case
 $\bar{\Bin} \to \bar{\Ks}_J(\to K \pi) \ga_{J_\ga} ( \to \lone \ltwobar)$.\footnote{We choose $\bar B \to \bar{\Ks}_J \lone \ltwobar$ transitions as the main template for the results in this paper.
Such transitions are governed by the $b \to s$ Hamiltonian, which is the standard in the theory literature and is used to define the Wilson coefficients. In the more conceptual sections we shall refer to $B \to \Ks \lonebar \ltwo$-transitions.}  
The relation 
$D_{\la_{\Bin}=0, \la_{\ga}-\la_{ \Ks }}^{J_{\Bin}=0}(\Omega) =\delta_{0, \la_{\ga}-\la_{ \Ks }}$  
 implies equality of helicities 
 \begin{equation}
\la \equiv \la_{\ga}=\la_{ \Ks } \;.
 \end{equation}
One may therefore reduce $ {\cal \Had}_{\la_{\ga} \la_{\Ks}} \to H_\la  $, which is the quantity known as the HA in the $\bar{B} \to \bar{K}^* \ell^+ \ell^-$-literature and carries the non-trivial dynamic information. 
 The HA ${\cal \Ks}_{\la_{\Kone} ,\la_{\Ktwo}}$ reduces to a scalar constant  (denoted by 
 $g_{K_J K \pi}$)
 since $\Kone \to K$, $\Ktwo \to \pi$ are both scalar particles. The third HA ${\cal L}_{\la_{1} \la_{2}}$ depends on the interaction vertex of the leptons, but is trivial to calculate once the interaction is known. We may rewrite the amplitude \eqref{eq:helamp}
 as
\begin{equation}
\label{eq:amp2}
\amp(\bar{\Bin} \to \bar{\Ks}_{J_{\Ks}} (\to K \pi) \ga_{J_\ga} ( \to \lone \ltwobar)) \sim  
\sum_{\la = -J_{\Ks} }^{J_{\Ks}}  \WignerD{J_K}{\la}{ 0}{\Omega_{\Ks}} 
\WignerD{J_\ga}{\la}{\la_\ell}{\Omega_\ell}  \amp^{J_\ga}_{\la, \la_1,\la_2}  \;,
\end{equation}
where the angles, depicted in figure \ref{fig:decay_planes}, 
are $\Omega_{\Ks} = (0,\theta_K,0)$ and $\Omega_\ell = (\hel_\ell,\thetal,-\hel_\ell)$.  Note, the passage from 
$\bar{D}$ to $D$-functions from \eqref{eq:helamp} to \eqref{eq:amp2}  is related to passing from $B$ to $\bar B$.

In the lepton-pair factorisation approximation, defined more explicitly in the following section, the amplitude $\amp^{J_\ga}_{\la, \la_1,\la_2} \sim H_\la {\cal L}_{\la_1\la_2}|_{J_\ga}$ is the product
of the hadronic and leptonic matrix elements. 
The angle $\hel_\ell$ is the helicity angle and is usually called simply $\hel$.  Before
 commenting on different conventions of the angles we quote  
the fourfold differential decay 
\begin{eqnarray}
\label{eq:dG4}
& &  \frac{d^4 \Gamma}{dq^2  d\cL d \cK d\phi} \sim  \sum_{\la_1 \la_2} |\amp|^2 \sim \\[0.1cm] 
& &   
\sum_{\la_1,\la_2 = -1/2 }^{1/2}   \sum_{\la,\la' = -J_\ga}^{J_\ga}
 {\amp}^{J_\ga}_{\la, \la_1,\la_2}  \bar{\amp}^{J_\ga}_{\la', \la_1,\la_2} 
 \WignerD{J_{K_J}}{\la}{0}{\Omega_{\Ks}} \bWignerD{J_{K_J}}{\la'}{0} {\Omega_{\Ks}}  
   \WignerD{J_\ga }{\la}{\la_\ell}{\Omega_\ell}  \bWignerD{J_\ga }{\la'}{\la_\ell}{\Omega_\ell}  \;,
\end{eqnarray}
in terms of amplitudes and Wigner $D$-functions. 
For the angles we use the $\bar{B} \to \bar{K}^* \ell \ell$ decay as a reference and use the same conventions
as the LHCb collaboration  \cite{LHCbKmumu2013} (appendix A), which differ from those 
used by the theory community. More precise statements, including a conversion diagram, 
can be found in appendix~\ref{app:ang-conventions}.

\subsection{Effective Theories rewritten as a coherent Sum of Sequential Decays}
\label{sec:effective}

In this section we give the formal steps to derive the expression of the angular distributions. 
The reader interested in the final result can directly proceed to section~\ref{sec:angularD}.

The amplitude \eqref{eq:amp2} is of a completely general form for the decay 
 where $\ga_{J_\ga}$ is an actual particle of spin $J_\ga$. In $\bar B \to \bar K^*(\to \bar K  \pi ) \lone \ltwobar$ 
 a part of the amplitude is in this form where the photon corresponds to the intermediate state ($\ga_1 = \ga$). 
 In general there are effective vertices, so-called contact terms, 
 where the intermediate particles are not present. 
In the interest of clarity we quote the  effective Hamiltonian for $b \to s \ell \bar \ell$:\footnote{
The adaptation from $\ell \bar \ell \to \lone \ltwobar$ is trivial and will not be spelled out explicitly.}
\begin{eqnarray}
\label{eq:Heffexplicit}
H^{\rm eff} =  c_H \hat{H}^{\rm eff} \;,  \quad 
c_H \equiv - \frac{4 G_F}{\sqrt{2}} \frac{\alpha}{4 \pi} V_{\rm ts}^*  V_{\rm tb} 
\;, \quad \hat{H}^{\rm eff}  =  \!\!\!\!\!\!\!\!\! \sum_{i=V,A ,S,P,\cal T} ( C_i O_i + 
C'_i O'_i) 
 \;. \quad  
\end{eqnarray}
Above $G_F$ is Fermi's constant, $\alpha$ the fine structure constant, $V_{\rm tD}$ are 
Cabibbo-Kobayashi-Maskawa (CKM) elements and the operators are
\begin{eqnarray}
 \label{eq:operators}
 O_{S(P)} = \bar s_L   b \;  \bar\ell (\gamma_5) \ell\;,  \quad  
 O_{V(A)} = \bar s_L  \gamma^\mu b  \; \bar\ell \gamma_\mu   (\gamma_5) \ell\;, \quad  
 O_{\cal T} = \bar s_L \sigma^{\mu\nu}  b \;  \bar\ell\sigma_{\mu\nu} \ell  \;,  
 \end{eqnarray}
 where $O' = O|_{s_L \to s_R}$, the labels refer to the lepton interaction vertex,   
$q_{L,R}  \equiv 1/2(1 \mp \ga_5) q$, $\bar \ell , \ell \to \lonebar, \ltwo$ 
for different lepton flavours   and a few additional relevant remarks 
deferred to appendix~\ref{app:Hamiltonian}. In passing we add that the notation
 $O_{9(10)}= O_{V(A)}$ is more common throughout the literature.  
 In the case where electroweak corrections are neglected at the matrix element
 level one may factorise the hadronic from the leptonic part. 
 We refer to this as the lepton-pair factorisation approximation (LFA) ($B$-particle factorisation approximation in the introduction).
 Schematically   \eqref{eq:Heffexplicit} is written
as a product of a hadronic part $H$ and a leptonic part ${\mathcal L}$ with a certain number of Lorentz contractions between them: 
\begin{equation}
\label{eq:Heff}
H^{\rm eff} \sim  \sum_{a = 1}^{N_0} \Had^a {\mathcal L}_a +  \sum_{b = 1}^{N_1}  \Had^b_{ \mu} {\mathcal L}_{b}^\mu +   \sum_{c=1}^{N_2} \Had^c_{ \mu_1 \mu_2} {\mathcal L}_c^{\mu_1 \mu_2}  \;.
\end{equation} 
The sum over $a$, $b$ and $c$  extends over operators with $0$, $1$ and $2$  Lorentz contractions between quark and lepton operators.
 In the example of $C_V O_V =  H_\mu \cal L^\mu$ we would have $H_\mu = C_V \bar s_L \ga_\mu b$  and $\cal L^\mu = \bar \ell \ga^\mu \ell$. On a formal level we might think of $O_V (O_9)$ as originating from integrating out a vector and a scalar particle, in the sense that the Lorentz contraction over index $\mu$ can be written as the sum of products of a spin-one and a timelike spin-$0$ polarisation vector.
This is expressed by the well-known completeness relation (e.g. \cite{Tub02,LW2007,HZ2013})
 \begin{equation}
\label{eq:c-relation}
 g^{\mu\nu} = \sum_{\la,\la' \in \{t, \pm,0\}  }  \gpol^{\mu}(\la) \bar \gpol^{\nu}(\la')  G_{\la \la' }
 \;, \quad G_{\la \la' } = \text{diag}(1,-1,-1,-1) \;,
\end{equation}
where the first entry in $G_{\la \la' }$ refers to $\la = \la' = t$ and an 
explicit parametrisation is given by
\begin{eqnarray}
\label{eq:gpol}
\gpol^\mu(\pm)  =  (0, \pm 1, i,0)/\sqrt{2}  \;, \quad   \gpol^\mu(0) =  (q_z,0,0,q_0)/\sqrt{q^2} \;, \quad   \gpol^\mu(t) =  (q_0,0,0,q_z)/\sqrt{q^2} \;,
\end{eqnarray}
which is consistent with the parameterisation $q^\mu = ( q_0 ,0,0,q_z)$.
The polarisation vectors $\gpol(\pm,0)$ are compatible with the 
 Jacob-Wick phase convention  \cite{JW1959} (cf. appendix \ref{app:Conventions} 
 and the corresponding footnote for further remarks). 
Let us pause a moment and emphasise that intermediate results do depend on the convention, which enters the definition of the HAs, and this dependence has to be taken into account when comparing to HAs appearing in the literature. 
We choose the convention in \cite{HZ2013}, since it is compatible 
with the Condon-Shortly convention that is standard for Clebsch-Gordon coefficients and Wigner 
matrices (e.g. \cite{PDG}).
\begin{figure}
\centering
\includegraphics[scale=0.75,clip=true,trim= 0 600 50 50]{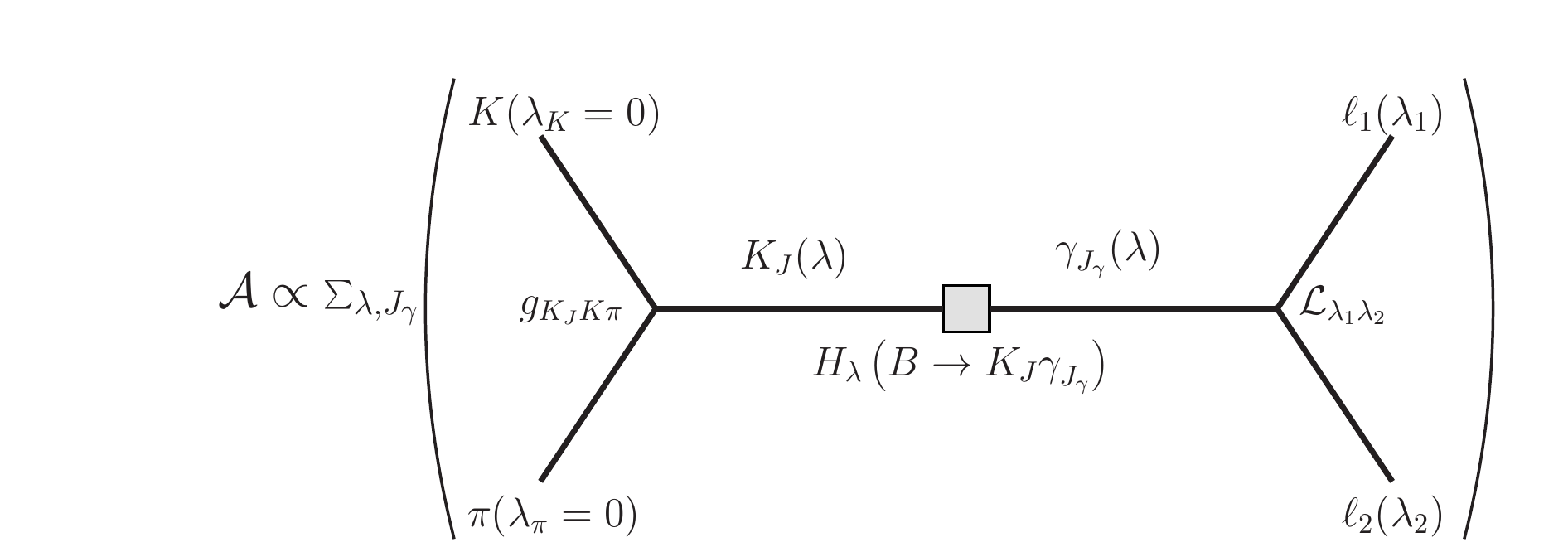}
\caption{\small A diagrammatic interpretation of the process, equation~\eqref{eq:amp3}, used to set up the formalism. The decay to two leptons is treated as being mediated by an effective particle $\ga_{J_\ga}$ of spin $J_\ga$. The factor $g_{K_J K \pi}$ has no dependence on helicities and depends only on the dynamics of the $K^{*}$ decay.} 
\label{fig:schematic}
\end{figure}

 We may think of $\gpol$ as being associated with the Lorentz group $SO(3,1)$. 
In the rest frame $q_z =0$ the timelike polarisation tensor transforms as a scalar under the restriction of $SO(3,1)$ to spatial rotations $SO(3)$.\footnote{Formally the branching rule for the Lorentz four vector $(1/2,1/2)$ is
$(1/2,1/2)_{SO(3,1)}|_{SO(3)} \to (\mathbf{1} + \mathbf{3})_{SO(3)}$.} 
For an effective operator with $n$ Lorentz indices the
relation \eqref{eq:c-relation} can be inserted $n$ times to obtain a HA 
with $n$ helicity indices. More precisely, the direct product of $SO(3,1)$ polarisation tensors decomposes 
into irreducible representations of $SO(3)$ polarisation tensors $\gpolt^{j,\la}_{\mu_1 \dots \mu_n}$ 
of spin $j = 0, \dots ,n$ and helicities $\la = -j, \dots ,j$.
 Using the expressions in Eqs.~\ref{eq:eps2} and \ref{eq:eps1} 
the  analogue of  $\amp^{J_\ga}_{\la, \la_1,\la_2}$ in \eqref{eq:amp2} 
on each spin component can be written as\footnote{In the notation used throughout the literature  $\HTime = \vev{\Had_b^\mu}  \gpolt^{0,0}_\mu   = \vev{\Had_b^\mu}  \gpol_\mu(t) $ is known as the timelike HA \cite{Tub02,LW2007}. 
By virtue of the equation of motion the timelike HAs can be absorbed into the scalar 
and pseudoscalar HAs, cf. appendix \ref{app:HadronHAs}.}
\begin{eqnarray}
\label{eq:cool}
 \amp^{J_\ga}_{\la, \la_1,\la_2}
 =  \left\{ \begin{array}{llll}       
 \;  \vev{\Had_a} \vev{{\mathcal L}_a}  \;\; +&    \vev{\Had_b^\mu} \vev{{\mathcal L}_b^\al}  \gpolt^{0,0}_\mu   \bar \gpolt_{\al}^{0,0}     & +   \;\;
  \vev{\Had_c^{\mu \nu}} \vev{{\mathcal L}_c^{\al \be}}  \gpolt^{0,0}_{\mu \nu} \cdot  \bar \gpolt^{0,0}_{\al \be}                
 &  \quad J_\ga =0  \\ 
   &   \vev{\Had_b^\mu} \vev{{\mathcal L}_b^\al}  \gpolt^{1,\la}_{\mu} \bar \gpolt^{1,\la_\ell}_{\al}\;  &+  \;\; \vev{\Had_c^{\mu \nu}} \vev{{\mathcal L}_c^{\al \be}}  \gpolt^{1,\la}_{\mu \nu}  \cdot \bar  \gpolt^{1,\la_\ell}_{\al \be}     & \quad J_\ga = 1 \\
  & & \phantom{+}\;\;  \vev{\Had_c^{\mu \nu}} \vev{{\mathcal L}_c^{\al \be}}  \gpolt^{2,\la}_{\mu \nu}  \cdot \bar \gpolt^{2,\la_\ell}_{\al \be}     & \quad  J_\ga = 2
   \end{array} \right.
\end{eqnarray}
where summation over Lorentz indices and the number of operators 
in \eqref{eq:Heff} are both implied, the scalar product ``$\cdot$" is detailed in \eqref{eq:eps2} and 
\begin{equation}
\label{eq:helmatel}
\vev{\Had_a^{\mu_1 \dots \mu_n}} \equiv \matel{\bar{K}_J(\la)}{\Had_a^{\mu_1 \dots \mu_n}}{\bar{B}} \;, \quad \vev{{\mathcal L}_a^{\mu_1 \dots \mu_n}} \equiv \matel{\lone(\la_1) \ltwobar(\la_2)}{{\mathcal L}_a^{\mu_1 \dots \mu_n}}{0} \;,
\end{equation}
are the leptonic and hadronic matrix elements. 
The helicities in \eqref{eq:cool} are the helicities of the outgoing particles of the HAs, with 
$\la$ for $K_J(\la)$ in $H^{B \to K_{J_K}}$ and $\la_\ell = \la_1 - \la_2$ for $\lone(\la_1) \ltwobar(\la_2)$ 
in $\mathcal L^{\ga_{J_\ga} \to \lone \ltwobar}$.  This is the main idea of the formalism: the angular dependence 
from the ingoing to outgoing particle is governed by the Wigner $D$-function, e.g. 
$\bar \gpolt^{J_{\ell ,\la }} = \bWignerD{J_\ga}{\la}{\la_\ell}{\Omega_\ell } \bar \gpolt^{J_{\ell ,\la_\ell} }$ for 
$\mathcal L^{\ga_{J_\ga}(\la) \to \lone(\la_1) \ltwobar(\la_2)}$, which is inherent in \eqref{eq:one}.
The generalised HA then becomes essentially a sum over all spin 
components $J_\ga$ necessary to saturate the Lorentz indices in the effective Hamiltonian,
\begin{eqnarray}
\label{eq:amp3}
 \amp(\bar{\Bin} \to \bar{\Ks}_{J_{\Ks}} (\to K \pi)   \lone \ltwobar) =  \frac{\sqrt{2 J_{K} + 1}}{4 \pi}    \;\; \sum_{J_\ga=0}^n 
\sum_{\la = - {\rm min}(J_\ga, J_{\Ks} )}^{ {\rm min}(J_\ga, J_{\Ks} ) } 
 \WignerD{J_K}{\la}{ 0}{\Omega_{\Ks}} 
\WignerD{J_\ga}{\la}{\la_\ell}{\Omega_\ell}  \amp^{J_\ga}_{\la, \la_1,\la_2}  \;,
\end{eqnarray}
where the overall factor follows from \eqref{eq:one}. 
A  schematic representation of equation~\eqref{eq:amp3} is given in figure \ref{fig:schematic}.
The differential decay distribution \eqref{eq:dG4} is replaced by a similar expression \begin{eqnarray}
& &  \frac{d^4 \Gamma}{dq^2  d\cL d \cK d\hel} \sim \sum_{\la_1 \la_2} |\amp|^2   
=\frac{2 J_{K} + 1}{4 \pi}  \sum_{\la_1 \la_2} \sum_{J_\ga \la  }  \sum_{J_\ga' \la'  }  
  \times  \nonumber  \\[0.1cm] 
& &   
 \amp^{J_\ga}_{\la, \la_1,\la_2} \bar{\amp}^{J_\ga'}_{\la' ,\la_1,\la_2} 
 \WignerD{J_K}{\la}{ 0}{\Omega_{\Ks}}   \bWignerD{J_K}{\la'}{ 0}{\Omega_{\Ks}} 
\WignerD{J_\ga}{\la}{\la_\ell}{\Omega_\ell} \bWignerD{J_\ga}{\la'}{\la_\ell}{\Omega_\ell}  \;,
\end{eqnarray}
with additional coherent sums over the spins $J_\ga$
\begin{equation}
  \sum_{\la_1 \la_2} = \sum_{\la_1,\la_2 = -1/2 }^{1/2}  \;, \quad  \sum_{J_\ga \la  } =   \sum_{J_\ga=0}^2   \sum_{\la = - \min(J_\ga,J_{K_J})}^{\min(J_\ga,J_{K_J})}
\end{equation}
and likewise for the sum over $J_\ga',\la'$.

\section{Angular distribution and Wigner $D$-functions}
\label{sec:angularD}

We now apply the method introduced in the previous section to decays governed by 
the  $ b \to s\lone \ltwobar$  effective Hamiltonian \eqref{eq:Heffexplicit}. First we consider the decay   $ \bar B \to \bar K^{*} \left( \to \bar K\pi \right) \lone \ltwobar$, and then in section~\ref{sec:BKll} we present similar results for $ \bar B \to \bar  K \lone \ltwobar$. The related decay $ \Lambda_b \to \Lambda \left( \to N\pi \right) \lone \ltwobar$, where $N = (p,n)$, can also be treated within this formalism, and will be briefly considered in appendix~\ref{app:baryons}.

\subsection{$ \bar B \to \bar K^{*} \left( \to \bar K\pi \right) \lone \ltwobar$ }

The use of the effective Hamiltonian  \eqref{eq:Heffexplicit} in the LFA 
restricts the partial waves to $J_\ga =0,1$ terms in equation~\eqref{eq:cool}. 
The discussion of higher partial  waves ( $J_\ga \geq 2$) is deferred to section~\ref{sec:limit}.
The   matrix element for \eqref{eq:Heffexplicit} is then given by the sum of an $S_\ell$- and $P_\ell$-wave amplitude (with the subscript $\ell$ 
 referring to the partial wave in the angle $\thetal$):
\begin{align}
\label{eq:MKs}
\hat{ \mathcal{M}}_{\la_1 , \la_2} &=  \matel{\bar K^*(\to \bar K\pi) \lone(\la_1) \ltwobar(\la_2)}{\hat{H}^{\rm eff}}{\bar B} \nonumber  \\[0.1cm]
&= \frac{\sqrt{3}}{4 \pi} \left[ \famp^0_{0, \la_1 , \la_2}  \WignerD{1}{0}{0}{\Omega_K} \delta_{\la_1 \la_2} + \sum_{\la=\pm,0}\famp^1_{\la, \la_1 , \la_2} \WignerD{1}{\la}{0}{\Omega_K}\WignerD{1}{\la}{\la_\ell}{\Omega_\ell} \right] \; ,
\end{align}
where the hat denotes the effective Hamiltonian without the $c_H$ prefactor 
\eqref{eq:Heffexplicit}.  
There is no $D$-wave since the two-indices in the tensor operator \eqref{eq:Heffexplicit} 
are antisymmetric and therefore in a spin $1$ representation (cf. discussion in section 
\ref{sec:higher} on higher spin operators).
The $K^*$ has spin $1$ and is therefore 
always in a $P_K$-wave in the $\theta_K$-angle, with analogous meaning for the $K$ subscript as before.
Above we have used $ \WignerD{0}{0}{\la_\ell}{\Omega} = \delta_{0\la_\ell}$ to impose $\delta_{\la_1 \la_2}$ on the scalar part of the matrix element. The principal objects to be calculated are the  amplitudes $\famp^{J_\ga}_{\la ,\la_1, \la_2}$. 
For $H^{\rm eff}$ \eqref{eq:Heffexplicit}  the $S_\ell$-  and $P_\ell$-wave  amplitudes (that is to say $\famp^0$ and $\famp^1$ respectively)  are written as
\begin{align}
\label{eq:ABKsll}
\famp^0_{0, \la_1 , \la_2}&=H^S\mathcal{L}^S_ {\la_1 , \la_2}+H^P \mathcal{L}^P_ {\la_1 , \la_2} \; ,\nonumber  \\
\famp^1_{\la , \la_1 , \la_2}&= - H_{\la}^V  \mathcal{L}^V_{\la_1 , \la_2}- H_{\la}^A \mathcal{L}
^A_ {\la_1 , \la_2}+ H^{T}_{\la}  \mathcal{L}^{T}_{\la_1 , \la_2} - 2H^{T_t}_{\la} 
\mathcal{L}^{T_t}_{\la_1 , \la_2} \; , 
\end{align}
with the relative signs and factor of 2 emerging from the (double) completeness relation \eqref{eq:double}, and the leptonic and the hadronic HAs are 
\begin{equation}
\label{eq:sketch}
 H^X_\la = \matel{\bar K^*(\la)}{\bar s \overline \Gamma^X b}{\bar B}  
  \;, \quad 
   {\mathcal L}^X_{\la_1 \la_2} \equiv 
  \matel{\lone(\la_1) \ltwobar(\la_2)}{ \bar \ell  \Gamma^X \ell   }{0}  \;,
  \end{equation}
 the expressions in \eqref{eq:helmatel} contracted with the corresponding polarisation vectors.
 Explicit results, as well as a more precise prescription concerning $\Gamma^X$, are given 
in Appendices \ref{app:leptonHAs} and \ref{app:HadronHAs} in  Eqs.~\eqref{eq:helmatellep} and \eqref{eq:HadronicHAdef}     respectively. 
\begin{table}[t]
\centering
\label{tab:Gammas}
\begin{tabular}{| c  | c || c | c | }
\hline
& $\Gamma^{S[P]}$     & $\Gamma^{V[A]}$  & $\Gamma^{T[T_t]}$  \\ \hline
$\Gamma^X$  & ${\mathbf 1}_4[ \ga_5] $   & $ \gamma^\mu[ \ga_5]  \gpol_\mu(\la_X)$ & $ \sigma^{\mu \nu}  \gpol^{1,\la_X}_{\mu \nu}[\gpol^{t,\la_X}_{\mu \nu}]$ \\ 
$J_\ga(X)$     & 0  & 1 & 1  \\ \hline
\end{tabular}
\caption{\small{The definitions of the $\Gamma^X$ and their associated spin $J_\ga(X)$. 
The contributions $J_\ga(X) = 0,1$ give rise to the $S_\ell$- and $P_\ell$-wave amplitudes respectively. 
The basic polarisation vector $\gpol_\mu$ is given in \eqref{eq:gpol} and the composed ones can be found in equation~\eqref{eq:completenessdecomposition}. 
The precise value of the helicity index $\la_X$ is specified when the leptonic and hadronic HAs are defined in Eqs.~(\ref{eq:helmatellep},\ref{eq:HadronicHAdef},\ref{eq:HXBK}).
Note that the additional structure $\Gamma^{T_5} =  \sigma_{\mu \nu} \gamma_5 $ can be absorbed into the other tensor structures due to the identity $\sigma^{\alpha \beta} \gamma_5 = -\frac{i}{2} \epsilon^{\alpha \beta \mu \nu} \sigma_{\mu \nu}$ 
(with the $\epsilon_{0123}=+1$  convention for the Levi-Civita tensor). Timelike contributions $\gamma^\mu [\ga_5] \gpol_\mu(t)$ can be absorbed into $\Gamma^{S,P}$ respectively, as detailed in appendix~\ref{app:HadronHAs}. Above $\sigma_{\mu \nu} = i/2 [\ga_\mu,\ga_\nu]$.} 
 } 
\end{table}
Squaring the matrix element in \eqref{eq:MKs}, summing over external helicities and averaging over final-state spins, one obtains an angular distribution 
\begin{align}
\label{eq:d4GKs}
I_{K^*}  \left( q^2, \Omega_K, \Omega_\ell \right) &\equiv \frac{32\pi}{3} \frac{d^4 \Gamma}{dq^2 \; d\textrm{cos}\thetal \; d\textrm{cos}\theta_K \; d \hel } = 
\frac{32\pi}{3} \NN  \sum_{\la_1, \la_2} \left |\hat{ \mathcal{M}}_{\la_1 , \la_2} \right|^2  \;, \quad  
\end{align}
with $I_{K^*}$ being a shorthand and $32 \pi/3$ is a convenient normalisation factor.  
The factor $\NN$,
\begin{equation}
\label{eq:NN}
 \NN \equiv  |c_H |^2  \kappa_{\rm kin} \;, \quad 
\kappa_{\rm kin}   \equiv \frac{\sqrt{\KallenB}\sqrt{\Kallengas}}{2^{6} \pi^3 m_B^3 q^2}  \;,
\end{equation}
is the product of the prefactor resulting from the effective Hamiltonian $c_H$ \eqref{eq:Heffexplicit}   and the kinematic phase space factor.
The  matrix element is defined in \eqref{eq:MKs}.
Above  $\KallenB \equiv \la(m_B^2, m_{K^*}^2, q^2) $ and $\Kallengas \equiv \la(q^2, m_{\ell_1}^2,m_{\ltwo}^2) $ where $\la (a,b,c)$ is the K\"all\'en-function defined in \eqref{eq:Kallen} and related to 
the absolute value of the three-momentum of the $K^*$ and the lepton pair by \eqref{eq:Kallenmomentum}.

\subsubsection{Angular Distribution}
\label{sec:angdistKstar}
The squared matrix element initially contains a plethora of different products of four Wigner functions. However, these correspond to  pairs of direct products that can be reduced to single Wigner functions by the Clebsch-Gordan series 
\begin{equation}
\label{eq:CGseries}
\WignerD{j}{m}{n}{\Omega} \WignerD{l}{p}{q}{\Omega} =  
\sum_{J=|j-l|}^{j+l} \sum_{M=-J}^J \sum_{N=-J}^J C^{Jjl}_{Mmp}  
C^{Jjl}_{Nnq} \WignerD{J}{M}{N}{\Omega} \; .
\end{equation}
Applied separately over the angles $\Omega_K = (0, \theta_K, 0)$ and $\Omega_\ell = (\hel, \thetal, - \hel)$, along with the identity
$\bWignerD{l}{m}{m'}{\Omega}= (-1)^{m'-m} \WignerD{l}{-m}{-m'}{\Omega}$, this allows the angular distribution to be written in the compact form
\begin{align}
\label{eq:angdistG}
 \IKsF \left( q^2, \Omega_K, \Omega_\ell \right) =  \Rea  \Big[
&   \GKstar{0}{0}{0}(q^2) \DD{0}{0}{0}  +\GKstar{0}{1}{0}(q^2)  \DD{0}{1}{0}   +  \GKstar{0}{2}{0}(q^2) \DD{0}{2}{0}  +  {}\nonumber \\
  & \GKstar{2}{0}{0}(q^2) \DD{2}{0}{0}  + \GKstar{2}{1}{0}(q^2)  \DD{2}{1}{0}  +  \GKstar{2}{1}{1}(q^2) \DD{2}{1}{1}  + {}\nonumber \\
   &   \GKstar{2}{2}{0}(q^2) \DD{2}{2}{0}   +  \GKstar{2}{2}{1}(q^2)  \DD{2}{2}{1} + \GKstar{2}{2}{2}(q^2)  \DD{2}{2}{2}  \quad \Big] \; , 
\end{align} 
where the superscript $(0)$ is a reminder that only $S_\ell$- and $P_\ell$-wave contributions were used 
to describe the amplitude \eqref{eq:MKs}.
The angular functions $\Omega$  are given in terms of Wigner $D$ functions
\begin{equation} 
\label{eq:Omega}
\DD{l_K}{l_\ell}{m} \equiv \DD{l_K}{l_\ell}{m}\left( \Omega_K, \Omega_\ell \right) \equiv \WignerD{l_K}{m}{0}{\Omega_K}\WignerD{l_\ell}{m}{0}{\Omega_\ell} = 
\WignerD{l_K}{m}{0}{\Omega_K'}\WignerD{l_\ell}{m}{0}{\Omega_\ell'}  \;.
\end{equation}
The variables $\Omega_K' = (\hel, \theta_K, -\hel)$ and  $\Omega_\ell' = (0, \thetal, 0)$ form an angular reparametrisation that will prove convenient when we discuss partial moments. 
The label $l_K$ corresponds to the $(K\pi)$-system, $l_\ell$ to the dilepton system, and the common index $m$ is the azimuthal component $\hel$ of either partial wave. 
The observables  $\GKstar{l_K}{l_\ell}{m}$ are functions of $q^2$ and the relation 
to  the standard observables in the literature is given in section~\ref{sec:relation}.
The explicit Wigner $D$-functions used above are given by
\begin{alignat}{3}
\label{eq:Wex}
& \WignerD{0}{0}{0}{\Omega} = 1 \;, \quad & & \WignerD{2}{0}{0}{\Omega} = \frac{1}{2} \left( 3 \cos^2 \theta -1 \right)\; ,  \quad 
 & & \WignerD{2}{2}{0}{\Omega} = \sqrt{\frac{3}{8}} e^{-2 i \hel} \sin^2 \theta  \; , \nonumber \\
& \WignerD{1}{0}{0}{\Omega} = \cos \theta \; , \quad  & & 
\WignerD{1}{1}{0}{\Omega} = -\frac{1}{\sqrt{2}} e^{-i \hel} \sin  \theta\; , \quad
& & \WignerD{2}{1}{0}{\Omega} = -\sqrt{\frac{3}{8}} e^{-i \hel} \sin 2\theta \; ,
\end{alignat}
and can be related to 
spherical harmonics $Y_{lm}\left( \theta, \hel \right)$ or  
associated Legendre polynomials $P_{lm}(x)$ as
\begin{equation}
\WignerD{l}{m}{0}{\hel, \theta, -\hel} = \sqrt{\frac{4 \pi}{ 2l+1} } \overline Y_{lm} \left( \theta, \hel \right) = \sqrt{\frac{(l-m)!}{(l+m)!}} P_{lm}( \cos \theta ) e^{-i m\hel} \; .  
\end{equation}

We comment briefly on four features of the angular distribution \eqref{eq:angdistG},  
all of which are encoded by the double Clebsch-Gordan series  
\eqref{eq:CGseries} but which can also be seen to emerge from the  
underlying physics:
\begin{itemize}
\item{The second helicity index of all Wigner $D$-functions in the  
angular distribution is zero.  The latter is the difference of the helicities 
of the final-state particles, which is zero since these helicities are summed incoherently, $(\la_1 - \la_2) - (\la_1 - \la_2) =0$.}
\item{The first helicity index $m$ is identical in all pairs of Wigner  
$D$-functions appearing in the angular distribution. This index  
contains the helicities of the internal particles, summed coherently.  
One can also see this as a property of the freedom of defining the  
reference plane for the angle $\hel$.}
\item{The range of the indices $l_K$ and $l_\ell$ is fixed 
between the range $0, \dots, 2 \; {\rm max}[ J_{K,\ell}]$. Including only  
$J_\ga \leq1 $ contributions emerging from the dimension-six effective  
Hamiltonian \eqref{eq:Heffexplicit} hence imposes $0 \leq l_\ell \leq  
2$, and likewise $J_{K}  = 1$ imposing  $0 \leq l_K \leq  2$. }
\item{The absence of angular structures with $l_K = 1$ is specific to  
this decay, due to the final state consisting of (pseudo)scalar mesons.}
\end{itemize}
The first three features are universal to such decay chains  
and apply even if some of the particles involved are fermions, for  
example in the decay $\Lambda_b \to \Lambda \left( \to (p,n) \pi \right)  \lone \ltwobar $, see appendix~\ref{app:baryons}.

\subsection{Relation of the $\GKstar{l_K}{l_\ell}{m}$ to standard literature observables}
\label{sec:relation}

The functions $\GKstar{l_K}{l_\ell}{m}$, omitting the explicit $q^2$-dependence hereafter,  are defined in terms of the standard basis of observables $\J_i (q^2)$ parametrised 
in  \eqref{eq:d4GJi} by
\begin{alignat}{3}
\label{eq:GasJ}
& \GKstar{0}{0}{0}= \frac{4}{9} \left( 3 \left( \J_{1c} + 2 \J_{1s} \right) - \left( \J_{2c} + 2 \J_{2s} \right) \right) \; , \quad
& & \GKstar{0}{1}{0} = \frac{4}{3} \left( \J_{6c} + 2 \J_{6s}  \right)\; , \quad
& & \GKstar{0}{2}{0} = \frac{16}{9} \left(   \J_{2c} + 2 \J_{2s} \right) \; , \nonumber \\
& \GKstar{2}{0}{0} = \frac{4}{9} \left( 6 \left(\J_{1c} - \J_{1s}\right) - 2 \left(\J_{2c} -  \J_{2s} \right) \right)\; , \quad
& & \GKstar{2}{1}{0} = \frac{8}{3} \left(  \J_{6c} - \J_{6s}  \right)\; , \quad
& & \GKstar{2}{2}{0} = \frac{32}{9} \left( \J_{2c} - \J_{2s} \right) \; , \nonumber \\
& \GKstar{2}{1}{1} = \frac{16}{\sqrt{3}} \mathcal{G}_5 \; , \quad
& & \GKstar{2}{2}{1} = \frac{32}{3} \mathcal{G}_4 \; , \quad
&  & \GKstar{2}{2}{2} = \frac{32}{3} \mathcal{G}_3 \; , 
\end{alignat}
where we have defined
$\mathcal{G}_{3,4,5} \equiv  \left(\J_{3,4,5} + i \J_{9,8,7} \right)$.

The twelve quantities \eqref{eq:GasJ}, keeping in mind that the last three are complex, have been rewritten 
in several ways in the literature. A frequently-used form is the set of observables given in \cite{DHMV-Observables}, constructed to be insensitive to form factors. 
In the notation of LHCb \cite{LHCb-P5p}, which includes their, and therefore our, angular conventions, the observables are given in 
terms of $\GKstar{l_K}{l_\ell}{m}$ by:\footnote{The extension of these relations to CP-odd and CP-even 
combinations, in the spirit of \cite{ABBBSW2008}, is straightforward, see section 4 of  \cite{DHMV-Observables}.}
\begin{alignat}{2}
\label{eq:O1}
& \binavg{P_1}  \Big|_{\rm LHCb} = \frac{\binavg{\Rea \left[\GKstar{2}{2}{2} \right] } }{ \Nbinz} \; , \quad
&& \binavg{P_2} \Big|_{\rm LHCb}   = \frac{ \binavg{2\GKstar{0}{1}{0} -
\GKstar{2}{1}{0}}}{ 3\Nbinz}  \; ,  
\nonumber \\
& \binavg{P_3} \Big|_{\rm LHCb}   = \frac{\binavg{\Ima \left[\GKstar{2}{2}{2} \right] } }{2 \Nbinz} \;,  & &  \nonumber \\
& \binavg{P_4'} \Big|_{\rm LHCb}   =\frac{\binavg{\Rea \left[  
\GKstar{2}{2}{1} \right]}  }{4 \Nbin    } \; , \quad
&& \binavg{P_8'} \Big|_{\rm LHCb}  = \frac{\binavg{ \Ima \left[  
\GKstar{2}{2}{1} \right]} }{4 \Nbin  } \; , \nonumber \\
&\binavg{ P_5' } \Big|_{\rm LHCb}  = \frac{\binavg{ \Rea\left[  
\GKstar{2}{1}{1} \right]} }{2\sqrt{3}\Nbin} \; , \quad
&& \binavg{ P_6' }  \Big|_{\rm LHCb}  = \frac{\binavg{\Ima\left[  
\GKstar{2}{1}{1} \right]} }{2\sqrt{3}\Nbin} \; , 
\end{alignat}
where we defined
\begin{equation*}
\binavg{f(q^2)} = \intbin f(q^2) \;,
\end{equation*}
as the integral over $q^2$ bins of the observable of interest, and\footnote{In terms of the $\J_i(q^2)$ basis, $\Nbinz = \frac{64}{3}  \binavg{\J_{2s}} $ and $\Nbin = \frac{16}{3} \sqrt{- \binavg{\J_{2c}} \binavg{\J_{2s}}} $.}
\begin{equation}
\label{eq:O2}
\Nbinz \equiv 4  \binavg{\GKstar{0}{2}{0} -  
\frac{1}{2} \GKstar{2}{2}{0}}  \;,\quad   \Nbin= \sqrt{- \binavg{\GKstar{0}{2}{0} - \frac{1}{2} \GKstar{2}{2}{0}}  
\binavg{\GKstar{0}{2}{0} +  
\GKstar{2}{2}{0}} } \; .
\end{equation}
Three other combinations of the $\GKstar{l_K}{l_\ell}{m}$ can be related to the branching fraction $\frac{d\Gamma}{dq^2}$, the forward-backward asymmetry $A_{\rm FB}$ and the longitudinal polarisation fraction $F_L$ \cite{BDH2013}: 
\begin{alignat}{2}
\label{eq:O3}
&\binavg{\frac{d\Gamma}{dq^2}} = \frac{3}{4}  \binavg{\GKstar{0}{0}{0}} \; , \quad 
&& \binavg{A_{\rm FB}}  \Big|_{\rm LHCb}  =  \frac{1}{2} \frac{ \binavg {\GKstar{0}{1}{0}} }{\binavg{\GKstar{0}{0}{0}}   }  \; , \nonumber \\
& \binavg{F_L} = \frac{\binavg{\GKstar{0}{0}{0}}+\binavg{\GKstar{2}{0}{0}}}{3\binavg{\GKstar{0}{0}{0}}} \; . \quad
&&
\end{alignat}
The observables in Eqs.~(\ref{eq:O1},\ref{eq:O2},\ref{eq:O3}) 
correspond to the twelve $\J_i$. The definitions of the $P_i'$ above correspond to those used by LHCb \cite{LHCb-P5p}; we give the  correspondence to the observables defined in \cite{DHMV-Observables} in appendix~\ref{app:ang-conventions}.

\subsection{$ \bar B \to \bar K  \lone \ltwobar $ }
\label{sec:BKll}
Having shown the $ \bar B \to \bar K^*  \lone \ltwobar $ HA analysis in detail we are going 
to be rather brief on $ \bar B \to \bar K  \lone \ltwobar $.
Skipping the step in \eqref{eq:helamp} we directly write down  
the $S_\ell$- and $P_\ell$-wave amplitudes (analogue of equation~\eqref{eq:ABKsll}):
\begin{align}
\label{eq:ABKll}
\famp^0_{0, \la_1 , \la_2}&=\HSK \mathcal{L}^S_{\la_1 , \la_2}+ 
\HPK \mathcal{L}^P_{\la_1 , \la_2} 
 \; , \nonumber  \\
\famp^1_{0, \la_1 , \la_2}&=- \HVK \mathcal{L}^V_{\la_1 , \la_2}- \HAK \mathcal{L}^A_{\la_1 , \la_2}+ \HTensK  \mathcal{L}^{T}_{\la_1 , \la_2} - 2\HTenstK  \mathcal{L}^{T_t}_{\la_1 , \la_2} \; ,
\end{align}
where the $\mathcal{L}^X_{\la_1 , \la_2}$ are the same as in the $ \bar B \to \bar K^*  \lone \ltwobar $ decay, and the hadronic HAs are taken over the same set of operators, but defined instead for $ \bar B \to \bar K$ transitions. We again refer the reader to appendix~\ref{app:branch} for a clarification of the signs and factor of 2 that emerge from the (double) completeness relation.

The reduced matrix element is then the sum of the $S_\ell$- and $P_\ell$-wave amplitude 
\begin{eqnarray}
\label{eq:MK}
   \hat{\mathcal{M}}_{\la_1 , \la_2} =  \frac{1}{\sqrt{4 \pi}} \left( \famp^0_{0, \la_1 , \la_2}  \delta_{\la_1 \la_2} + \famp^1_{0, \la_1 , \la_2} \WignerD{1}{0}{\la_\ell}{\Omega_\ell} \right) \; ,
\end{eqnarray}
where $\Omega_\ell = (0,\thetal,0)$ in this case.
 The angular distribution (with $0 \leq \thetal \leq \pi$) is given by squaring the matrix element
\begin{eqnarray}
\label{eq:angdistg}
I_K(q^2, \thetal) \equiv   \frac{d^2 \Gamma}{dq^2 \; d\textrm{cos}\thetal} &=& \NN \sum_{\la_1, \la_2}  \left | \hat{\mathcal{M}}_{\la_1 , \la_2} \right|^2 \;.
\end{eqnarray}
Using \eqref{eq:MK} one obtains
 \begin{alignat}{3}
\label{eq:angdistg2}
\IKF &= \GK{0}(q^2) &\;+\;& \GK{1}(q^2) \WignerD{1}{0}{0}{\Omega_\ell} &\;+\;& \GK{2}(q^2)  \WignerD{2}{0}{0}{\Omega_\ell} \nonumber \\[0.1cm]
   &= \GK{0}(q^2)  &\;+\;& \GK{1}(q^2) P_1(\cos\thetal)  &\;+\;&
\GK{2}(q^2) P_2(\cos\thetal) \nonumber \\[0.1cm]
&= \GK{0}(q^2)  &\;+\;& \GK{1}(q^2) \cos \thetal &\;+\;& \GK{2}(q^2) \frac{1}{2} \left( 3 \cos^2 \thetal -1 \right) \;,
\end{alignat}
where we used $P_l(\cos \thetal) = \WignerD{l}{0}{0}{\Omega_\ell}$ and 
 $\WignerD{0}{0}{0}{\Omega_\ell} =1$. For convenience, we have 
given results in terms of the explicit angle $\thetal$ using  equation~\eqref{eq:Wex}.
The superscript $(0)$ is  again a reminder that the restriction to $l_\ell  \leq 2$ 
is a consequence of only including  $S_\ell$- and $P_\ell$-waves  in \eqref{eq:MK}.
The explicit functions $\GK{0,1,2}$, whose $q^2$-dependence we omit hereafter,
 are given in appendix~\ref{app:BtoK} in equation~\eqref{eq:gi} in terms of 
HAs.\footnote{The observables $\GK{l_\ell}$ and the angular coefficients used in the literature \cite{BHP2007} are related by
$a(q^2) = \GK{0} - \frac{1}{2}\GK{2} $, $b(q^2) = \GK{1}$ and  $ c(q^2) = \frac{3}{2} \GK{2} $ 
where $\IKF = a + b \cos \thetal + c \cos^2 \thetal$.} 

With respect to the parametrisation of the angular distribution used in the experimental community, \cite{angularK14}
\begin{equation}
\label{eq:exp}
\frac{1}{\Gamma} \frac{d \Gamma}{ d \cos \theta_\ell} = \frac{3}{4} \left ( 1 - F_H \right )(1- \cos^2 \theta_\ell )+ \frac{1}{2} F_H + A_{\rm FB} \cos \theta_\ell \; ,
\end{equation}
the relationship to the $\GK{i}$ in \eqref{eq:angdistg2} is given by 
\begin{equation}
\Gamma = 2 \vev{ \GK{0}} \;, \quad A_{\rm FB} = \sigma_X  \frac{\vev{ \GK{1}}}{2 \vev{ \GK{0}}} \;,
\quad F_H = \frac{ \vev{  \GK{0}}  +  \vev{\GK{2}} } {  \vev{\GK{0}} } \;,
\end{equation}
where $\vev{X} = \int dq^2 X $ denotes the integration or appropriate binning over $q^2$ 
and $\sigma = \pm 1$ depending on the conventions.\footnote{In our conventions by definition $\sigma_{\rm GHZ} =1$ and the translation to the
 LHCb conventions \cite{angularK14} are as follows
$\sigma_{\rm GHZ} = \sigma(B^{\pm})$ and $\sigma_{\rm GHZ} = -\sigma(B^0, \bar{B}^0)$. 
The charged and neutral decays are different because the neutral mode, being observed in $K_S$, 
is not self-tagging.  Comparing with the theory paper \cite{BHP2007} we find 
$\sigma_{\rm GHZ} = - \sigma_{\rm BHP } $ for both charged and neutral modes.}

\section{Method of total and partial  Moments}
\label{sec:Moments}

The MoM is a powerful tool to extract the angular observables  
$\GKstar{l_K}{l_\ell}{m}$ by the use of orthogonality relations. 
In $B$ physics, for example, the method has been applied to 
$B \to J/\Psi(\to \bar \ell \ell) K^*(\to K \pi)$ type decays \cite{Dighe:1998vk} 
during the first $B$-factory era. 

In experiment the angular information on  $B \to K^* \ell \ell$  has been extracted through 
the likelihood fit method, at the level of  
$\IKsF$ \cite{LHCbBKmumu2015}, and it has also been suggested for analysis at the amplitude level \cite{Egede:2015kha}.
A possible advantage of the MoM over the likelihood fit
is that it is less sensitive to theoretical assumptions. 
More precisely, one can test each angular term independent of the rest of the distribution.
Generically  the fourfold angular distribution can be expanded
 over the complete set of functions $ \DD{l_K}{l_\ell}{m}$ \eqref{eq:Omega}
\begin{equation}
\label{eq:Igen}
  I_{K^*} \left( q^2, \Omega_K, \Omega_\ell \right) = 
 \sum_{l_K , l_\ell \geq 0} \sum_{m =0}^{\text{min}(l_K, l_\ell)}
  \Rea  \left[  \GKstar{l_K}{l_\ell}{m}  \DD{l_K}{l_\ell}{m}(\theta_K,\thetal,\phi) \right] \;,
\end{equation}
of which the distribution  $\IKsF$ \eqref{eq:angdistG} is a subset. 
Note that the sum over $m$ does not need to be continued for negative values since 
 $I_{K^*}$ is real-valued.
By using the orthogonality properties of the Wigner $D$-functions  (e.g. \cite{Hecht})
with $\Omega = (\al,\be,\ga)$
\begin{equation}
\label{eq:Schur}
\int_{-1}^1 d\cos\beta \int_{0}^{2\pi}  \textrm{d}\alpha \int_{0}^{2\pi}    
\textrm{d}\gamma\; \WignerD{j}{m}{n}{\Omega}  
\bWignerD{l}{p}{q}{\Omega}  = \frac{8 \pi^2}{2 j+1} \delta_{jl}  
\delta_{mp} \delta_{nq} \;,
\end{equation}
the MoM allows to extract
the observables $\GKstar{l_K}{l_\ell}{m} $ from the angular distribution.
In particular one can test for the absence of all higher moments
and therefore test very specifically the assumptions made when deriving the distribution 
$\IKsF$ \eqref{eq:angdistG}. 
We refer to this method as the method of (total) moments or simply MoM with 
results given in  section~\ref{sec:MotM}. 
Integrating over a subset of angles, referred to as partial moments, is discussed in section 
\ref{sec:pm}. In the latter case orthogonality does not hold in the generic case and different 
$\GKstar{l_K}{l_\ell}{m} $ enter the same moment. 

Elements of the MoM have previously been applied to $\Lambda_b \to \Lambda \left( \to (p,n) \pi \right) \lonebar \ltwo $ \cite{BDF2015} and more systematically to the other channels discussed in this paper, crucially including a study of how to account for detector-resolution acceptance effects, in  \cite{BCDS2015}. 
Our study differs from the latter in that we start at the level of the HAs, and obtain the distribution 
\eqref{eq:Igen} through a direct computation, whereas the other studies proceed backwards and  directly expand 
the decay distribution in the orthogonal basis of associated Legendre polynomials. Our approach is therefore advantageous in that it provides additional insight, by clarifying the structure of the decay distribution  \eqref{eq:angdistG} and what 
type of physics goes beyond it. This is an aspect we return to in section~\ref{sec:limit}.

\subsection{Method of total Moments}
\label{sec:MotM}

In order to condense the notation slightly we define the scalar product
\begin{alignat}{3}
\label{eq:SPfull}
&\SPththc{f(\Omega) } {g(\Omega) }  & &\equiv \qquad&  &\frac{1}{8 \pi} \int_{-1}^{1}  d\cos    \theta_K \;\int_{-1}^{1}   d\cos \thetal \;\int_0^{2\pi}   d\hel \bar f(\Omega) g(\Omega)  \; , 
\end{alignat}
 normalised such that $\SP{1}{1} =1$. 
Using $\SPththc{f(\Omega) } {g(\Omega) } $ we can thus extract all observables $ \GKstar{l_K}{l_\ell}{m}$ separately from each other, by taking moments\footnote{The moments 
$\MM{l_K}{l_\ell}{m} $ and the quantities $S_{l_\ell,l_K,m}$ introduced in \cite{BCDS2015} are related 
as follows: $ 8 \pi  \GKstar{0}{0}{0} S_{l_\ell, l_K, m} =  \GKstar{l_K}{l_\ell}{m}  =  
 \MM{l_K}{l_\ell}{m}/ \cc{l_K}{l_\ell}{m} $.}
\begin{equation}
\label{eq:M}
\MM{l_K}{l_\ell}{m}   \equiv \SPththc{\DD{l_K}{l_\ell}{m} } {I_{K^*}(q^2,\Omega_K, \Omega_\ell) } = 
\cc{l_K}{l_\ell}{m}  \GKstar{l_K}{l_\ell}{m}  \;,
\end{equation}
where 
\begin{equation}
\label{eq:c}
 \cc{l_K}{l_\ell}{m}= \frac{1+\delta_{m0}}{2( 2 l_K+1) (2 l_\ell+1)}  \;.
\end{equation}
Using the equation above  the terms in \eqref{eq:angdistG} are  given in Tab.~\ref{tab:moments}.
\begin{table}
\begin{center}
\begin{tabular}{| l |  l |  l |  l |  l |  l |  l |  l |  l |  l | }
\hline
 $\MM{l_K}{l_\ell}{m} $   &     $ \GKstar{0}{0}{0}$  &  $\frac{1}{3} \GKstar{0}{1}{0}$ & $\frac{1}{5} \GKstar{0}{2}{0}$ & 
  $ \frac{1}{5} \GKstar{2}{0}{0}$  &  $\frac{1}{15} \GKstar{2}{1}{0}$ & $\frac{1}{25} \GKstar{2}{2}{0}$ &
   $ \frac{1}{30} \GKstar{2}{1}{1}$  &  $\frac{1}{50} \GKstar{2}{2}{1}$ & $\frac{1}{50} \GKstar{2}{2}{2}$ 
\\ \hline
\end{tabular}
\end{center}
\caption{\small Moments $\MM{l_K}{l_\ell}{m} $ in terms of  $\GKstar{l_K}{l_\ell}{m}$ as defined by equation~\eqref{eq:M} with 
factor of proportionality  $\cc{l_K}{l_\ell}{m}$ evaluated with \eqref{eq:c}.}
\label{tab:moments}
\end{table}
Furthermore, the orthogonality condition also implies that
\begin{alignat}{3}
\label{eq:highermoments}
&\MM{j}{j'}{m} &=& \; 0 \; , \quad  & & \forall m \text{ and } j \geq 3 \text{ or }  j' \geq 3 \; , \nonumber \\
&\MM{1}{j'}{m} &=& \;0 \; ,   & & \forall j',m \;.
\end{alignat}
Hence the higher and $l_K=1$ moments vanish, providing a very specific test of the theoretical assumptions behind $\IKsF$.

\subsection{Partial Moments}
\label{sec:pm}

The results given previously show how to extract the individual $ \GKstar{l_K}{l_\ell}{m}$. 
We propose the method of partial moments whereby one integrates only over a subset of angles.  
The distributions might be  regarded as generalisations of uni- and double-angular distributions as these in turn can be viewed as  partial moments with respect to unity.
The method is effectively a hybrid between the likelihood fit and the
total MoM. 
To this end we define the further scalar products 
\begin{alignat}{3}
\label{eq:SPpartial}
&\SPthc{f(\Omega) } {g(\Omega) }{} & &\equiv \qquad&  &\frac{1}{4 \pi}  \int_{-1}^{1}  d\cos    \theta \;\int_0^{2\pi}   d\hel \bar f(\Omega) g(\Omega) \; , \nonumber \\ 
&\SPthth{f(\Omega) } {g(\Omega) }  & &\equiv \qquad&  &\frac{1}{4}  \int_{-1}^{1}  d\cos    \theta_K \;\int_{-1}^{1}   d\cos \thetal \bar f(\Omega) g(\Omega) \; ,
\end{alignat}
again normalised such that $\SP{1}{1} =1$. The orthogonality relation \eqref{eq:Schur} can then be rewritten as  
\begin{equation}
\label{eq:orthogonalSP}
 \SPthc{\WignerD{l}{p}{0}{\Omega_\ell} } {\WignerD{j}{m}{0}{\Omega_\ell} }{\ell}= \frac{1}{2 l+1} \delta_{jl}  \delta_{mp} \; .
\end{equation}

\subsubsection{Integrating over $\thetal,\phi$:  $\kpartial{l_\ell}{m}$-moments}
The partial moment over $\thetal$ and $\phi$ is defined and given by
\begin{equation}
\label{eq:partial1}
\kpartial{l_\ell}{m}    
\equiv \SPthc{\WignerD{l_\ell}{m}{0}{\Omega_\ell}}{I_{K^*}(q^2, \Omega_K, \Omega_\ell)}{\ell} 
=  \frac{1+\delta_{m0}}{2 \left(2 l_\ell+1\right)}\sum_{l_K \geq 0} \WignerD{l_K}{m}{0}{\Omega_K} \GKstar{l_K}{l_\ell}{m} \;
\end{equation}
 Assuming the distribution  \eqref{eq:angdistG} ($ l_K = 0,2$) 
there are six non-vanishing moments
\begin{alignat}{2}
\label{eq:partialKopen}
\kpartial{0}{0}  &=   \GKstar{0}{0}{0} + \GKstar{2}{0}{0} \WignerD{2}{0}{0}{\Omega_K}\;\; &=&\;  \GKstar{0}{0}{0}  +  \frac{1}{2} \left( 3 \cos^2 \theta_K -1 \right) \GKstar{2}{0}{0}   \; , \nonumber \\
\kpartial{1}{0}   &= \frac{1}{3} \left(\GKstar{0}{1}{0} + \GKstar{2}{1}{0} \WignerD{2}{0}{0}{\Omega_K}\right)\;\;  &=&\;  \frac{1}{3} \left(\GKstar{0}{1}{0}  +  \frac{1}{2} \left( 3 \cos^2 \theta_K -1 \right)  \GKstar{2}{1}{0}  \right) \; , \nonumber \\ 
\kpartial{2}{0}  &=\frac{1}{5}\left(\GKstar{0}{2}{0}  + \GKstar{2}{2}{0} \WignerD{2}{0}{0}{\Omega_K}  \right) \;\; &=&\;  \frac{1}{5} \left(\GKstar{0}{2}{0}  +  \frac{1}{2} \left( 3 \cos^2 \theta_K -1 \right)  \GKstar{2}{2}{0}  \right) \; , \nonumber \\
\kpartial{1}{1}  &= \frac{1}{6} \GKstar{2}{1}{1} \WignerD{2}{1}{0}{\Omega_K}\;\; &=&\; \frac{-1}{6} \sqrt{\frac{3}{8}}  \sin 2\theta_K \;
\GKstar{2}{1}{1}  \; ,  \nonumber \\
\kpartial{2}{1} &= \frac{1}{10} \GKstar{2}{2}{1} \WignerD{2}{1}{0}{\Omega_K}\;\; &=&\; \frac{-1}{10} \sqrt{\frac{3}{8}} \sin 2\theta_K  \; 
\GKstar{2}{2}{1} \; , \nonumber \\
\kpartial{2}{2} &= \frac{1}{10} \GKstar{2}{2}{2} \WignerD{2}{2}{0}{\Omega_K} &=&\; \frac{1}{10} \sqrt{\frac{3}{8}}  \sin^2 \theta_K \; \GKstar{2}{2}{2} \; ,
\end{alignat}
where we used $\WignerD{0}{0}{0}{\Omega_K} =1$.
As was the case in the MoM, with respect to the distribution $\IKsF$
higher partial moments vanish
\begin{equation}
\kpartial{l_\ell}{m} = 0 \; \quad  \forall l_\ell \geq 3 , \forall m \;.
\end{equation}

\subsubsection{Integrating over $\theta_K,\phi$:  $\lpartial{l_\ell}{m}$-moments} 

The partial moment over $\theta_K$ and $\phi$ is defined  in complete analogy with the previous partial moment \eqref{eq:partial1} by,
\begin{equation}
\label{eq:partial2}
l^{l_K}_m(\thetal)    
\equiv \SPthc{\WignerD{l_K}{m}{0}{\Omega_K'}}{I_{K^*}(q^2, \Omega_K', \Omega_\ell')}{K}  
=  \frac{1+\delta_{m0}}{2 \left(2 l_K+1\right)}\sum_{l_\ell \geq 0} \WignerD{l_\ell}{m}{0}{\Omega_K} \GKstar{l_K}{l_\ell}{m} \;,
\end{equation}
where we make use of the reparametrisation of angles given in \eqref{eq:Omega}.
Again assuming the distribution  \eqref{eq:angdistG} ($l_\ell=0,1,2$) 
there are four non-vanishing moments
\begin{align}
\label{eq:partialLopen}
\lpartial{0}{0} &= \GKstar{0}{0}{0} +  \GKstar{0}{1}{0} \WignerD{1}{0}{0}{\Omega_\ell'}+  \GKstar{0}{2}{0} \WignerD{2}{0}{0}{\Omega_\ell'} \nonumber \\
 &  = 
\GKstar{0}{0}{0} + \cos \thetal \GKstar{0}{1}{0}  + \frac{1}{2} \left( 3 \cos^2 \thetal -1 \right)  \GKstar{0}{2}{0} 
\;, \nonumber \\[0.1cm]
\lpartial{2}{0} &=\frac{1}{5} \left(\GKstar{2}{0}{0}  +  \GKstar{2}{1}{0} \WignerD{1}{0}{0}{\Omega_\ell'}+  \GKstar{2}{2}{0} \WignerD{2}{0}{0}{\Omega_\ell'}   \right)  \nonumber \\[0.1cm]
 &=\frac{1}{5} \left(  \GKstar{2}{0}{0} + \cos \thetal \GKstar{2}{1}{0}  +\frac{1}{2} \left( 3 \cos^2 \thetal -1 \right)  \GKstar{2}{2}{0}       \right) \;, \nonumber \\[0.1cm]
\lpartial{2}{1} &=  \frac{1}{10} \left(\GKstar{2}{1}{1} \WignerD{1}{1}{0}{\Omega_\ell'} +  \GKstar{2}{2}{1} \WignerD{2}{1}{0}{\Omega_\ell'}\right) \nonumber \\[0.1cm]
 & =   \frac{-1}{10\sqrt{2}}  \left( \sin  \thetal  \GKstar{2}{1}{1} + \sqrt{\frac{3}{4}}  \sin 2\thetal  \GKstar{2}{2}{1}  \right) \;,  \nonumber \\[0.1cm]
\lpartial{2}{2} &= \frac{1}{10}\GKstar{2}{2}{2} \WignerD{2}{2}{0}{\Omega_\ell'} =  
\frac{1}{10} \sqrt{\frac{3}{8}}  \sin^2 \thetal  \GKstar{2}{2}{2}   \; ,   
\end{align}
where we used $\WignerD{0}{0}{0}{\Omega_\ell'}=1$.
With respect to the distribution $\IKsF$
higher partial moments vanish
\begin{equation}
 \lpartial{l_\ell}{m}=0 \;, \quad  \forall l_K \geq 3 , \; \forall m \quad  \text{and} \quad  l_K =1, \; \forall m \;.
 \end{equation}
 
\subsubsection{Integrating over $\theta_K,\thetal$: $\partialphi{l_K}{l_\ell}{m}{m'}$-moments}

Finally, we can consider  projecting on to moments of the form $\WignerD{l}{m}{0}{\Omega_K}\WignerD{l'}{m'}{0}{\Omega_\ell'}$ with respect to $\theta_K$, $\thetal$. In this case the full orthogonality relation \eqref{eq:Schur} no longer holds, but due to \eqref{eq:CGseries} there exist selection rules as to which of the $\GKstar{l_K}{l_\ell}{m}$ can contribute to the partial moments 
\begin{equation}
\partialphi{l_K}{l_\ell}{m}{m'}\equiv \SPthth{\WignerD{l_K}{m}{0}{0,\theta_K,0}\WignerD{l_\ell}{m'}{0}{0,\thetal,0}}{I_{K^*}(q^2, \Omega_K, \Omega_\ell)} \; .
\end{equation}
Assuming  $\IKsF$ a few non-vanishing moments are
\begin{alignat}{4}
&\partialphi{0}{0}{0}{0}&& = \frac{1}{6} \left(6 \GKstar{0}{0}{0} + \Rea[  e^{-2i \hel} \GKstar{2}{2}{2}  ] \right)\; , \qquad
&& \partialphi{0}{1}{0}{0}&&= \frac{1}{3} \GKstar{0}{1}{0} \;,  \nonumber \\
&\partialphi{0}{2}{0}{0}&&= \frac{1}{30} \left(6 \GKstar{0}{2}{0}  - \Rea[  e^{-2i \hel} \GKstar{2}{2}{2}  ] \right)\;, 
&& \partialphi{2}{0}{0}{0}&&= \frac{1}{30} \left(6 \GKstar{2}{0}{0}  - \Rea[  e^{-2i \hel} \GKstar{2}{2}{2}  ] \right)\;,  \nonumber \\
&\partialphi{2}{1}{0}{0}&&= \frac{1}{15} \GKstar{2}{1}{0} \;,
&&\partialphi{2}{2}{0}{0}&&= \frac{1}{150} \left(6 \GKstar{2}{2}{0}  + \Rea[  e^{-2i \hel} \GKstar{2}{2}{2}  ] \right)\;, 
\nonumber \\
&\partialphi{2}{1}{1}{1}&&=\frac{1}{15} \Rea[  e^{-i \hel} \GKstar{2}{1}{1}  ]  \; , 
&&\partialphi{2}{2}{1}{1}&&=\frac{1}{25} \Rea[  e^{-i \hel} \GKstar{2}{2}{1}  ]   \; . 
\end{alignat}
A consequence of the fact that the full orthogonality of the Wigner functions has been lost 
is that higher moments contain lower $G$-functions. As an interesting example we quote
\begin{equation}
\partialphi {4} {1} {2} {0}    = \frac{1}{9 \sqrt{10}} \left(\GKstar{0}{1}{0} + \GKstar{2}{1}{0} \right) = \frac{4}{9 \sqrt{10}} \J_{6c}  \; .
\end{equation}
This quantity is of some interest since $\J_{6c} = 0$ in the SM,
 as it involves scalar and tensor operators at the level of the dimension-six effective Hamiltonian \eqref{eq:Heffexplicit}.

\section{Including higher Partial Waves}
\label{sec:limit}

The compact form of the angular distribution  $\IKsF$ \eqref{eq:angdistG} is a consequence of 
the LFA and the restriction to the $P_K$-wave in 
the $(K\pi)$-channel.  In this section we elaborate on the consequences of going beyond these approximations. 
The double partial wave expansion is outlined in section~\ref{sec:double} followed by 
a qualitative discussion of the effect of higher spin operators and the inclusion of electroweak effects in sections   
 \ref{sec:higher} and \ref{sec:beyond} respectively. 
In section~\ref{sec:aktuell} we emphasise how testing for higher moments can be used 
to diagnose the size of  QED corrections.
Throughout this section we change the notation from $\lone \ltwobar \to \ell^+ \ell^-$ 
for the sake of familiarity and simplicity.

\subsection{Double Partial Wave Expansion}
\label{sec:double}

In order to discuss the origin of 
generic terms in the full distribution \eqref{eq:Igen}, it is advantageous to return to the amplitude level.
Somewhat symbolically we may rewrite the amplitude \eqref{eq:amp3}, omitting the sum over $J_\ga$, as
\begin{equation}
{\cal A}(B \to K_J(\la)(\to K \pi) \ell^+(\la_1)\ell^-(\la_2)) = {\cal A}^{J_\ga,J_K}_{\la,\la_\ell}
\bWignerD{J_K}{\la}{0}{\Omega_K} \bWignerD{J_\ga}{\la}{\la_\ell}{\Omega_\ell}
\end{equation}
with $\la_\ell = \la_1 - \la_2$ as defined in \eqref{eq:lal}. The two opening angles $\theta_K$ and $\theta_\ell$ 
allow for two separate partial wave expansions. The partial waves in the $\theta_K$- and $\thetal$-angles   
are denoted by  $S_K,P_K, \dots$ and $S_\ell, P_\ell, \dots$ respectively. 
 
Throughout this work we mostly restricted ourselves to $K_J = K^*$ thereby imposing  
$J_K = 1$ i.e. a $P_K$-wave. The signal of $K^*$ is part of the  $(K \pi)$ $P_K$-wave.
The importance of considering the $S_K$-wave interference through $K^*_0(800)$ (also known as $\kappa(800)$) was emphasised a few years ago in \cite{Becirevic:2012dp}. 
The separation of the various partial waves in the $(K\pi)$-channel is a problem that can be solved experimentally e.g. \cite{Blake:2012mb}. We refer the reader to Ref.~\cite{DHJS14} 
for a generic study of the lowest partial waves at high $q^2$.

The second partial wave expansion originates from the lepton angle $\thetal$, which will be the main focus hereafter.
By restricting ourselves to the dimension-six effective Hamiltonian equation~\eqref{eq:Heffexplicit} 
as well as the lepton-pair facorisation approximation (LFA)\footnote{We remind the reader that in the LFA no electroweak gauge bosons are exchanged between the lepton pair and other particles when calculating the matrix element. 
This is the same approximation that is relevant for the endpoint relations \cite{Zwicky2013,HZ2013}.} only $S_\ell$- and 
$P_\ell$-waves were allowed (cf. equation~\eqref{eq:MKs}), bounding  
$l_\ell \leq 2$ in \eqref{eq:Igen}.
This pattern is broken by the inclusion of higher spin operators and non-factorisable corrections 
between the lepton pair and the quarks. It is therefore important to be able to distinguish these two effects from each other.

 \subsection{Qualitative discussion of Effects of higher Spin Operators in $H^{\rm eff}$}
 \label{sec:higher}
 
 Operators of higher dimension are suppressed and neglected in the standard analysis. 
 Operators of higher spin in the lepton and quark parts are necessarily of higher dimension 
 and bring in new features.  An operator of (lepton- and quark-pair) spin $j$ is given by 
 \begin{equation}
 \label{eq:spinj}
 \Op{j} = \bar s_L \Gamma^{(j-)}_{\mu_1 \dots \mu_j} b \; \bar \ell  \Gamma^{(j+) \;\mu_1 \dots \mu_j}  \ell
 \end{equation}
 with $\Gamma^{(j \pm )}_{\mu_1 \dots \mu_j}  \equiv \ga^{\phantom{+}}_{ \{ \mu_1} D^\pm_{\vphantom{\{ } \mu_2} \dots D^\pm_{ \mu_j \} }$, ${D^\pm}   \equiv \overset{\leftarrow} D \pm \overset{\rightarrow} D $, with $\overset{\rightarrow} D$ 
 the directional covariant derivative and curly brackets denoting symmetrisation in the Lorentz indices. 
 In passing let us note that in this notation $ \Op{1}  = O_V \equiv O_{9}$ with $O_V$ defined in \eqref{eq:Heffexplicit}.
 The  operator \eqref{eq:spinj} is of dimension  $d_{\Op{j}} = 4+2 j$ and the corresponding Wilson coefficients 
 are suppressed  by powers of $m_W$.  
 Neglecting electroweak corrections  and including  the dimensional 
estimate of the matrix elements the leading relative contributions are given by  $(m_b/m_W ) ^{2(j-1)}$ where
$ (m_b/m_W) \sim 6 \cdot 10^{-3}$. 
 
Operators of the form \eqref{eq:spinj}  present new opportunities to test physics beyond the SM 
provided that their contribution is larger than that of the breaking of lepton factorisation through 
electroweak corrections.
The operator $\Op{2}$, for example, gives rise to non-vanishing moments  
of the type 
$\GKstar{2}{4}{2}$ in $B \to K^*(\to K\pi) \bar \ell^+ \ell^-$ and
 $\GKstar{4}{4}{4}$ in $B \to K_2(\to K\pi) \bar \ell^+ \ell^-$, 
  \cite{preparation} both of which are absent in the LFA.

\subsection{Qualitative discussion of QED Corrections}
\label{sec:beyond}

 \begin{figure}
\includegraphics[scale=0.5,clip=true,trim= 20 300 20  
80]{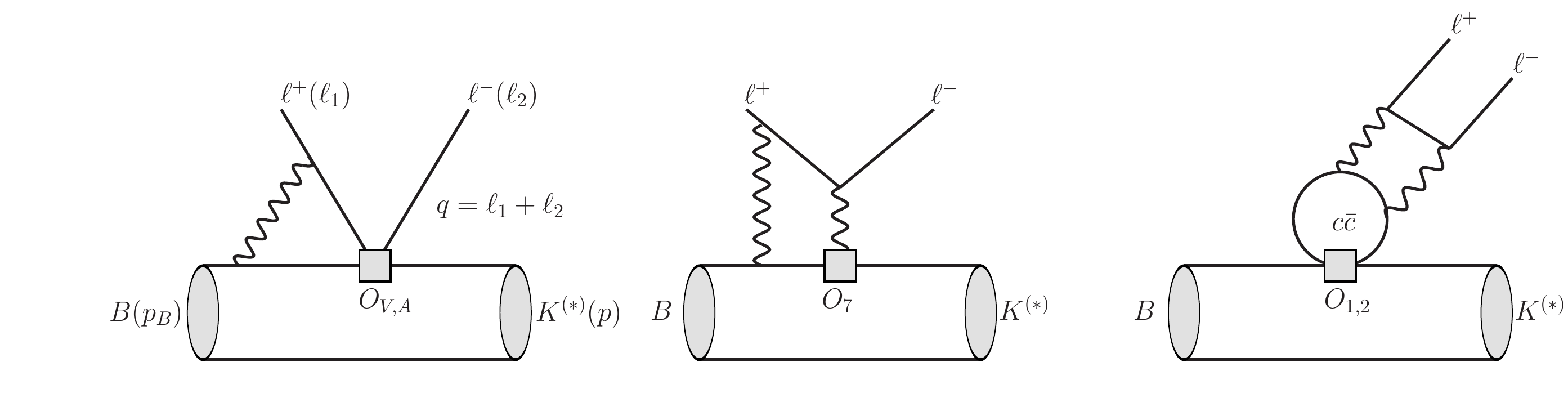}
 \caption{\small Examples of virtual QED corrections to $B \to K \ell^+ \ell^-$, where either a photon is exchanged 
 between the decaying $b$-quark and a final state lepton, with effective operators $O_{V,A}$ (left) and $O_7$ (middle), or a second photon is emitted by the charm loop (right). Other topologies relevant 
 for higher moments include the interaction of the leptons with the spectator 
 as well as the $B$- and $K^{(*)}$-meson.}
 \label{fig:BKllQED}
 \end{figure}

The  $B \to K \ell^+ \ell^-$ channel  allows the discussion of the consequences of going 
beyond the LFA in a simplified setup, and is of particular relevance because of a 
recent LHCb measurement \cite{RKLHCb}.

In the LFA  \eqref{eq:angdistg} the single opening angle $\thetal$ of the decay
 is restricted to 
$l_\ell \leq 2$ moments  in  $\IKF$ \eqref{eq:Igen}. 
More precisely, $l_\ell  \leq  2 j$ with $\Op{j}$ as in \eqref{eq:spinj} (see also the discussion following equation~\eqref{eq:angdistG}).
From the viewpoint of a generic $1 \to 3$ decay there is 
no reason for this restriction, as it is only the sum of the total (orbital and spin) angular momentum that is conserved.  However, in 
the LFA the $B \to K [ \ell^+ \ell^-]$ decay mimics a $1 \to 2$ process, imposing this constraint. This pattern is broken  by exchanges of photons and $W$- and $Z$-bosons, as depicted 
in figure \ref{fig:BKllQED} for a few operators relevant to the decay.
The $W$ and $Z$ are too heavy to impact on the matrix elements, but their effect is included 
in the Wilson coefficient.

As stated above QED corrections  turn the decay into a true  $1 \to 3$ process, and this
necessitates a reassessment of the kinematics. 
 By crossing the process can be written as  a $2 \to 2$ process 
\begin{equation}
B(p_B) + \ell^-(-\ell_1) \to K(p) + \ell^-(\ltwo) \;,
\end{equation}
with Mandelstam variables $s = (p+ \ltwo)^2$, $t=(\ell_1 + \ltwo)^2 = q^2$ and 
$u = (p+ \ell_1)^2$,
\begin{equation}
\label{eq:ET}
s[u]  =  \frac{1}{2} \left[ (m_B^2 +  m_K^2  + 2m_\ell^2  -q^2) \pm
\beta_\ell \sqrt{ \lambda \left(m_B^2, m_{K}^2 , q^2 \right)} \cos \thetal \right] \;,
\end{equation}
 obeying the Mandelstam constraint $s + t + u = m_B^2 +  m_K^2  + 2m_\ell^2 $.
Crucially, the kinematic variables $s$ and $u$  become \emph{explicit} functions of the angle $\theta_\ell$. 
In a generic computation these variables enter (poly)logarithms, which when expanded give contributions 
to any order $l_\ell$ in the Legendre polynomials. This statement applies at the amplitude 
level and therefore also 
to  the decay distribution 
\eqref{eq:angdistg2} 
\begin{eqnarray}
\label{eq:angdistg3}
  \frac{d^2 \Gamma(B \to K \ell^+ \ell^-) }{dq^2 \; d\textrm{cos}\thetal}  &=&  \sum_{l_\ell \geq 0 } \GK{l_\ell } P_{l_\ell}(\cos\thetal)  \;.
\end{eqnarray}
The $B \to K \ell \ell$ moments are simply given by
\begin{equation}
\label{eq:Mll}
M_{\bar \ell\ell}^{(l_\ell)} = \int_{-1}^{1} d\cos \thetal  P_{l_\ell}( \cos \thetal)  \frac{d^2 \Gamma(B \to K \ell^+ \ell^-) }{dq^2 \; d\textrm{cos}\thetal}
= \frac{1}{2 l_\ell+1} \GK{l_\ell }_{\bar \ell \ell} 
\end{equation}
where we have introduced a lepton-subscript for further reference.
In the SM the effects  are dependent on the lepton mass, 
for example  through logarithms of the form $\ln (m_\ell /m_b)$ times the fine structure constant.
There are new qualitative features of which we would like to highlight the following two:
\begin{itemize}
\item Both vector and axial couplings $O_{V(A)} = O_{9(10)}$ \eqref{eq:Heffexplicit} contribute 
to  any moment $l_\ell \geq 0$.  
In the  LFA $l_\ell$-odd terms (forward-backward asymmetric) 
arise from broken parity through interference of $O_V$ and $O_A$ \eqref{eq:Heffexplicit}.
The physical interpretation is that there is a preferred direction for charged leptons in the presence of the
charged quarks of the decay. In the specific diagram figure \ref{fig:BKllQED} (left) it is the charge of the 
$b$-quark which attracts or repels the charged lepton(s) with definite preference. 
It is possible that one can establish a higher degree of symmetry by using charge-averaged 
 forward-backward asymmetries.
\item A key question is how the moments vary in $l_\ell$. In the absence of a computation a precise 
answer is not possible. 
Nevertheless we can assess the question semi-quantitatively by considering for example the triangle graph between 
the photon, a lepton and the $b$-quark in figure \ref{fig:BKllQED} (left) and the corresponding one with the $s$-quark.
Neglecting the Dirac structures the triangle graph is given by
$C_0(m_\ell^2, p_B^2, s[\cos\thetal] , m_\ell^2,0,m_b^2)$.\footnote{We use conventions for the Passarino-Veltman function 
$C_0(p_1^2, p_2^2, p_3^2 , m_1^2,m_2^2,m_3^2)$ such that the two-particle cuts begin at 
$p_1^2  \geq (m_1+m_2)^2$, $p_2^2  \geq (m_2+m_3)^2$ and $p_3^2 
 \geq (m_3+m_1)^2$.}\footnote{We have refined this analysis by taking into account that the $b$- and $s$-quark only carry 
a fraction of the momentum of the corresponding mesons. This amounts to the substitution 
$p_B^2 \to (p_B - x p)^2 $ and $s[u] \to ( \ltwo[\ell_1] + x p)^2$ with $x$ being the momentum fraction carried by the $s$-quark. For the vertex diagrams one expects the Feynman mechanism (i.e. $ x \simeq 0$)  to dominate. This  changes when 
spectator corrections are taken into account. \label{foot:refine}} 
Expanding this function in partial waves $C_0 = \sum_{l_\ell \geq 0} C_0^{(l_\ell)} P_{l_\ell}(\cos\thetal)$ we find that 
$|C_0^{(l_\ell)}|$ does fall off in $l_\ell$. 
Averaging over several configurations (cf. footnote \ref{foot:refine}) we conclude that the $l_\ell =2$ ($D$-wave) contribution
is suppressed by approximately a  factor of $2$ with respect to $l_\ell = 0$, with a slightly steeper fall-off with increasing $l_\ell$ for the $b$-quark versus $s$-quark vertex correction.
Note the graph where the photon couples to the other lepton comes with a different Dirac structure 
and is not obtainable through a straightforward symmetry prescription. 
We therefore think that it is sensible to consider those graphs separately. 
We stress that this semi-quantitative analysis does not replace a complete QED computation, which would include corrections to Wilson coefficients, all virtual corrections and importantly also the real photon emission. 
\end{itemize}

We now turn to the most important consideration,  the relative size of the QED corrections versus higher spin operators. 
For effective field theories of the type $\vev{H^{\rm eff}} \sim C^{(j)}(\mu_F)  \vev{\Op{j}(\mu_F)}$ \eqref{eq:Heffexplicit}, 
the precise separation scale $\mu_F$ is arbitrary to a certain degree and effects 
are therefore contained in the Wilson coefficients as well as the matrix elements. 
We find it convenient to discuss the effect at the level of the Wilson coefficients. 
For the latter QED corrections arising from modes 
from $m_W$ to $\mu_F \simeq m_b$ can be absorbed into a tower of the  higher spin operators  $\Op{j}$ \eqref{eq:spinj}. 
The leading contribution to the corresponding Wilson coefficients 
from the initial matching procedure  and the mixing due to QED behaves parametrically as
\begin{equation}
\label{eq:Cj}
C^{(j)} =  \frac{{\cal O}(1)}{(m_W^2)^{ j}} + 
  \al f_j  \cdot \left(  \frac{m_W^2}{m_b^2} \right)^{(j-1)}   \frac{{\cal O}(1)  }{(m_W^2)^{ j} } \;,  \quad \text{ for } j \geq 1 \;,
\end{equation}
where we have implicitly used $\mu_F = m_b$ in   $\vev{H^{\rm eff}} \sim C^{(j)}(\mu_F)  \vev{\Op{j}(\mu_F)}$. 
Above $\al$ is the fine structure constant and  $f_j$ parametrises the comparatively moderate fall-off of the higher moments 
due to QED.
In the SM one therefore expects QED effects to dominate over those due to higher spin operators,
except for $j=2$ where they could be comparable \cite{preparation}.
At the level of matrix elements this hierarchy could even shift further towards QED as a result of infrared enhancements through  $\ln (m_\ell /m_b)$-contributions.

The discussion of $B \to K^*(\to K \pi) \ell^+ \ell^-$ is similar, but involves the kinematics of a $1 \to 4$ decay.
The decay distribution  becomes a generic function of all three angles  
$\thetal$, $\theta_K$ and $\hel$. 
It should be added that the selection of the $K^*\to K \pi$ signal ($P_K$-wave) 
restricts   $l_K = 0,2$.

\subsection{On the Importance of testing for higher Moments for $B \to
K^{(*)} \bar \ell^+ \ell^-$}
\label{sec:aktuell}

We have stressed throughout the text that it is of importance to probe
for moments that are vanishing in
the decay distributions $\IKsF$ \eqref{eq:angdistG} of $\bar{B} \to \bar{K}^*(\to
K\pi) \ell \ell)$ and $\IKF$ \eqref{eq:angdistg} of
$\bar{B} \to \bar{K} \ell \ell$ respectively.
In this section we highlight specific cases of current
experimental anomalies in exclusive decay modes where their nature
might be clarified using an analysis of (higher) moments.

\subsubsection{Diagnosing QED background to $R_K$}
\label{sec:diagnose}

In the SM the decays $B^+ \to K^+ e^+e^-$ and $B^+ \to K^+ \mu^+\mu^-$ are identical up to phase-space lepton mass effects
and electroweak corrections.
 The observable 
\begin{equation}
\label{eq:RKdef}
R_K|_{[q^2_{\rm min},q^2_{\rm max}]} \equiv \frac{{\cal B}(B^+ \to K^+ \mu^+\mu^-)}{{\cal B}(B^+ \to K^+  e^+e^-)}\Big|_{[q^2_{\rm min},q^2_{\rm max}]} 
\end{equation}
has been put forward in Ref.~\cite{RK} as an interesting test of lepton flavour universality (LFU). Above $q^2_{\rm min/max}$ stands for the bin boundaries. 
Neglecting electroweak corrections the SM prediction is 
$R_K|_{[1,6]\GeV^2} \simeq 1.0003(1)$ \cite{RKSM}, which is at $2.6 \sigma$-tension with 
the LHCb measurement at $3\ifb$ \cite{RKLHCb} 
\begin{equation}
\label{eq:RK}
R_K = 0.745^{+0.090}_{-0.074}(\text{stat}) \pm 
0.036 (\text{syst}) \;.
\end{equation}
Previous measurements  \cite{Wei:2009zv,Lees:2012tva}, with much larger uncertainties, 
 were found to be consistent with the SM as well as \eqref{eq:RK}.
This led to investigations of physics beyond the SM with 
$C_9^{\bar ee} \neq C_9^{\bar \mu\mu}$ 
(where $O^{\bar \ell\ell}_9  \equiv \bar b \ga_\al s \bar \ell \ga^\al \ell $) amongst other variants for
which we quote a few recent works \cite{Gripaios:2014tna,Becirevic:2015asa,Niehoff:2015bfa,Crivellin:2015era,Greljo:2015mma,Alonso:2015sja,Varzielas:2015iva,Crivellin:2015lwa,Calibbi:2015kma,Crivellin:2015mga}  as well as the general  review  
\cite{Blake:2015tda} for further references.

Let us summarise the aspects of QED corrections which are of relevance for the discussion below:
i) they break lepton factorisation and therefore give rise to higher moments, and ii)  
they  depend on the lepton mass, for example through  logarithmic terms   of $\ln (m_\ell /m_b)$. 
In view of the lack of a full QED computation\footnote{A partial result, photon emission from initial 
and final state, was reported in \cite{Mexico}.}
 we  suggest diagnosing the size of QED corrections, as well 
as their lepton dependence, by experimentally assessing higher moments.\footnote{Collinear photon emission 
in the inclusive case was studied recently in \cite{HHL2015}. The additional photon of course leads to terms 
which go beyond the $\IKF$ angular distribution. 
Note, in view of the presence of these terms through virtual corrections they also have to be present in real emission 
by virtue of the Bloch-Nordsieck QED infrared cancellation theorem \cite{BN1937}.  The authors \cite{HHL2015} find within their approximation that the third and fourth moment are two orders of magnitude smaller than the leading contributions. This is in the  expected parametric range but one cannot draw precise 
conclusions on the size of this effect for the exclusive channels discussed in this paper.}
The latter is directly relevant for $R_K$.
Let us be slightly more concrete and define the normalised angular functions as follows
$\GKh{l_\ell }_{\bar \ell \ell}  \equiv \GK{l_\ell }_{\bar \ell \ell} / (2 \GK{0}_{\bar \ell \ell})$ \eqref{eq:angdistg3} 
(in this convention  $2 \GK{0}_{\bar \ell \ell}=  d\Gamma(B \to K \ell \ell)/dq^2$, $ \GKh{1 }_{\bar \ell \ell} = A_{\rm FB}$ and 
$ \GKh{2 }_{\bar \ell \ell} = (F_H - 1)/2$ in the notation of \cite{BHP2007}).
We would like to stress the following points:
\begin{itemize}
\item \emph{How to distinguish QED corrections from higher dimensional operators:} 
both contributions give rise to higher moments but crucially 
the QED corrections dominate for moments of increasing $l_\ell$, cf. the discussion at the end
of  section~\ref{sec:beyond} and specifically equation~\eqref{eq:Cj}. A $J_\ga$-wave at the amplitude level contributes 
to a $l_\ell = J_\ga +1$ moment through interference with the SM $P_\ell$-wave. 
We conclude that  QED and higher spin operators could be comparable for $\GKh{3 }_{\bar  \ell \ell}$  but for  $\GKh{ l_\ell > 3 }_{\bar  \ell \ell}$ one would expect the former to dominate.\footnote{Another criterion could be that corrections from higher spin operators are  uniform in the lepton mass provided that  lepton flavour universality is unbroken. 
This is though delicate since the measurement of $R_K$ questions this aspect.}
\item \emph{Lepton-flavour dependence  of QED corrections:}
differences between $\GKh{l_\ell \geq 3 }_{\bar  \mu \mu}$ and  
$\GKh{l_\ell \geq 3 }_{\bar  ee}$ in the range above $q^2 > 1 \GeV^2$ 
indicate the importance of the flavour dependence. This gives an indication on how much 
the branching fractions (zeroth moments) and therefore 
$R_K$ is affected  by QED through lepton mass effects.
Note that due to $\ln(m_\ell/m_b)$-effects it is conceivable that  $\GKh{l_\ell \geq 3 }_{\bar  \mu \mu}$ is small, say  $O(1\%)$, but that $\GKh{l_\ell \geq 3 }_{\bar  ee}$ is larger. 
Note for example that  $\GKh{1}_{\mu\mu} = A_{\rm FB}$ is consistent with the 
SM prediction excluding QED, which is $O(m_\mu),$ within errors in the few percent range \cite{angularK14}. 
\end{itemize}

\subsubsection{Combinatorial background in $B \to K^{*} \mu^+\mu^-$ below
the narrow charmonium resonance region}
\label{sec:combinatorial}

A characteristic feature of  $B \to K^{(*)} \ell^+ \ell^-$ transitions is  the
large contribution to the branching fraction
through the intermediate narrow charmonium states $J/\Psi$ and $\Psi(2S)$.
For example
 ${\cal B}( B \to K J/\Psi)  {\cal B}(J/\Psi \to \mu^+\mu^-) \simeq (8
\cdot 10^{-4}) (6 \cdot 10^{-2}) \simeq 5 \cdot 10^{-5}$    is three
orders of magnitude larger than the measured differential branching fraction,
$d{\cal B}(B \to K^{*} \mu^+ \mu^-)/dq^2 \simeq 2 \cdot 10^{-8} /\GeV^2$
\cite{Aaij:2014pli}, well below the narrow charmonium resonances region.
It is therefore legitimate to be concerned with possible combinatorial
backgrounds in this region.

Assuming that such backgrounds are relevant this raises the question as to how they
can be distinguished from the signal event. 
In the case where they can be absorbed into the background fit-function they would not
impact on the analysis. Whether or not this is the case is a non-trivial question.
Pragmatically, however,  background events can be expected to perturb the
hierarchy of the moments as compared to the true signal event.
 One would expect the background events to fall off only slowly 
 for higher moments in the lepton partial wave.\footnote{Similar things can be said about the hadronic partial wave, but as the detection of the $P_K$-wave is part of the signal selection the presence of such higher waves would have less influence. However, the
remaining background might impact
 on the $S_K$-wave, which does matter since the $S_K$-wave enters the analysis.}
 Hence the size of
these effects can be diagnosed
 through the measurement of higher moments as a function of $q^2$, 
independent of  model assumptions. By the latter we mean that higher moments peaking 
below the charmonium resonances will be indicative of the type of combinatorial background 
mentioned above.

A possible example of such backgrounds is the process $B \to K    \mu^+ \mu^- \ga $
where the photon is not detected but  energetic enough to cause a
significant downward shift in $q^2 = (\ell_1 + \ltwo)^2$. 
Such an event would be rejected  as a $B \to K  \mu^+ \mu^- $ signal
because the reconstructed $B$-mass $m_{K\mu\mu}$ would fall outside the
signal window (i.e. $m_{K\mu\mu} < m_B - \Delta$ and $\Delta \simeq O(100 \MeV)$).
If additionally  a $\pi$-meson from the underlying event is detected, the event
     could conspire to enter the signal window of  $B \to K^{*}(\to K\pi)
\mu^+ \mu^-$
  (i.e. $m_{K\pi\mu\mu} \simeq m_B  $ and   $m_{K\pi} \simeq m_{K^*} $).
It is therefore conceivable that  the small chance of the events described above is overcome by 
 the enhancement by three orders of magnitude of the $J/\Psi$ transition.
 If such events are present and not rejected then this leads to a  bias
in $B \to K^{*}(\to K\pi) \mu^+ \mu^-$ transitions below the narrow charmonium resonances. 
More precisely, denoting the momentum of the undetected photon by $r$, the shift in $q^2$ is as follows
$q^2 \simeq m_{J/\Psi}^2 = (\ell_1 +\ltwo +r)^2 \to q^2_{\rm signal} \equiv (\ell_1 +\ltwo )^2   < m_{J/\Psi}^2$.

This is particularly relevant as some of the anomalies from the LHCb
measurements,  in particular the angular observable $P_5'$, 
are most pronounced in bins just below the $J/\Psi$-resonance \cite{LHCb-P5p,LHCbBKmumu2015}.
To what extent such operators correspond to new physics in $O_9 \equiv O_V$ 
\cite{Descotes-Genon:2013wba,Altmannshofer:2013foa} or effects from charm resonances 
\cite{LZ2014} is a difficult question since they contribute to the same helicity amplitude. 
They can be distinguished from each other by 
analysing the $q^2$-spectrum of the observables and by  
the determination of the strong phases which can 
originate from the charm resonances \cite{LZ2014}.
This could be through the determination of the complex-valued residues of 
the resonance poles \cite{LZ2014}, or simply the strong phase in the region 
below the $q^2$-resonance through $\Ima[ \GKstar{2}{1}{1} ] \sim P_6'$, which corresponds 
to the imaginary part of $\Rea[ \GKstar{2}{1}{1} ] \sim P_5'$ \eqref{eq:O1}.


\section{Conclusions}
\label{sec:conclusions}

In this work we have generalised the standard helicity formalism to effective field theories  
of the $b \to s \ell \ell$-type. The framework applies  to any semi-leptonic and radiative decay. 
The formalism has been used to derive  the angular distributions $\IKsF$ \eqref{eq:angdistG} and $\IKF$ \eqref{eq:angdistg2} for non-equal lepton masses with the full dimension-six effective Hamiltonian, including in particular scalar and tensor operators. 
Explicit results for $\bar B \to \bar K^* \lone \ltwobar$  and for $\bar B \to \bar K \lone \ltwobar$ can be found 
in appendices \ref{app:BtoKs}  and \ref{app:BtoK} respectively as well 
as a Mathematica notebook (notebookGHZ.nb) provided in the arXiv version.
Comments on differences conversion of  observables between theory and experiment with the literature
are reported in appendix \ref{app:Pp-conversion}. Minor discrepancies  in tensor contributions 
with respect to previous results are discussed in 
appendix ~\ref{sec:comparison}.

The approach clarifies how the lepton factorisation approximation determines the specific 
 form of the angular distributions $\IKsF$  and $\IKF$, and how these distributions are extended 
 by the inclusion of virtual and real QED corrections, as well as  higher-spin operators in the effective Hamiltonian.
 Higher-dimensional spin operators provide new opportunities to search for physics beyond the 
 SM.
 We have argued that, within the SM, QED effects and higher-spin operators can be distinguished from each other
 by their differing fall-off behaviour in increasingly higher moments in the $\thetal$-angle.\footnote{
 In addition higher-spin operators   can be distinguished from QED corrections by universality in the lepton flavour. However, it should be kept in mind that   lepton-universality  is questioned by the $R_K$ measurement.}   
 
 Assessing higher moments can shed light on current anomalies with respect to  the SM.
We have argued (cf. section~\ref{sec:diagnose}) that higher moments in  $B \to K \ell^+\ell^-$ ($\ell = e,\mu$)  are a window into QED corrections and therefore of importance with regard to the $R_K$ measurement  \cite{RKLHCb}. 
In view of  tensions of angular predictions in 
$B \to K^* \mu^+\mu^-$ with experiment  \cite{LHCb-P5p,LHCbBKmumu2015}, the higher moments can be of help in assessing their origin, such as the possible leakage of $J/\Psi$ events into the lower nearby $q^2$-bins (cf. section~\ref{sec:combinatorial}).
As another example we mention the  $R(D^{(*)}) = {\cal B}(B \to D^{(*)} \tau \nu)/{\cal B}(B \to D^{(*)} \mu \nu)$  ratio measurement \cite{BelleRD2010,BaBarRD2013,LHCbRD},  suggestive of some tension with the SM.
A higher moment analysis could again be useful in assessing the impact of QED, lepton mass or cross channel 
backgrounds on these results. 

To measure and bound higher moments is  relevant as their contributions can bias  likelihood fits.  
We  therefore encourage the investigation of higher moments in several  experimental channels 
from the various perspectives discussed above.

\appendix
\numberwithin{equation}{section}

\subsection*{Acknowledgements}

We are grateful to Simon Badger, Damir Becirevic, Tom Blake, Christoph Bobeth, 
Marcin Chrzaszcz, Peter Clarke,  Sebastian Descotes-Genon, Martin Gorbahn, Enrico Lunghi, 
Joaquim Matias, Matthias Neubert, Kostas Petridis, Alexey Petrov, 
Maurizio Piai, Steve Playfer, Nico Serra, David Straub, Olcyr Sumensari, Javier Virto, Renata Zukanov  and in particular to  Greig Cowan for many useful discussions.  
JG acknowledges the support of an STFC studentship (grant reference ST/K501980/1). MH acknowledges support from the Doktoratskolleg ``Hadrons in Vacuum, Nuclei and Stars'' of the Austrian Science Fund, FWF DK W1203-N16.

\paragraph{Note added}
{\small While this paper was in its final phase a paper using the helicity formalism for $B \to K^* \ell^+ \ell^-$ appeared \cite{Dey:2015rqa}.
The paper uses the standard Jacob-Wick formalism and therefore includes HAs of definite spin. 
This is an approximation that holds up to lepton mass corrections in the SM and does not allow
the inclusion of  scalar operators for example.}

\paragraph{Changes in conventions and presentation}
{\small Notational changes with respect to  the arXiv version 1, aimed at clarifying  the underlying 
structure, are as follows: i) results are presented for $\bar B \to \bar K^{(*)}\ell_1 \bar \ell_2$  
rather than the conjugate decay, ii) we use $C_{\cal T}^{(')}$ Wilson coefficients 
in place of $C_{T(5)}$ for the tensor operators cf. appendix~\ref{app:Hamiltonian} 
for details, iii) the angular distribution \eqref{eq:d4GJi} 
is presented in terms of $\J_i$ in place of $J_i$  in order to emphasise the differences 
of angular convention of this paper and the theory community (as discussed in appendix~\ref{app:ang-conventions}), iv) lepton HAs are presented in the $A,V$ rather than $L,R$ basis and v) timelike HAs are absorbed into scalar and pseudoscalar HAs. In addition we provide a Mathematica notebook, entitled notebookGHZ.nb, containing the results presented in appendix \ref{sec:explicitGdifferentmass} for the decay mode $ \bar{B} \to \bar{K}^* ( \to K \pi) \lone \ltwobar $ for non-equal lepton masses.}

\section{Results relevant for all decay modes}

\subsection{Decomposition of $SO(3,1)$ into $SO(3)$ up to spin $2$}
\label{app:branch}

The aim of this appendix is to give some more detail about the decomposition 
\eqref{eq:c-relation} and in particular extend it to the two-index case, which 
includes the discussion of spin $2,1,0$.

In section~\ref{sec:effective} it was shown that insertion of the completeness relation 
\eqref{eq:c-relation} corresponds to the decomposition, or branching rule, 
\begin{equation}
(1/2,1/2)_{SO(3,1)}\Big|_{SO(3)} \to (\mathbf{1} + \mathbf{3})_{SO(3)} \;, 
\end{equation}
where $(1/2,1/2)$ is the irreducible vector Lorentz representation. 
We remind the reader that the irreducible Lorentz representations, denoted by 
$(j_1,j_2)$, are characterised by the eigenvalues of the two Casimir operators of $SO(3,1)$. Inserting the completeness relation twice therefore corresponds to taking
the tensor product $(1/2,1/2)  \otimes (1/2,1/2)$ which decomposes as
\begin{eqnarray}
\label{eq:kron}
& & ( (1/2,1/2) \otimes (1/2,1/2) )_{SO(3,1)} = (\left[ (1,1)  \right] \oplus \left[ (1,0) \oplus (0,1) \right] \oplus (0,0))_{SO(3,1)}\Big|_{SO(3)}  \to \nonumber \\
& &( \left[  1 \cdot \mathbf{5} \oplus 1 \cdot \mathbf{3} \oplus 1 \cdot  \mathbf{1} \right] \oplus 
\left[   2 \cdot \mathbf{3} \right] \oplus 1 \cdot \mathbf{1} )_{SO(3)} = 
( 1 \cdot \mathbf{5} \oplus 3 \cdot \mathbf{3} \oplus 2 \cdot  \mathbf{1})_{SO(3)}  \;.
\end{eqnarray}
The double completeness relation 
\begin{equation}
\label{eq:double}
g_{\al \be} g_{\ga \de} = \delta_{\al \be \ga \de} +  \delta^t_{\al \be \ga \de}  +  \delta^{tt}_{\al \be \ga \de} 
\end{equation}
can be decomposed
\begin{equation}
\label{eq:deltas}
\delta_{\al \be \ga \de} =  \sum^2_{J=0} \sum_{\la =-J}^J \gpol^{J,\la}_{\al \ga}  \overline \gpol^{J,\la}_{\be \de}   \;, \quad 
\delta_{\al \be \ga \de}^t =   - \sum_{\la =-1}^1 \gpol^{t,\la}_{\al \ga}    \overline\gpol^{t,\la}_{\be \de}  - \sum_{\la =-1}^1 \gpol^{t,\la}_{\ga \al}    \overline\gpol^{t,\la}_{\de \be } \;, \quad 
\delta_{\al \be \ga \de}^{tt} =  \gpol^{tt}_{\al \ga}    \overline\gpol^{tt}_{\be \de}  \;,
\end{equation}
 into parts containing zero, one and two timelike polarisation vectors
\begin{eqnarray}
\label{eq:completenessdecomposition}
& &  \gpol^{t,\la}_{\al \ga} =\gpol_\al(t) \gpol_\ga(\la)  \;, \quad
 \gpol^{tt}_{\al \ga} = \gpol_\al(t) \gpol_\ga(t)  \;, \nonumber \\
& &  \gpol^{J,\la}_{\al \ga} =  \!\!\!\!\! \sum_{\la_1,\la_2 =-1}^1 
 C^{J11}_{\la \la_1 \la_2} \gpol_\al(\la_1) \gpol_\ga(\la_2)  \;, \quad 
\end{eqnarray}
with $ \la=\la_1+\la_2$ in the first term and the polarisation vectors $\gpol_\al(\la)$ are parametrised as
\begin{eqnarray}
\label{eq:gpol2}
\gpol^\mu(\pm)  =  (0, \pm 1, i,0)/\sqrt{2}  \;, \quad   \gpol^\mu(0) =  (q_z,0,0,q_0)/\sqrt{q^2} \;, \quad   \gpol^\mu(t) =  (q_0,0,0,q_z)/\sqrt{q^2} \;,
\end{eqnarray}
which we reproduce from \eqref{eq:gpol} for the reader's convenience.  A few explanations seem in order.  
The minus sign in front of $\delta_{\al \be \ga \de}^t$  in \eqref{eq:deltas} is due to there being an odd number of timelike polarisation vectors. 
The first, second and third term in \eqref{eq:double} correspond respectively to 
the $(1,1)$-, $[(1,0) \oplus (0,1)]$- and $(0,0)$-terms in \eqref{eq:kron}. 
It is convenient to rewrite the double completeness relation  \eqref{eq:double} in a form that makes
the decomposition into the different spins $j$ explicit 
\begin{equation}
\label{eq:double_compact}
g_{\al\be} g_{\ga\de} =  \sum^2_{J=0} \sum_{\la =-J}^J 
\sca{ \gpolt^{J,\la}_{\al \ga}}{  \overline\gpolt^{J,\la}_{\be \de} } \;.
\end{equation}
Above the scalar product ``$\cdot$" stands for
\begin{equation}
\label{eq:eps2}
\sca{ \gpolt^{\,\la}_{\al \ga} }{  \overline\gpolt^{J,\la'}_{\be \de}}  
 = \delta_{J0}\Big[    \gpol^{0,0}_{\al \ga}  \overline\gpol^{0,0}_{\be \de} + \gpol^{tt}_{\al \ga}  \overline \gpol^{tt}_{\be \de}  \Big] +
   \delta_{J1}\Big[  
  \gpol^{1,\la}_{\al \ga}  \overline\gpol^{1,\la'}_{\be \de} -     \gpol^{t,\la}_{\al \ga }   
   \overline\gpol^{t,\la'}_{\be \de}-     \gpol^{t,\la}_{ \ga \al}   
   \overline\gpol^{t,\la'}_{\de \be }    \Big] + 
    \delta_{J2}\Big[     \gpol^{2,\la}_{\al \ga}  \overline \gpol^{2,\la'}_{\be \de}    \Big]  \;.
      \end{equation}
The single completeness relation \eqref{eq:c-relation} in the analogous notation of \eqref{eq:double_compact} reads 
\begin{equation}
\label{eq:eps1} 
    g_{\al \be}  = \sum_{J=0}^1 \sum_{\la = -J}^J \gpolt_\al^{J,\la}  \overline\gpolt^{J,\la}_\be  \;,
\end{equation}
with $\gpolt^{J,\la}_\al = \delta_{J1} \omega_\al(\la)  + \delta_{J0} \omega_\al(t)$.

When applying the double completeness relation to generic decay structures, it can be seen from \eqref{eq:deltas} that in general one expects two distinct contributions to the amplitude from $ \delta^t_{\al \be \ga \de}$, 
\begin{equation*}
H_{\mu\nu} L^{\mu \nu} \to - \left ( \mathcal{H}_{t\la} \mathcal{L}_{t\la}+  \mathcal{H}_{\la t} \mathcal{L}_{\la t} \right) + \dots \; ,
\end{equation*}
where $ \mathcal{H}_{t\la} = H_{\mu \nu} \overline\gpol_{t,\la}^{\mu \nu}$, and analogous notation for $\mathcal{H}_{\la t}$, $\mathcal{L}_{t \la }$ and $\mathcal{L}_{\la t}$. If, however, the objects $H_{\mu\nu}$ and $L^{\mu \nu}$ are both symmetric or antisymmetric in the Lorentz indices, then $\mathcal{H}_{\la t} \mathcal{L}_{\la t} = \mathcal{H}_{t\la} \mathcal{L}_{t\la}$ and the two contributions can be combined. We have used this simplification in defining the generalised HAs for the $\bar{B} \to \bar{K}^* \lone \ltwobar$ \eqref{eq:ABKsll} and $\bar{B} \to \bar{K} \lone \ltwobar$ \eqref{eq:ABKll} decays respectively, resulting in the extra factor of 2 associated with the terms $H^{T_t}_{\la} \mathcal{L}^{T_t}_{\la_1 , \la_2}$, $\HTenstK \mathcal{L}^{T_t}_{\la_1 , \la_2}$ relative to other contributions in the generalised HAs.

\subsection{Additional Remarks on  effective Hamiltonian}
\label{app:Hamiltonian}

Here we collect a few additional remarks to the effective 
$b \to s \ell \ell$ Hamiltonian quoted in Eqs.(\ref{eq:Heffexplicit},\ref{eq:operators}).
Contributions proportional to $V_{\rm us} V_{\rm ub}^*$ have been neglected. 
The chromoelectric  and chromomagnetic operators $O_7$ and $O_8$, along with the contributions of 
the four-quark operators $O_{1, \dots, 6}$, 
can be absorbed into $O_V$ through defining an effective Wilson coefficient $C_V^{\rm eff} = C_9^{\rm eff}$.
We can rewrite $O_{\cal T}^{(')} = 1/2(O_{T} \pm O_{T_5})$, with the latter defined as
\begin{equation}
O_T = \bar s \sigma_{\mu\nu}b \bar \ell   \sigma_{\mu\nu} \ell \;,  \quad 
O_{T5} = \bar s \sigma_{\mu\nu} \ga_5 b \bar \ell   \sigma_{\mu\nu} \ell  \;, 
\end{equation}
(note: $ O_{T5} =  \bar s \sigma_{\mu\nu} b \bar \ell   \sigma_{\mu\nu}  \ga_5 \ell      =  - \frac{i}{2} \epsilon^{\al\be \mu\nu} \bar s \sigma_{\al\be}b \bar \ell   \sigma_{\mu\nu} \ell     $  with the last equality depending on conventions)  
and the relation between the Wilson coefficients is therefore 
\begin{equation}
C_{\cal T}^{(')} = C_T \pm C_{T5}   \;,  \quad   C_{T(5)} = \frac{1}{2}(C_{\cal T} \pm C_{\cal T}')
\end{equation}
in the sense that 
$  C_T O_{T} + C_{T5} O_{T5}  = C_{\cal T} O_{\cal T} +  C_{{\cal T}}' O'_{{\cal T}} $. 

\subsection{Definitions and Results of Leptonic Helicity Amplitudes}
\label{app:leptonHAs}
The calculation of the Leptonic Helicity amplitudes is an important part of the generalised helicity formalism described in this paper, and the method for their calculation is outlined in \cite{Haber1994}. 
In the Dirac basis of the Clifford algebra, with $\sigma^i$ as the usual $2\times 2$ Pauli matrices,
\begin{equation}
\label{eq:Dirac}
 \ga^0 = \begin{pmatrix} 1&0 \\ 0& -1 \end{pmatrix}\;, \qquad
  \ga^i = \begin{pmatrix} 0 & \sigma^i\\ -\sigma^i & 0 \end{pmatrix}\;, \qquad
  \ga_{5} = 
\begin{pmatrix} 0& 1\\ 1& 0 \end{pmatrix}  \;,
\end{equation}
the particle $u$ and anti-particle $v$ spinor are given by
\begin{alignat*}{3}
& u\left(\frac{1}{2}\right) & &= \left(\sqrt{E_1 + m_{\lone}}, 0,\sqrt{E_1 - m_{\lone}},0\right)^T & & = \left(\beta_1^{+}, 0, \beta_1^{-},0\right)^T \; , \nonumber \\
& u\left(-\frac{1}{2}\right) & &= \left(0,\sqrt{E_1 + m_{\lone}}, 0,-\sqrt{E_1 - m_{\lone}},\right)^T & & = \left(0,\beta_1^{+}, 0, -\beta_1^{-}\right)^T \; , \nonumber \\
& v\left(\frac{1}{2}\right) & &= \left(\sqrt{E_2 - m_{\ltwo}}, 0,-\sqrt{E_2 + m_{\ltwo}},0\right)^T & & = \left(\beta_2^{-}, 0, -\beta_2^{+},0\right)^T \; , \nonumber \\
& v\left(-\frac{1}{2}\right) & &= \left(0,\sqrt{E_2 - m_{\ltwo}}, 0,\sqrt{E_2 + m_{\ltwo}}\right)^T & & = \left(0,\beta_2^{-}, 0, \beta_2^{+}\right)^T \; ,
\end{alignat*}
with implicit definition of $\beta_i^\pm \equiv \sqrt{E_ i \pm m_{\ell_i}}$. 
The spinors are normalised as  $\bar u(\la_1) u(\la_2)   = \delta_{\la_1\la_2} 2m_{\lone}$ and 
 $\bar v(\la_1) v(\la_2)   = - \delta_{\la_1\la_2} 2m_{\ltwo}$. The leptonic HAs \eqref{eq:helmatel}
 contracted with polarisation vectors give rise to the HAs ${\mathcal L}_{\la_1 \la_2}$
 \begin{equation}
\label{eq:helmatellep}
 {\mathcal L}^X_{\la_1 \la_2} \equiv 
  \matel{\lone(\la_1) \ltwobar(\la_2)}{ \bar \ell   \; \Gamma^X \ell   }{0}  = 
  \bar u(\la_1) \Gamma^X v(\la_2) \;,
\end{equation}
(where $\lone = e^-$ for example) and the  $\Gamma^X|_{\la_X \to \la_\ell} $ ($\la_\ell  = \la_1 - \la_2$) as defined in Tab.~\ref{tab:Gammas}.
 Using all the equations above the evaluation of the lepton HAs is then straightforward and the results are presented below, for lepton masses $m_{\lone} \neq m_{\ltwo}$ in the first set of matrices and $m_{\lone} = m_{\ltwo} \equiv m_\ell$ in the 
 second set.\footnote{The expressions for $m_{\lone} \neq m_{\ltwo}$ can be applied to studies of lepton flavour-violating processes in all the decay modes considered in this paper within the lepton factorisation approximation, and are also applicable to decays involving an $l \bar{\nu}$ in the final state e.g. $ B \to D^* \ell \bar \nu$.}
 The first row (column) corresponds to $\la_1 (\la_2) = -\frac{1}{2}$ and the second row (column) corresponds to $\la_1 (\la_2) = +\frac{1}{2}$.
For the $\bar B \to \bar K^* \ell_1 \bar \ell_2$ decay mode, ie $\ell_1 = \ell^-$, 
the lepton HAs are given by   
\begin{alignat}{2}
\label{eq:LHAs}
&\mathcal{L}^{V}(\la_1, \la_2) = \left(
\begin{array}{cc}
\bonep \btwop - \bonem \btwom &-  \sqrt{2} \left ( \bonep \btwop + \bonem \btwom \right )  \\
- \sqrt{2} \left ( \bonep \btwop + \bonem \btwom \right )  & \bonep \btwop - \bonem \btwom \\
\end{array}
\right) &&\to \left(
\begin{array}{cc}
2 \ml &-  \sqrt{2 q^2}  \\
- \sqrt{2 q^2} & 2\ml \\
\end{array}
\right)   \; , \nonumber \displaybreak[0]\\
&\mathcal{L}^{A}(\la_1, \la_2) = \left(
\begin{array}{cc}
\bonep \btwom - \bonem \btwop & \sqrt{2} \left ( \bonep \btwom + \bonem \btwop \right )   \\
- \sqrt{2} \left ( \bonep \btwom + \bonem \btwop \right )  & \bonem \btwop - \bonep \btwom \\
\end{array}
\right) &&\to  \left(
\begin{array}{cc}
0 & \sqrt{2 q^2} \beta_\ell  \\
- \sqrt{2 q^2} \beta_\ell & 0 \\
\end{array}
\right)   \; , \nonumber \displaybreak[0]\\
&\mathcal{L}^{S}(\la_1, \la_2) =\left(
\begin{array}{cc}
\bonep \btwom + \bonem \btwop & 0  \\
0 & \bonep \btwom + \bonem \btwop \\
\end{array}
\right) &&\to \left(
\begin{array}{cc}
\sqrt{q^2} \beta_\ell & 0  \\
0 & \sqrt{q^2} \beta_\ell \\
\end{array}
\right)   \; , \nonumber \displaybreak[0]\\
&\mathcal{L}^{P}(\la_1, \la_2) = \left(
\begin{array}{cc}
 \bonep \btwop + \bonem \btwom & 0  \\
0 & -\bonep \btwop - \bonem \btwom \\
\end{array}
\right) &&\to \left(
\begin{array}{cc}
\sqrt{q^2} & 0  \\
0 & -\sqrt{q^2} \\
\end{array}
\right)   \; , \nonumber \displaybreak[0]\\
&\mathcal{L}^{T}(\la_1, \la_2) =  \left(
\begin{array}{cc}
- i \sqrt{2} \left(\bonep \btwom + \bonem \btwop \right) & - 2 i \left(\bonep \btwom - \bonem \btwop \right)  \\
2 i \left(\bonep \btwom - \bonem \btwop \right) & i \sqrt{2} \left(\bonep \btwom + \bonem \btwop \right) \\
\end{array}
\right) &&\to  \left(
\begin{array}{cc}
-i \sqrt{2 q^2} \beta_\ell & 0  \\
0 & i \sqrt{2 q^2} \beta_\ell \\
\end{array}
\right)   \; , \nonumber \displaybreak[0]\\
&\mathcal{L}^{T_t}(\la_1, \la_2) =  \left(
\begin{array}{cc}
i \left(\bonep \btwop + \bonem \btwom \right) & -  i \sqrt{2} \left(\bonep \btwop - \bonem \btwom \right)  \\
- i \sqrt{2} \left(\bonep \btwop - \bonem \btwom \right) & i \left(\bonep \btwop + \bonem \btwom \right) \\
\end{array}
\right) &&\to  \left(
\begin{array}{cc}
i \sqrt{q^2} & - 2i \sqrt{2} m_\ell  \\
- 2i \sqrt{2} m_\ell & i \sqrt{q^2} \\
\end{array}
\right)   \;, \displaybreak[0]
\end{alignat}
where $\beta_{1,2}^{\pm} = \sqrt{ E_{1,2} \pm m_{\ell_{1,2}}}$ as before.
Above we have used $\beta_i^+ \beta_i^- \to  E  \beta_{\ell} $ for $m_{\ell_{1,2}} \to m_{\ell}$ 
since $E^2 = q^2/4$, where $E$ is the energy of either lepton in the rest frame of the lepton pair.
Note that the scalar transitions $S$ and $P$ are necessarily diagonal since $\la_\ell = \la_1 -\la_2 =0$.
Timelike vector and axial Lepton HAs are integrated into the Hadron HAs 
\eqref{eq:HHAs}.

\section{Details on Kinematics for Decay Modes}
\label{app:Conventions}

While within the formalism described in this paper it is not essential to consider the full kinematics of the decay, as the evaluation of the hadronic and leptonic HAs can be performed within their respective rest frames, we collect here the kinematics used in calculating the angular distribution using 
the Dirac trace technology  approach \cite{KSSS99,Tub02}  in order to facilitate comparison. 
The K\"all\'en function that often appears in our results is given by
\begin{equation}
\label{eq:Kallen}
 \lambda(a,b,c) \equiv a^2 + b^2 + c^2 - 2(ab+ac+bc) \;.
\end{equation}
For a decay $A \to B +C$, in the rest-frame of $A$,  it is related to the absolute value 
the spatial momentum of the $B$ and $C$ particles as
\begin{equation}
\label{eq:Kallenmomentum}
|\vec{p_B}| = |\vec{p_C}|  =  \frac{\sqrt{\la \left(m_A^2,m_B^2,m_C^2 \right)}}{2 m_A} \; . 
\end{equation}

\subsection{Basis-dependent kinematics  for 
$\bar B \to  \bar K^* \lone \ltwobar $   }

We parametrise the kinematics of the ($\ell_1 = \ell^-$ and $\ell_2=\ell^-$)
\begin{equation}
\bar B \to  \bar K^* \! \left( \to \bar K(p_K) \pi(p_\pi) \right) \ell_1(\ell_1)   \bar \ell_2(\ell_2) 
\end{equation}  
decay mode. 
To do so  we need all four momenta $p_\pi$, $p_K$ ($p = p_\pi+p_K$), 
 $\ell_1$ and $\ell_2$ ($q = \ell_1 + \ell_2$) in a specific frame for which we choose the 
 $\bar{B}$-restframe. It is simplest to first obtain $\ell_{1,2}$ and $p_{\pi,K}$ in the restframe 
 of the lepton pair and the $\bar{K}^*$-meson respectively:
\begin{alignat}{3}
& \ell_{1,2}\text{-restframe}: \qquad  & &  \ell_1^\mu = (E_1 ,  \left|\vec{p}_\ell \right| \, \hat{\ell}) \; , \quad 
& & \ell_2^\mu = (E_2 , -  \left|\vec{p}_\ell \right| \, \hat{\ell}) \; , \nonumber  \\
& p_{\pi,K}\text{-restframe}: & & p_K^\mu = (E_K , \left|\vec{p}_K \right| \, \hat{k}) \; , \quad  
& & p_\pi^\mu = (E_\pi ,- \left|\vec{p}_K \right| \, \hat{k}) \; ,
\end{alignat}
 and the definitions
\begin{alignat}{2}
& \hat{\ell} = (\cos \phi \sin \thetal, - \sin \phi \sin \thetal, \cos \thetal) \; , \quad  
& &  \left|\vec{p}_\ell \right| = \frac{\sqrt{\Kallengas}}{2\sqrt{q^2}} \;, \nonumber \\
& \hat{k} = (-\sin \theta_K, 0, -\cos \theta_K) \; , 
& &  \left|\vec{p}_K \right| = \frac{\sqrt{\KallenKs}}{2 m_{K^*}} \; , 
\end{alignat}
where 
\begin{equation}
\label{eq:Kex}
\Kallengas \equiv  \lambda(q^2,m_1^2,m_2^2) \;, \quad  \KallenKs \equiv \lambda(m_{K^*}^2,m_K^2,m_\pi^2) \;, \quad 
\KallenB \equiv \lambda(m_{B}^2,m_{K^*}^2,q^2) \;,
\end{equation}
are the explicit K\"all\'en function used throughout.
The lepton and hadron energies are then given by 
$E_{1,2} = \sqrt{ m_{\ell_{1,2}}^2+  \left|\vec{p}_\ell \right|^2}$,  $E_{\pi,K} = \sqrt{ m_{\pi,K}^2+  \left|\vec{p}_K \right|^2}$ 
and obey  $E_1 + E_2 = \sqrt{q^2}$ and $E_\pi + E_K = m_{K^*}$.

The polarisation vectors $\eta^\mu (\la)$ of the $\bar{K}^*$-meson in its restframe, using the 
convention in \cite{HZ2013}, are\footnote{The polarisation vector $\eta$ 
corresponds to $\gamma$ in \cite{HZ2013} (c.f. appendix A  therein). The exact correspondence 
between the convention used in \cite{HZ2013}, and also in this paper,
and the Jacob-Wick convention \cite{JW1959,Haber1994} is 
$\eta(\pm)_\mu |_{\mbox{\cite{HZ2013}}}  = -  \eta(\pm)_\mu|_{\mbox{\cite{JW1959}}}$, 
$\eta(0)_\mu |_{\mbox{\cite{HZ2013}}}  =   \eta(0)_\mu|_{\mbox{\cite{JW1959}}}$. 
The final distributions remain the same but the off-diagonal elements 
of the lepton HAs (or matrices) change sign \eqref{eq:LHAs}. 
Note in particular that the hadron HAs \eqref{eq:HHAs} remain unchanged.}
\begin{equation}
 \eta^\mu(0) = (0,0,0,1) \; , \quad
 \eta^\mu(\pm) = (0,\mp 1,i,0)/\sqrt 2 \ .
\end{equation}
In the $\bar{B}$-restframe, $p_B = (m_B,0,0,0)$,  the momenta take the following form
\begin{alignat}{2}
& (\ell_1)^\mu &=\;& (f_\ell(E_1,q_0,q_z) , \left|\vec{p}_\ell \right|  \sin \theta_\ell \cos \phi,  - \left|\vec{p}_\ell \right|  \sin \theta_\ell \sin \phi,f_\ell(E_1,q_z,q_0)) \; , \nonumber  \\
&(\ell_2)^\mu &=\;& (f_\ell(E_2,q_0,-q_z), -\left|\vec{p}_\ell \right| \sin \theta_\ell \cos \phi,   + \left|\vec{p}_\ell \right|  \sin \theta_\ell \sin \phi,f_\ell(E_2,q_z,-q_0)) \; ,\nonumber  \\
 &(p_K)^\mu &=\;&  (f_{K^*}(E_K,p_0,q_z) ,- \left|\vec{p}_K \right| \sin \theta_K, 0, -f_{K^*}(E_K,q_z,p_0)) \; , \nonumber \\
& (p_\pi)^\mu &=\;& (f_{K^*}(E_\pi,p_0,-q_z),\left|\vec{p}_K \right|  \sin \theta_K, 0,-f_{K^*}(E_\pi,q_z,-p_0)) \; ,
\end{alignat}
with $f_\ell(a,b,c)= (a b + c \left|\vec{p}_\ell \right| \cos \theta_\ell)/\sqrt{q^2}$ and $f_{K^*}(a,b,c)= (a b + c \left|\vec{p}_K \right| \cos \theta_K)/{m_K^*}$, and it is easily verified that 
\begin{equation}
\label{eq:small}
q^\mu = (\ell_1+\ell_2)^\mu = (q_0, 0,0,q_z) \;,\quad p^\mu = (p_K + p_\pi )^\mu = (p_0 ,0,0,-q_z) \;,
\end{equation}
$(p_0 = E_{K^*}$) while the polarisation vectors of the $\bar{K}^*$ in the $\bar{B}$-restframe are
\begin{eqnarray}
\eta^\mu(0) = (-q_z,0,0,p_0)/m_{K^*} \; , \quad
\eta^\mu(\pm)= (0,\mp 1,i,0)/\sqrt 2 \; ,
\end{eqnarray}
where $p_0 + q_0 = m_B$ and $q_z =\sqrt{\KallenB}/(2m_B)$, in accordance with 
\eqref{eq:Kallenmomentum}, is the three-momentum of the lepton pair.

For completeness let us add that in the case of:
\begin{itemize}
\item $B \to K^* \lonebar(\lone) \ltwo(\ltwo) $ 
the replacement rule $\hat{\ell} \to \hat{\ell}_{\phi \to -\phi}  = (\cos \phi \sin \thetal,   + \sin \phi \sin \thetal, \cos \thetal)$ applies. Note this is coherent with figure \ref{fig:conventions} in the next section;
\item identical lepton masses the following replacements are in order
\begin{equation}
E_{1,2} \to \sqrt{q^2}/2 \;, \quad \sqrt{\Kallengas} \to (q^2)  \beta_\ell
\end{equation}
where we recall that $\beta_\ell \equiv \sqrt{1-\frac{4 m_\ell^2} {q^2}}$.
\end{itemize}

\subsection{Basis-independent kinematics  for 
$\bar B \to  \bar K^* \lone  \ltwobar $   }

Introducing the notation 
\begin{equation}
\label{eq:big}
Q^\mu = (\ell_1-\ell_2)^\mu    \;,\quad P^\mu = (p_K - p_\pi  )^\mu   \;,
\end{equation}
in addition to \eqref{eq:small}. the invariants that can be formed out of
$p,P,q$ and $Q$ are given by
\begin{alignat}{4}
\label{eq:kinINV}
& q \cdot Q &\;=\;&  \mlone^2 - \mltwo^2 \; , \quad Q^2 = 2\left( \mlone^2 + \mltwo^2 \right) - q^2 \; , 
\quad q \cdot p  =  \frac{1}{2}\left( \mB^2 - \mKs^2 - q^2 \right) \;, 
\nonumber \\
 & p \cdot P &\;=\;& \mK^2 - \mpi^2 \; , \quad  P^2 = 2\left( \mK^2 + \mpi^2 \right) - \mKs^2 \; ,  \quad 
 q \cdot P =  \frac{2 p \cdot P \, q \cdot p + \cos \theta_K \sqrt{\KallenB \KallenKs}}{2 \mKs^2} \; , \nonumber \\
 & Q \cdot P &\;=\;&  \frac{p \cdot P \sqrt{ \KallenB \Kallengas  } \cos \theta_\ell + 2 q \cdot p \sqrt{ \KallenKs \Kallengas  } \cos \theta_K \cos \theta_\ell }{2 \mKs^2 q^2 }+ \frac{ \sqrt{ \KallenKs \Kallengas  } \sin \theta_K \sin \theta_\ell \cos \phi }{\mKs \sqrt{ q^2 } }     \nonumber  \\
 & & \;+\;&  \frac{ q \cdot Q \,   q \cdot P}{q^2}  \;,
 \nonumber  \\
 & Q \cdot p &\;=\;&  \frac{2 q \cdot Q \,  q \cdot p + \cos \theta_\ell \sqrt{\KallenB \Kallengas}}{2 q^2} \; , 
\quad  \nonumber \\
& & & \!\!\!\!\!\!\!\!\!\!\!\! \!\!\!\!\!\!\!\!\!   \epsilon\left(P,  p , Q, q\right) =  - \frac{\sin \theta_K \sin \theta_\ell \sin \phi \sqrt{ \KallenB \KallenKs \Kallengas} }{2 \mKs \sqrt{ q^2 }  }
 \end{alignat}
with $p^2 = m_{K^*}^2$,  the $\la$'s defined in equation~\eqref{eq:Kex}, 
$\epsilon\left(P, p, Q, q\right) =   \epsilon_{\al\be\ga\de} 
P^\al p^\be  Q^\ga q^\de$ 
and the $\epsilon_{0123} = 1$ convention for the Levi-Civita tensor.
Note that the kinematic invariants for $ B \to  K^* \lonebar(\ell_1) \ltwo(\ell_2)$
are the same up to $\epsilon\left(P, p, Q, q\right) \to - \epsilon\left(P, p, Q, q\right)$
which originates from the only change in angles $\phi \to -\phi$.

\section{Specific Results for $\bar B \to \bar K^*(\to \bar K \pi) \lone \ltwobar$}
\label{app:BtoKs}

\subsection{Fourfold Differential Decay Rate}
\label{app:dGammaKstar}

The angular distribution for $\bar B \to \bar K^*(\to \bar K \pi) \lone \ltwobar$ is usually presented in the form (e.g. \cite{BDH2013}) 
\begin{alignat}{1}
\label{eq:d4GJi}
\frac{8 \pi}{3} \frac{d^4 \Gamma}{dq^2 \; d\textrm{cos}\thetal \; d\textrm{cos}\theta_K \; d \hel }  = 
\frac{I^{(0)}_{K^*}}{4}  =  & \left(\J_{1s} + \J_{2s}\cos 2\thetal + \J_{6s} \cos \thetal \right) \sin^2 \theta_K \,+  \nonumber \\
&{} \left(\J_{1c} + \J_{2c}\cos 2\thetal + \J_{6c} \cos \thetal \right) \cos^2 \theta_K \,+  \nonumber \\
&{} \left(\J_3 \cos 2 \hel +\J_9  \sin 2\hel\right)\sin^2 \theta_K \sin^2 \thetal \, +  \nonumber \\
&{} \left(\J_4 \cos  \hel +\J_8  \sin \hel\right) \sin 2 \theta_K \sin 2\thetal  \,+  \nonumber \\
&{} \left(\J_5 \cos  \hel +\J_7  \sin \hel\right) \sin 2 \theta_K \sin \thetal \; , 
\end{alignat}
which can be condensed as
\begin{alignat}{1}
\frac{8 \pi}{3} \frac{d^4 \Gamma}{dq^2 d\textrm{cos}\thetal \; d\textrm{cos}\theta_K \; d \hel }  = 
    \Rea   \left[   \right.  &
       \left(\J_{1s} + \J_{2s}\cos 2\thetal + \J_{6s} \cos \thetal \right) \sin^2 \theta_K \,  + 
   \nonumber \\
& \left(\J_{1c} + \J_{2c}\cos 2\thetal + \J_{6c} \cos \thetal \right) \cos^2 \theta_K \,+  \nonumber \\
&   e^{-2i\hel} \mathcal{G}_3 \sin^2 \theta_K \sin^2 \thetal  \,+ \nonumber  \\
&  e^{-i\hel} \sin 2\theta_K  \left(\mathcal{G}_4 \sin 2 \thetal +  \mathcal{G}_5  \sin \thetal \right) 
 \left.  \right] \; ,
\end{alignat}
where we have defined
\begin{equation}
\label{eq:calJ}
\mathcal{G}_{3,4,5} =  \left(\J_{3,4,5} + i \J_{9,8,7} \right) \; .
\end{equation}
We have introduced the notation $\J_i$ rather than $J_i$ in order to minimise the potential of confusion due to the angular conventions discussed in 
appendix~\ref{app:ang-conventions}.
The relationship between the $\J_i (q^2)$ and the $\GKstar{l_K}{l_\ell}{m}(q^2)$ was given in \eqref{eq:GasJ} but is repeated here for convenience:
\begin{alignat}{3}
\label{eq:GasJappendix}
& \GKstar{0}{0}{0}= \frac{4}{9} \left( 3 \left( \J_{1c} + 2 \J_{1s} \right) - \left( \J_{2c} + 2 \J_{2s} \right) \right) \; , \quad
& & \GKstar{0}{1}{0} = \frac{4}{3} \left( \J_{6c} + 2 \J_{6s}  \right)\; , \quad
& & \GKstar{0}{2}{0} = \frac{16}{9} \left(   \J_{2c} + 2 \J_{2s} \right) \; , \nonumber \\
& \GKstar{2}{0}{0} = \frac{4}{9} \left( 6 \left(\J_{1c} - \J_{1s}\right) - 2 \left(\J_{2c} -  \J_{2s} \right) \right)\; , \quad
& & \GKstar{2}{1}{0} = \frac{8}{3} \left(  \J_{6c} - \J_{6s}  \right)\; , \quad
& & \GKstar{2}{2}{0} = \frac{32}{9} \left( \J_{2c} - \J_{2s} \right) \; , \nonumber \\
& \GKstar{2}{1}{1} = \frac{16}{\sqrt{3}}\underbrace{(\J_5 + i \J_7)}_{= \mathcal{G}_5}\; , \quad
& & \GKstar{2}{2}{1} = \frac{32}{3}\underbrace{(\J_4 + i \J_8)}_{= \mathcal{G}_4}  \; , \quad
&  & \GKstar{2}{2}{2} = \frac{32}{3}\underbrace{(\J_3 + i \J_9)}_{= \mathcal{G}_3} \; . 
\end{alignat}
Explicit results for the $\GKstar{l_K}{l_\ell}{m}$ are presented in section~\ref{sec:explicitGsamemass} for the case of identical final-state leptons $\mlone = \mltwo$ and section~\ref{sec:explicitGdifferentmass} for the more general case $\mlone \neq \mltwo$.

\subsubsection{Kinematic endpoint relations in terms of $\GKstar{l_K}{l_\ell}{m}$ }
\label{app:endpoint}

In Ref.~\cite{HZ2013} it was shown that the HAs obey symmetry relations at the kinematic endpoint due
to symmetry enhancement. This is due to the $K^*$ being at rest resulting in symmetry in all space directions i.e. helicity directions. The relations for the HAs in equation~(13) in \cite{HZ2013} lead  
to the following equivalent of equation~(21) in \cite{HZ2013}
\begin{equation}
\label{eq:endpoint}
\GKstar{0}{0}{0} \neq 0 \;, \quad \GKstar{2}{2}{0} \to  \Rea[\GKstar{2}{2}{0}]    \;, \quad  \GKstar{2}{2}{1}  \to -2   \Rea[\GKstar{2}{2}{0}]  \; ,
\quad \GKstar{2}{2}{2} \to 2   \Rea[\GKstar{2}{2}{0}]  \;,
\end{equation}
with all other five $\GKstar{l_K}{l_\ell}{m}$ vanishing. 
Recall that $\GKstar{0}{0}{0}$ is proportional to the total decay rate. 
The relations between the $ \GKstar{2}{2}{m}$  are not accidental but have to do with the symmetries 
of a multiplet. The factor of two between $ \GKstar{2}{2}{0}$ and  $\GKstar{2}{2}{1(2)} $, once more,  
originates from absorbing $\GKstar{2}{2}{-1(2)} $ into $\GKstar{2}{2}{1(2)} $.
 The results of the threshold expansion, linear in the $K^*$ momentum $\kappa \sim \la(m_B^2, m_{K^*}^2, q^2) $, can be inferred from equation~(30) in \cite{HZ2013} taking into account the different angular conventions detailed 
 in figure \ref{fig:conventions}.

\subsubsection{Comparison of angular distribution with the literature}
\label{sec:comparison}

The angular distribution \eqref{app:dGammaKstar}  was first computed in 
the SM for massless leptons in \cite{KSSS99}, extended to include equal lepton masses in \cite{Tub02,KM05}.  A full dimension-six operator basis was considered 
in \cite{KT07}.
The basis was extended to lepton mass corrections for   (pseudo)scalar 
operators in  \cite{ABBBSW2008},  enforcing the $\J_{6c}$-structure,  and tensor operators 
by the authors in \cite{Gosh09,BDH2013}. 
We compare our results with regard to \cite{BDH2013}, which is the latest reference.

Taking into account the change  $g_{4,6,7,9} \to - J_{4,6,7,9}$ (cf. figure \ref{fig:conventions}) 
and comparing at the level of form factors (naive factorisation) only we find agreement 
except for tensor interference terms.  
Agreement is established when  $C_{T5} \to - C_{T5}$ in  \cite{BDH2013}. 
The latter might be related to the fact that  the relations 
${\rm tr}[ \ga^\al \ga^\be \ga^\ga \ga^\de \ga_5 ] =  4 \la i \epsilon^{\al\be \ga\de}$ and 
$\sigma^{\al\be} \ga_5 = - \la \frac{i}{2} \epsilon^{\al\be \ga\de} \sigma_{\ga\de} $ 
(with $\la = \pm 1$ depending on conventions - $\la =1$ in this paper) are not consistent with  equation C.16 \cite{BDH2013} (v3). 

A minor difference is that the authors of  \cite{BDH2013} have chosen not to present 
the tensor contribution in $J_{8,9} (\J_{8,9})$, since such contributions vanish in the 
narrow-width approximation.\footnote{In v3 of \cite{BDH2013} it is stated 
that agreement with v4 of \cite{Gosh10} is found up to a sign of an interference term 
between a scalar and a tensor HA. This suggests that we agree with \cite{BDH2013} 
but disagree with \cite{Gosh10} on that sign, as well as the sign of $C_{T5}$.}
In addition, we find that a few of the HAs in \cite{BDH2013} equation C.13
do not agree with their definitions. These disagreements do, however, drop out in the final expression.

\subsection{Angular conventions}
\label{app:ang-conventions}

In this section  we discuss and compare the  LHCb and  theory angular conventions.  
The main result is shown in form of a commutative diagram in figure \ref{fig:conventions}.  
We proceed by first discussing the CP-conjugate modes in 
each case and then link the conventions with each other.

The LHCb conventions \cite{LHCbKmumu2013}, which are the same as adapted in this paper, are shown 
in figure \ref{fig:decay_planes}. The rationale behind the definition of the conjugate mode 
 is as follows. 
Firstly, particles are mapped into antiparticles, corresponding to a C-transformation. Then the momenta of all particles are reversed, changing the angle 
$\phi \to 2 \pi - \phi$. This leads to sign changes in $\J_{7,8,9}$.
Hence the conjugate mode corresponds to a full CP-transformation 
$$ \frac{ d^4  \bar{\Gamma}}{dq^2  d\cL d \cK d\phi  } \Big|_{\rm LHCb} = 
\frac{d^4   \Gamma}{dq^2  d\cL d \cK d\phi  } \Big|_{\rm CP} \;,   $$
and  the quantity
$$
\frac{d^4 ( \Gamma \pm \bar{\Gamma}) }{dq^2  d\cL d \cK d\phi}\Big|_{\rm LHCb} 
\;, \qquad 
$$ 
is therefore CP-even (-odd). Above $\Gamma = \Gamma( \bar B \to \bar K^* \lone \ltwobar)$ and $\bar \Gamma = \Gamma(  B \to  K^* \lonebar \ltwo)$.

The theory conventions for CP conjugates are such that they facilitate the implementation of decays which are not self-tagging 
(such as $\bar B_s, B_s \to \phi(\to K^+ K^-) \ell^+ \ell^-$ at hadron colliders). 
When going between conjugate modes the conventions are that the angles transform as $( \thetal,\theta_K,\phi)   \to (\pi - \thetal,\pi - \theta_K,2 \pi -\phi),$\footnote{
Equivalently one can use the angular redefinitions $(\pi - \thetal,\theta_K,\pi-\phi)$  
and $(  \thetal - \pi ,\theta_K,2 \pi -\phi )$, which are sometimes stated in the literature.}
which leads to sign changes in $\J_{5,6,8,9}$. This transformation rule corresponds to 
a full CP-conjugation, but with the angles $\thetal,\theta_K$ associated to the same particle rather than the anti-particle.

To find the transformation between the theory and LHCb conventions is not straightforward because it is difficult 
 to find a theory paper that  resolves the four-fold ambiguity of defining the angle $\phi$ and/or shows a figure consistent with the definitions used in 
 the corresponding work.
We have taken a pragmatic route in verifying that the results in  \cite{KM05,ABBBSW2008,BDH2013} 
agree with each other for common contributions, and crucially that our results are in agreement with these contributions 
for $\bar B \to \bar K^* \lone \ltwobar$ if  $J_{4,6,7,9} = - \J_{4,6,7,9}$ and  
$J_{1,2,3,5,8} =  \J_{1,2,3,5,8}$.  
This completes the diagram in figure \ref{fig:conventions}.\footnote{One can come to the same conclusion in another way \cite{Serra}.  
Let us again consider $\bar B \to \bar K^*  \lone \ltwobar$.  
In general $\theta_K| _{\rm LHCb} = \theta_K| _{\rm theory}$ 
is chosen to be the same angle and the theoretical community chooses 
$\theta_\ell| _{\rm LHCb}  = \pi- \thetal|_{\rm theory}$.
The only unknown remains the angle $\phi$, for which one may use the scalar- and cross-product definitions. Using Appendix A in 
\cite{LHCbKmumu2013} and likewise in \cite{KM05}, we infer that 
$\cos \phi|_{\rm LHCb} = \cos \phi|_{\rm theory} $ 
and $\sin \phi|_{\rm LHCb} = - \sin \phi|_{\rm theory}$. Taking all angular changes into 
account this results in sign changes in  $\J_{4,6,7,9}$ which is consistent with our 
explicit computations mentioned above.}

\begin{figure}[h!]
\begin{center}
\label{fig:conventions}
\includegraphics[scale=1.1,clip=true,trim= 100 330 100 285]{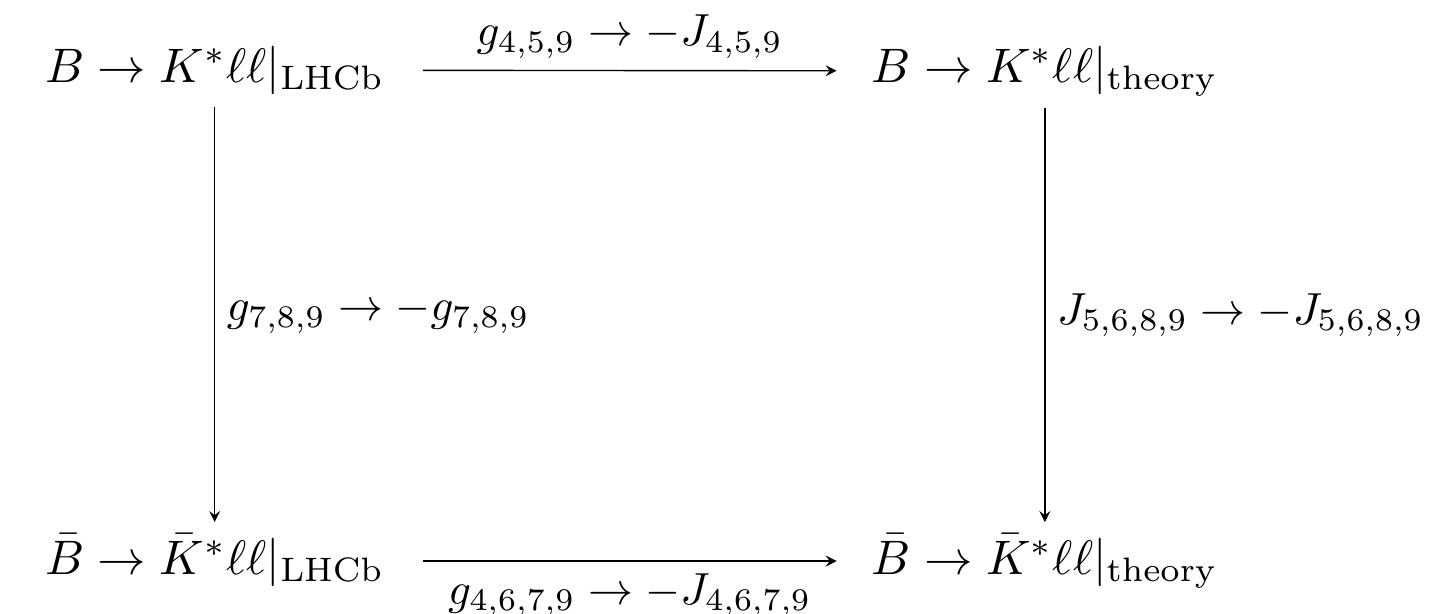} 
 \end{center}
 \caption{\small Changes of angular functions when going from one mode to the other. 
 For CP conjugates the conjugation of the CP-odd (weak) phases are suppressed. 
 Angular functions whose signs do not change are not indicated. }
\end{figure}

In summary,
\begin{alignat}{2}
\label{eq:sum}
& \frac{d^4 ( \Gamma \pm \bar{\Gamma}) }{dq^2  d\cL d \cK d\phi}\Big|_{\rm LHCb}   & \quad\Leftrightarrow \quad &  S[A]_{1,2,3,4,5,6,7,8,9} \;, \nonumber \\
& \frac{d^4 ( \Gamma \pm \bar{\Gamma}) }{dq^2  d\cL d \cK d\phi}\Big|_{\rm theory} & \quad \Leftrightarrow \quad &  S[A]_{1,2,3,4,7}, A[S]_{5,6,8,9} \;,
\end{alignat}
where the CP-even (-odd) quantities are
\begin{equation}
\label{eq:SA}
S_i [A_i] = \frac{ \J_i \pm \J_i^{\rm CP}}{\Gamma + \Gamma^{\rm CP}} \;
\end{equation}
with adapted notation from  \cite{ABBBSW2008}.
Written in yet another way  \eqref{eq:sum} is equivalent to 
\begin{alignat}{2}
\label{eq:short}
& (\J,A,S)_{1,2,3,5,8}|_{\rm LHCb} &\; = \;&  + (J,A,S)_{1,2,3,5,8}|_{\rm theory} 
\;,  \nonumber \\
& (\J,A,S)_{4,6,7,9}|_{\rm LHCb} &\; = \;&  - (J,A,S)_{4,6,7,9}|_{\rm theory}  \;.
\end{alignat}
 In order to understand \eqref{eq:sum} and \eqref{eq:short} one has to keep in mind that $\bar B \to \bar K^*  (\lone \ltwobar)$ 
rather than its conjugate is the reference decay.  Note that at the LHCb (hadron collider) $B_s \to \phi \mu^+ \mu^-$ is untagged and 
therefore, setting aside the issue of production asymmetry, only $S_{1,2,3,4,7}, A_{5,6,8,9} $ are experimentally accessible.

 \subsubsection{Angular observables in the literature and conventions}
 \label{app:Pp-conversion}
 
 We aim to find the relation between  angular $P_i^{(')}$ observables 
 as defined  by the theorists 
 \cite{DHMV-Observables} and their adaptation by LHCb \cite{LHCb-P5p}. 
In matching the results and creating the dictionary one needs to pay 
attention to the fact that   \cite{LHCb-P5p} and \cite{DHMV-Observables}
define the $P_i'$ in terms  of $\J_i$ and  $J_i$  differently, as well as the 
 different angular conventions for $\J_i$ and $J_i$  per se (shown in figure \ref{fig:conventions})!

Amongst the twelve observables discussed in section \ref{sec:relation}, eight of them, $P_{1,2,3}, P'_{4,5,6,8}$ and $A_{\rm FB}$, depend on angles and definitions.
 
The $P_i'$ and $A_{\rm FB}$  are defined by LHCb \cite{LHCb-P5p} as
 \begin{equation}
 P_{4,5,6,8}'|_{\rm LHCb} = \frac{S_{4,5,7,8}|_{\rm LHCb}}{\sqrt{F_L(1-F_L)}}  \;, \quad 
  A_{\rm FB}|_{\rm LHCb} = \frac{3(S_{6s}  |_{\rm LHCb})}{4 ( \Gamma + \bar \Gamma)}  \;, 
 \end{equation}
 where $S_i$ is defined in equation~\eqref{eq:SA} and 
  $2 \Nbin = \sqrt{F_L(1-F_L)}$ in our notation used in section \ref{sec:relation}.
 LHCb has not defined $P_{1,2,3}$  and we shall  assume the same functional 
 form  as in the theory paper \cite{DHMV-Observables}. 
 
In \cite{DHMV-Observables} the eight equivalent angular observables are defined as follows\footnote{Note that \cite{ABBBSW2008,BDH2013} define 
 $A_{\rm FB} =  \frac{3 S_{6s} }{4 ( \Gamma + \bar \Gamma) }  $ which results 
 in $A_{\rm FB}|_{\mbox{\cite{ABBBSW2008,BDH2013}}} = - A_{\rm FB}|_{\rm LHCb}$.}
 \begin{alignat}{4}
 & P_1 &\;=\;&  \frac{1}{2 S_{2s}} S_3  &\;=\;&  \frac{1}{2 S_{2s}} (S_3|_{\rm LHCb})     &\;=\;& + P_1|_{\rm LHCb}  \nonumber  \;,  \\
 & P_2 &\;=\;& \frac{1}{8 S_{2s}} S_{6s}   &\;=\;&  \frac{1}{8 S_{2s}} (- S_{6s}|_{\rm LHCb})     &\;=\;& - P_2|_{\rm LHCb}  \nonumber  \;,  \\
  & P_3&\;=\;& \frac{- 1}{4 S_{2s}} S_{9}    &\;=\;&  \frac{- 1}{4 S_{2s}} (-S_{9}|_{\rm LHCb})  &\;=\;& 
  - P_3|_{\rm LHCb}    \nonumber \;,  \\
 & P_4' &\;=\;& \frac{1}{\Nbin} S_4 &\;=\;&  \frac{1}{\Nbin} ( - S_4 |_{\rm LHCb})  &\;=\;&   
 -2P_4'|_{\rm LHCb}   \nonumber \;,  \\
 & P_5' &\;=\;& \frac{1}{2 \Nbin} S_5 &\;=\;&  \frac{1}{2 \Nbin} ( S_5  |_{\rm LHCb})  &\;=\;& + P_5'|_{\rm LHCb} \nonumber \;, \\
  & A_{\rm FB} &\;=\;& - \frac{3 S_{6s} }{4 ( \Gamma + \bar \Gamma) }  &\;=\;& 
   - \frac{3(-S_{6s}  |_{\rm LHCb})}{4 ( \Gamma + \bar \Gamma)}  &\;=\;& +A_{\rm FB}|_{\rm LHCb}
    \nonumber \;, \\
  & P_6' &\;=\;& \frac{- 1}{2 \Nbin} S_7 &\;=\;&  \frac{-1}{2 \Nbin} ( - S_7 |_{\rm LHCb})  &\;=\;& 
  + P_6'|_{\rm LHCb}  \nonumber \;,  \\
 & P_8' &\;=\;& \frac{- 1}{ \Nbin} S_8 &\;=\;&  \frac{-1}{\Nbin} (  S_8 |_{\rm LHCb})  &\;=\;&  
   - 2 P_8'|_{\rm LHCb}    \nonumber \;,
 \end{alignat}
 where we have directly translated into the LHCb conventions. 
 It seems that we differ from the theory community in the sign of the observables 
 $S[A]_{7,8,9}$.  
 For example, both  
$ P_6' = P_6'|_{\rm LHCb} $ and $ P_8' = -2P_8'|_{\rm LHCb}$ differ from 
the relations given in the caption of table 1 in \cite{Descotes-Genon:2013wba}  
by the aforementioned sign. 
Our relation  $S[A]_9 = - S[A]_9|_{\rm LHCb} $ also differs from the one given by \cite{Altmannshofer:2013foa} 
in table 1 by a sign.

\subsection{$\GKstar{l_k}{l_\ell}{m} $ for $\bar B \to \bar K^* \lone  \ltwobar$ in terms of Helicity Amplitudes for $m_{\ell_i}  \equiv m_\ell$}
\label{sec:explicitGsamemass}
When the masses of the leptons are identical, we obtain for 
$\GKstar{l_K}{l_\ell}{m} = \NN q^2 \GKstaro{l_K}{l_\ell}{m}$ (with  $\NN$ defined in \eqref{eq:NN}), 
\begin{alignat}{2}
\label{eq:explicitGsamemass}
& \GKstaro{0}{0}{0} &&= \frac{4}{9}\left( 1 - \hat{\ml}^2 \right) \left(\left|\HVp\right|^2+\left|\HVm\right|^2 + \left|\HVzero\right|^2 +\left( V \to A \right) \right)  \nonumber \\
& && {} + \frac{4 \hat{\ml}^2}{3} \left(\left|\HVp\right|^2+\left|\HVm\right|^2 + \left|\HVzero\right|^2 -\left( V \to A \right) \right) + \frac{2}{3}  \beta_\ell^2 \left|\HS\right|^2 + \frac{2}{3}  \left|\HPS\right|^2  \nonumber \\
& && {} + \frac{8}{9} \left( 1 + 8 \hat{\ml}^2 \right) \left(  \left|\HTenstp\right|^2+\left|\HTenstm\right|^2+\left|\HTenstzero\right|^2\right) + \frac{4}{9}   \beta_\ell^2\left(  \left|\HTensp\right|^2+\left|\HTensm\right|^2+\left|\HTenszero\right|^2\right) \nonumber \\
& && {} + \frac{16}{3} \hat{\ml}  \, \Ima \left[  \HVp \HTenstpbar+\HVm \HTenstmbar +\HVzero \HTenstzerobar\right]   \; , \displaybreak[0]\nonumber \\[0.2cm]
&\GKstaro{0}{1}{0} &&= \frac{4  \beta_\ell }{3} \left( \Rea \left[\HVp \HApbar - \HVm \HAmbar \right] + \Ima \left[\sqrt{2} \HTenszero \HPSbar +2 \HTenstzero \HSbar\right]  \right.  \nonumber \\
& &&  \phantom{= \frac{4 \beta_\ell}{3} \Bigg(}\left.{}- 2 \hat{\ml}  \Rea \left[ \HVzero \HSbar \right]   + 4 \hat{\ml}  \Ima \left[\HAp \HTenstpbar -\HAm \HTenstmbar \right]  \right) \; , \displaybreak[0]\nonumber \\[0.2cm]
&\GKstaro{0}{2}{0} &&= -\frac{2}{9}   \beta_\ell^2 \left(2 \left|\HVzero \right|^2 - \left|\HVp \right|^2 - \left|\HVm \right|^2 + \left( V \to A \right) - 2 \left(2 \left|\HTenszero\right|^2 - \left|\HTensp\right|^2- \left|\HTensm\right|^2 \right) \right.  \nonumber \\
& && \phantom{=-\frac{2}{9}  q^2 \beta_\ell^2\Bigg(} \left.  - 4 \left(2 \left|\HTenstzero\right|^2 - \left|\HTenstp\right|^2- \left|\HTenstm\right|^2 \right)  \right) \;, \displaybreak[0]\nonumber \\[0.2cm]
&\GKstaro{2}{0}{0} &&=-\frac{4}{9}\left(1 - \hat{\ml}^2\right) \left(\left|\HVp\right|^2+\left|\HVm\right|^2 - 2\left|\HVzero\right|^2 +\left( V \to A \right) \right)  \nonumber \\
& && {} - \frac{4 \hat{\ml}^2}{3} \left(\left|\HVp\right|^2+\left|\HVm\right|^2 - 2 \left|\HVzero\right|^2 -\left( V \to A \right) \right) + \frac{4}{3}  \beta_\ell^2 \left|\HS\right|^2 + \frac{4}{3}   \left|\HPS\right|^2  \nonumber \\
& && {} - \frac{8}{9} \left( 1 + 8 \hat{\ml}^2 \right) \left(  \left|\HTenstp\right|^2+\left|\HTenstm\right|^2 - 2 \left|\HTenstzero\right|^2\right) - \frac{4}{9}  \beta_\ell^2  \left(  \left|\HTensp\right|^2+\left|\HTensm\right|^2 - 2 \left|\HTenszero\right|^2\right) \nonumber \\
& && {} - \frac{16}{3} \hat{\ml}  \,  \Ima \left[  \HVp \HTenstpbar+\HVm \HTenstmbar - 2 \HVzero \HTenstzerobar\right]  \; , \displaybreak[0]\nonumber \\[0.2cm]
&\GKstaro{2}{1}{0} &&= -\frac{4  \beta_\ell}{3} \Big( \Rea \left[\HVp \HApbar - \HVm \HAmbar \right]  -2 \, \Ima \left[\sqrt{2} \HTenszero \HPSbar +2 \HTenstzero \HSbar\right]    \nonumber \\ 
& &&  \phantom{= -\frac{4  \beta_\ell}{3} xx }  {} + 4\hat{\ml}  \left( \Rea \left[ \HVzero \HSbar \right] +   \Ima \left[\HAp \HTenstpbar -\HAm \HTenstmbar \right] \right) \Big) \; , \displaybreak[0]\nonumber \\[0.2cm]
&\GKstaro{2}{2}{0} &&=  -\frac{2}{9}   \beta_\ell^2 \left(4 \left|\HVzero \right|^2 + \left|\HVp \right|^2 + \left|\HVm \right|^2 + \left( V \to A \right) - 2 \left(4 \left|\HTenszero\right|^2 + \left|\HTensp\right|^2 + \left|\HTensm\right|^2 \right) \right.  \nonumber \\
& && \phantom{=-\frac{2}{9}  q^2 \beta_\ell^2 \Bigg(} \left.  - 4 \left(4 \left|\HTenstzero\right|^2 + \left|\HTenstp\right|^2 + \left|\HTenstm\right|^2 \right)  \right)\nonumber \;, \\[0.2cm]
&\GKstaro{2}{1}{1}&&=  \frac{4  \beta_\ell }{\sqrt{3} } \left( \HVp \HAzerobar + \HAp \HVzerobar - \HVzero \HAmbar - \HAzero \HVmbar + 2 \hat{\ml}  \left(\HVp \HSbar + \HS \HVmbar \right)  \right. \nonumber \\
& && \phantom{=- \frac{\sqrt{\Kallengas} }{\sqrt{3} } \Bigg( } {}  - \sqrt{2} i \left( \HPS \HTensmbar -  \HTensp \HPSbar  + \sqrt{2} \left( \HS \HTenstmbar - \HTenstp \HSbar \right) \right) \nonumber \\
& && \left.\phantom{= -\frac{\sqrt{\Kallengas} }{\sqrt{3} } \Bigg( }{} - 4 i  \hat{\ml}  \left( \HAp \HTenstzerobar + \HTenstzero \HAmbar - \HTenstp \HAzerobar - \HAzero \HTenstmbar \right)  \right)\; , \displaybreak[0] \nonumber \\[0.2cm]
&\GKstaro{2}{2}{1}&&=  \frac{4}{3}   \beta_\ell^2 \left( \HVp \HVzerobar + \HVzero \HVmbar +\left(V \to A \right) - 2 \left( \HTensp \HTenszerobar+\HTenszero \HTensmbar +2 \left(\HTenstp \HTenstzerobar+\HTenstzero \HTenstmbar \right)  \right)\right) \; , \displaybreak[0]\nonumber \\[0.2cm]
&\GKstaro{2}{2}{2}&&= - \frac{8}{3}  \beta_\ell^2 \left(\HVp \HVmbar +\HAp \HAmbar -2 \left( \HTensp \HTensmbar+ 2  \HTenstp \HTenstmbar\right) \right) \; , \displaybreak[0]
\end{alignat}
where $\hat{m}_\ell = m_\ell/\sqrt{q^2}$ and   we recall that $\beta_\ell \equiv \sqrt{1-4 \hat{m}_\ell^2}$. 
 The number $m$ in $\GKstar{l_k}{l_\ell}{m} $ corresponds to the units of plus helicities.
 The common factor of $q^2$ in all observables as compared with standard literature results is a consequence of our choice of normalisation, whereby all global factors are placed outside the HAs. The factors of $i$ where they appear (explicitly and implicitly) in $\GKstaro{2}{1}{1}$, $\GKstaro{2}{2}{1}$ and $\GKstaro{2}{2}{2}$ are not accidental, as the results given above are complex and one must take the real and imaginary parts of these results to recover the observables $\J_{3,4,5,7,8,9}$. 

Note that it is sometimes convenient to express results in terms of the transversity amplitudes, which possess a definite parity. The relations to the HAs used throughout this paper are 
\begin{alignat}{4}
&  H_{\parallel(\perp)}^{L/R} &\;\equiv\;& \frac{1}{\sqrt{2}}(H^{L/R}_+ \pm H^{L/R}_{-})  \;, & & & &   \nonumber \\[0.1cm]
& H_S &\; \equiv \;& \HSca \; , && H_t  &\; \equiv \; & \HTime \nonumber  \;, \\[0.1cm]
& H_{\parallel(\perp)}^{{\cal T}_t} &\;\equiv\;& \frac{1}{\sqrt{2}}( \HTenstp \pm \HTenstm) \;, \quad & &   
H_{0}^{{\cal T}_t} &\;\equiv\;& \HTenstzero  \nonumber \;,  \\[0.1cm]
&  H_{\parallel(\perp)}^{{\cal T}} &\;\equiv\;& \frac{1}{\sqrt{2}}(\HTensp \pm \HTensm )  \;, \quad & &   
H_{\parallel \perp}^{{\cal T}} &\;\equiv\;& \HTenszero  \; .
\end{alignat}
In \cite{BDH2013} the notation $A_{ij}$, with $i,j = \parallel,\perp,0$, is used for the transversity amplitudes. 
Note that when comparing to this paper the difference in the convention of the polarisation vectors 
has to be taken into account.

\subsection{$\GKstar{l_k}{l_\ell}{m} $ for $\bar B \to \bar K^* \lone  \ltwobar$  in terms of Helicity Amplitudes for $m_{\lone} \neq m_{\ltwo}$}
\label{sec:explicitGdifferentmass}

The formalism discussed in this paper allows a simple extension to the case $m_{\lone} \neq m_{\ltwo}$, so that the results presented in \eqref{eq:explicitGsamemass} can be adapted to test for possible lepton-flavour violating processes. Using the notation 
\begin{equation} 
\frac{\Kallengas}{ 4 q^2} = \bonep \bonem \btwop \btwom  \;,
\end{equation}
($\Kallengas$ given in \eqref{eq:Kex} and $\beta_{1,2}^{\pm} \equiv \sqrt{E_{1,2} \pm m_{\ell_{1,2}}} $), we obtain the following  expressions for $\GKstar{l_K}{l_\ell}{m} = \NN \GKstart{l_K}{l_\ell}{m}$ (with  $\NN$ defined in \eqref{eq:NN}) :
\begin{alignat}{2}
\label{eq:explicitGdifferentmass}
&\GKstart{0}{0}{0} &&= \frac{4}{9}\left( 3 E_1 E_2+ \frac{\Kallengas}{4q^2}\right) \left(\left|\HVp\right|^2+\left|\HVm\right|^2 + \left|\HVzero\right|^2 +\left( V \to A \right) \right)  \nonumber \\
& && {} + \frac{4 \mlone \mltwo}{3} \left(\left|\HVp\right|^2+\left|\HVm\right|^2 + \left|\HVzero\right|^2 -\left( V \to A \right) \right) \nonumber \\
& && {} + \frac{4}{3} \left(  E_1 E_2 - \mlone \mltwo  +\frac{\Kallengas}{4q^2} \right) \left|\HS\right|^2 + \frac{4}{3} \left(  E_1 E_2 + \mlone \mltwo  + \frac{\Kallengas}{4q^2} \right) \left|\HPS\right|^2  \nonumber \\
& && {} + \frac{16}{9} \left( 3 \left( E_1 E_2 + \mlone \mltwo \right) - \frac{\Kallengas}{4q^2} \right) \left(  \left|\HTenstp\right|^2+\left|\HTenstm\right|^2+\left|\HTenstzero\right|^2\right) \nonumber \\
& && {} + \frac{8}{9} \left( 3 \left( E_1 E_2 - \mlone \mltwo \right) - \frac{\Kallengas}{4q^2} \right) \left(  \left|\HTensp\right|^2+\left|\HTensm\right|^2+\left|\HTenszero\right|^2\right) \nonumber \\
& && {} + \frac{16}{3} \left( \mlone E_2 + \mltwo E_1 \right) \Ima \left[  \HVp \HTenstpbar+\HVm \HTenstmbar +\HVzero \HTenstzerobar\right] \nonumber \\
& && {} + \frac{8 \sqrt{2}}{3} \left(  \mlone E_2 - \mltwo E_1 \right) \Ima \left[  \HAp \HTenspbar+\HAm \HTensmbar +\HAzero \HTenszerobar \right]  \; , \displaybreak[0]\nonumber \\[0.2cm]
&\GKstart{0}{1}{0} &&= \frac{4\sqrt{\Kallengas}}{3} \left( \Rea \left[\HVp \HApbar - \HVm \HAmbar \right] + 2 \sqrt{2} \frac{\mlone^2-\mltwo^2}{q^2} \Rea \left[\HTensp \HTenstpbar - \HTensm \HTenstmbar \right] \right.  \nonumber \\ 
& &&  {} +2\frac{\mlone + \mltwo}{\sqrt{q^2}} \Ima \left[\HAp \HTenstpbar -\HAm \HTenstmbar \right]  +\sqrt{2}\frac{\mlone - \mltwo}{\sqrt{q^2}} \Ima \left[\HVp \HTenspbar -\HVm \HTensmbar \right] \nonumber \\
& && \left.{} - \frac{\mlone - \mltwo}{\sqrt{q^2}}  \Rea \left[ \HAzero \HPSbar \right]- \frac{\mlone + \mltwo}{\sqrt{q^2}} \Rea \left[ \HVzero \HSbar \right]   +  \Ima \left[\sqrt{2} \HTenszero \HPSbar +2 \HTenstzero \HSbar\right]  \right) \; , \displaybreak[0]\nonumber \\[0.2cm]
&\GKstart{0}{2}{0} &&= -\frac{2}{9} \frac{\Kallengas}{  q^2 } \left(2 \left|\HVzero \right|^2 - \left|\HVp \right|^2 - \left|\HVm \right|^2 + \left( V \to A \right) - 2 \left(2 \left|\HTenszero\right|^2 - \left|\HTensp\right|^2- \left|\HTensm\right|^2 \right) \right.  \nonumber \\
& && \phantom{=-\frac{2}{9} \frac{\Kallengas}{ q^2 }\Bigg(} \left.  - 4 \left(2 \left|\HTenstzero\right|^2 - \left|\HTenstp\right|^2- \left|\HTenstm\right|^2 \right)  \right) \;, \displaybreak[0]\nonumber \\[0.2cm]
&\GKstart{2}{0}{0} &&=-\frac{4}{9}\left( 3 E_1 E_2+ \frac{\Kallengas}{4q^2}\right) \left(\left|\HVp\right|^2+\left|\HVm\right|^2 - 2\left|\HVzero\right|^2 +\left( V \to A \right) \right)  \nonumber \\
& && {} - \frac{4 \mlone \mltwo}{3} \left(\left|\HVp\right|^2+\left|\HVm\right|^2 - 2 \left|\HVzero\right|^2 -\left( V \to A \right) \right) \nonumber \\
& && {} + \frac{8}{3} \left(  E_1 E_2 - \mlone \mltwo  +\frac{\Kallengas}{4q^2} \right) \left|\HS\right|^2 + \frac{8}{3} \left(  E_1 E_2 + \mlone \mltwo  + \frac{\Kallengas}{4q^2} \right) \left|\HPS\right|^2  \nonumber \\
& && {} - \frac{16}{9} \left( 3 \left( E_1 E_2 + \mlone \mltwo \right) - \frac{\Kallengas}{4q^2} \right) \left(  \left|\HTenstp\right|^2+\left|\HTenstm\right|^2 - 2 \left|\HTenstzero\right|^2\right) \nonumber \\
& && {} - \frac{8}{9} \left( 3 \left( E_1 E_2 - \mlone \mltwo \right) - \frac{\Kallengas}{4q^2} \right) \left(  \left|\HTensp\right|^2+\left|\HTensm\right|^2 - 2 \left|\HTenszero\right|^2\right) \nonumber \\
& && {} - \frac{16}{3} \left( \mlone E_2 + \mltwo E_1 \right) \Ima \left[  \HVp \HTenstpbar+\HVm \HTenstmbar - 2 \HVzero \HTenstzerobar\right] \nonumber \\
& && {} - \frac{8 \sqrt{2}}{3} \left(  \mlone E_2 - \mltwo E_1 \right) \Ima \left[  \HAp \HTenspbar+\HAm \HTensmbar - 2 \HAzero \HTenszerobar \right]  \; , \displaybreak[0]\nonumber \\[0.2cm]
&\GKstart{2}{1}{0} &&= -\frac{4\sqrt{\Kallengas}}{3} \left( \Rea \left[\HVp \HApbar - \HVm \HAmbar \right] + 2 \sqrt{2} \frac{\mlone^2-\mltwo^2}{q^2} \Rea \left[\HTensp \HTenstpbar - \HTensm \HTenstmbar \right] \right.  \nonumber \\ 
& &&  {} +\frac{2\left(\mlone + \mltwo\right)}{\sqrt{q^2}} \Ima \left[\HAp \HTenstpbar -\HAm \HTenstmbar \right]  +\frac{\sqrt{2}\left(\mlone - \mltwo\right)}{\sqrt{q^2}} \Ima \left[\HVp \HTenspbar -\HVm \HTensmbar \right] \nonumber \\
& && \left.{} + 2\frac{\mlone - \mltwo}{\sqrt{q^2}}  \Rea \left[ \HAzero \HPSbar \right] + 2\frac{\mlone + \mltwo}{\sqrt{q^2}} \Rea \left[ \HVzero \HSbar \right]   - 2 \, \Ima \left[\sqrt{2} \HTenszero \HPSbar +2 \HTenstzero \HSbar\right]  \right) \; , \displaybreak[0]\nonumber \\[0.2cm]
&\GKstart{2}{2}{0} &&=  -\frac{2}{9} \frac{\Kallengas}{ q^2 } \left(4 \left|\HVzero \right|^2 + \left|\HVp \right|^2 + \left|\HVm \right|^2 + \left( V \to A \right) - 2 \left(4 \left|\HTenszero\right|^2 + \left|\HTensp\right|^2 + \left|\HTensm\right|^2 \right) \right.  \nonumber \\
& && \phantom{=-\frac{8}{9} \frac{\Kallengas}{ 4 q^2 } \Bigg(} \left.  - 4 \left(4 \left|\HTenstzero\right|^2 + \left|\HTenstp\right|^2 + \left|\HTenstm\right|^2 \right)  \right)\nonumber \;, \\[0.2cm]
&\GKstart{2}{1}{1}&&= \frac{4\sqrt{\Kallengas} }{\sqrt{3} } \left( \left(  \HVp \HAzerobar + \HAp \HVzerobar - \HVzero \HAmbar - \HAzero \HVmbar \right) + \frac{\mlone + \mltwo}{\sqrt{q^2}} \left(\HVp \HSbar + \HS \HVmbar \right)  \right. \nonumber \\
& && \phantom{=- \frac{\sqrt{\Kallengas} }{\sqrt{3} } \Bigg( } {}  - \sqrt{2} i \left( \HPS \HTensmbar -  \HTensp \HPSbar  + \sqrt{2} \left( \HS \HTenstmbar - \HTenstp \HSbar \right) \right) \nonumber \\
& && \phantom{=- \frac{\sqrt{\Kallengas} }{\sqrt{3} }\Bigg( }  +  \frac{\mlone - \mltwo}{\sqrt{q^2} } \left(\HAp \HPSbar + \HPS \HAmbar \right)  \nonumber \\
& && \phantom{=- \frac{\sqrt{\Kallengas} }{\sqrt{3} } \Bigg( }{}   - 2 i  \frac{\mlone + \mltwo}{\sqrt{q^2}} \left( \HAp \HTenstzerobar + \HTenstzero \HAmbar - \HTenstp \HAzerobar - \HAzero \HTenstmbar \right)  \nonumber \\
& && \phantom{=- \frac{\sqrt{\Kallengas} }{\sqrt{3} } \Bigg( }{}   - \sqrt{2} i \frac{\mlone - \mltwo}{\sqrt{q^2}} \left( \HVp \HTenszerobar + \HTenszero \HVmbar - \HTensp \HVzerobar - \HVzero \HTensmbar \right) \nonumber \\
& && \left.\phantom{=- \frac{\sqrt{\Kallengas} }{\sqrt{3} } \Bigg( }{}   + 2\sqrt{2} \frac{\mlone^2 - \mltwo^2}{q^2} \left( \HTensp \HTenstzerobar + \HTenstp \HTenszerobar - \HTenszero \HTenstmbar - \HTenstzero \HTensmbar \right)  \right)  \; , \displaybreak[0] \nonumber \\[0.2cm]
&\GKstart{2}{2}{1}&&= \frac{4}{3} \frac{\Kallengas}{ q^2}  \left( \HVp \HVzerobar + \HVzero \HVmbar +\left(V \to A \right) - 2 \left( \HTensp \HTenszerobar+\HTenszero \HTensmbar +2 \left(\HTenstp \HTenstzerobar+\HTenstzero \HTenstmbar \right)  \right)\right) \; , \displaybreak[0]\nonumber \\[0.2cm]
& \GKstart{2}{2}{2} &&= - \frac{8}{3} \frac{\Kallengas}{ q^2} \left(\HVp \HVmbar +\HAp \HAmbar -2 \left( \HTensp \HTensmbar+ 2  \HTenstp \HTenstmbar\right) \right) \; . \displaybreak[0]
\end{alignat}

\subsection{Explicit Helicity Amplitudes in terms of Form Factors}
\label{app:HadronHAs}

We collect here the definitions of the Helicity Amplitudes in terms of which our results are expressed.
The hadronic HA is defined by 
\begin{equation}
\label{eq:HadronicHAdef}
H^X_\la = \matel{\bar K^*(\la)}{\bar s \overline\Gamma^X b}{\bar B} \;,
\end{equation}
with $\overline\Gamma^X|_{\la_X \to \la}$ as defined in Tab.~\ref{tab:Gammas} and the further replacement $\gpol \to \bar \gpol$. 
 The definitions of the hadronic matrix elements used in the calculations are standard (e.g. \cite{BSZ2015}).
Below we evaluate the HAs using form factors to make clear the relative signs between the various contributions, allowing for definite comparison with the literature.

Results for form factors for low $q^2$ 
can be found from Light-Cone Sum Rules (LCSR) with vector distribution amplitudes (DA) in \cite{Ball:2004rg,BSZ2015} and $B$-meson DA in \cite{Khodjamirian:2006st}, and for high $q^2$ from lattice QCD \cite{Horgan:2013hoa}.
Long-distance effects contribute to $H^V_\la $ only, and
 include quark loops (QL), the chromomagnetic operator $O_8$,
quark loop spectator scattering (QLSS) and weak annihilation (WA). 
At low $q^2$, effects have been evaluated in  QCD factorisation (QCDF)  in the leading $1/m_b$-limit and in LCSR.  
Results for $O_8$, WA and QLSS in QCDF are given in \cite{BFS2001}, and additional contributions 
for $O_8$ in \cite{FM02}. 
In Ref.~\cite{BFS2001} it was shown that quark loops can be integrated into the $1/m_b$ framework using the 
results from inclusive matrix element computations \cite{AAGW2001}.
Results for $O_8$ and WA, as well as a prescription for dealing with endpoint divergences of  QLSS,
can be found in \cite{DLZ2012} and \cite{LZ2013}. 
Results for charm loops beyond the $1/m_b$ approximation can be found in \cite{KMPW10}  
for LCSR with $B$-meson DA, and \cite{BJZ2006,MXZ2008} for LCSR (at $q^2=0$ only) 
for vector-meson DA.
At high $q^2$ many of the long-distance contributions are suppressed in the formulation in terms
of an OPE in $1/q^2$ (with $q^2 \simeq m_b^2$) \cite{GP2004,BBF2011}. 
It should be added that the large contribution of broad charm resonances in $B \to K \mu^+\mu^-$ 
observed by the LHCb collaboration \cite{Aaij:2013pta} demands a reassessment of duality violations \cite{LZ2014}.
Long-distance contributions can be found elsewhere.

Explicit results for the $\bar B \to \bar K^*( \lone  \ltwobar)$-mode are given by
\begin{align}
\label{eq:HHAs}
H^V_{0} &= \frac{4 i \mB \mKs \left[ (\WC{V} - \WC{V}') \left(\mB + \mKs \right) A_{12} + m_b  (\WC{7}-\WC{7}') T_{23} \right]}{\sqrt{q^2} \left(\mB + \mKs \right)} \;   , \nonumber  \\ 
H^A_0 &= \frac{4 i \mB \mKs }{\sqrt{q^2} } (\WC{A}-\WC{A}') A_{12} \; , \nonumber  \\ 
H^V_{\pm}&=\frac{i}{2\left(\mB + \mKs\right)} \left(\pm(\WC{V} + \WC{V}') \sqrt{\KallenB} \, V- \left(\mB + \mKs\right)^2 (\WC{V} - \WC{V}')A_1 \right)   \nonumber  \\
& \quad {} + \frac{i m_b}{q^2} \left(\pm (\WC{7}+\WC{7}') \sqrt{\KallenB} \, T_1 -  (\WC{7}-\WC{7}') \left( \mB^2 - \mKs^2 \right) T_2 \right) \; ,  \nonumber  \\ 
H^A_{\pm} &=\frac{i}{2\left(\mB + \mKs\right)} \left(\pm(\WC{A} + \WC{A}') \sqrt{\KallenB} \, V- \left(\mB + \mKs\right)^2 (\WC{A} - \WC{A}')A_1 \right) \; , \nonumber  \\
H^P &= \frac{i \sqrt{\KallenB}}{2} \left( \frac{ \WC{P} - \WC{P}'}{m_b + m_s} + 
\frac{ \mlone + \mltwo }{ q^2  } (\WC{A} - \WC{A}' ) \right) A_0 \; , \nonumber \\
H^S &= \frac{i \sqrt{\KallenB}}{2} \left( \frac{ \WC{S} - \WC{S}'}{m_b + m_s} + 
\frac{ \mlone - \mltwo }{ q^2  }  ( \WC{V} - \WC{V}' ) \right) A_0 \; , \nonumber \\
H^T_0 &= \frac{ 2 \sqrt{2} \mB \mKs}{\mB + \mKs} \left( \WC{\mathcal{T}} + \WC{\mathcal{T}}' \right) T_{23} \; , \nonumber \\
H^{T_t}_0 &= \frac{ 2 \mB \mKs}{\mB + \mKs} \left( \WC{\mathcal{T}} - \WC{\mathcal{T}}' \right) T_{23} \; , \nonumber \\
H^{T}_{\pm} & = \frac{1}{\sqrt{2 q^2} } \left(\pm \left( \WC{\mathcal{T}} - \WC{\mathcal{T}}' \right) \sqrt{ \KallenB} T_1 -  \left( \WC{\mathcal{T}} + \WC{\mathcal{T}}' \right) \left( \mB^2 - \mKs^2 \right) T_2 \right) \; , \nonumber \\
H^{T_t}_{\pm} & = \frac{1}{2 \sqrt{q^2} } \left(\pm \left( \WC{\mathcal{T}} + \WC{\mathcal{T}}' \right) \sqrt{ \KallenB} T_1 -  \left( \WC{\mathcal{T}} - \WC{\mathcal{T}}' \right)\left( \mB^2 - \mKs^2 \right) T_2 \right) \;,
\end{align}
where $\WC{V(A)} = \WC{9(10)}$  in the standard notation used in the literature and
the $q^2$-dependence of the form factors is suppressed. 
Furthermore we have used 
\begin{align*}
A_{12} &= \frac{\left( \mB + \mKs \right)^2 \left (\mB^2 - \mKs^2 - q^2 \right) A_1 - \KallenB A_2}{16 \mB \mKs^2 \left(\mB + \mKs \right)} \; , \\
T_{23} &= \frac{\left( \mB^2 - \mKs^2 \right) \left (\mB^2 +3 \mKs^2 - q^2 \right) T_2 - \KallenB T_3}{8 \mB \mKs^2 \left(\mB - \mKs \right)} \; , 
\end{align*}
the same shorthand for zero-helicity form factor combinations as in  \cite{Horgan:2013hoa,BSZ2015}. 

The so-called timelike HAs, often denoted by $H_t$ in the literature, have been 
absorbed into $H^S$ and $H^P$. This is exceptional and follows from the vector and axial Ward identities
 $ q^\mu  \bar u (\ell_1) \ga_\mu [\ga_5]   v(\ell_2) =  (m_{\lone} \mp m_{\ltwo}) 
  \bar u (\ell_1)[\ga_5]   v(\ell_2)   $. 
  A similar simplification procedure could be repeated by use of the equation 
of motion  $i \partial^\nu (\bar s i \sigma_{\mu \nu} b)   =
   - (m_s+m_b) \bar s \gamma_\mu b +i \partial_\mu (\bar s b) 
   - 2 \bar s i \!\stackrel{\leftarrow}{D}_{\mu}  b$
(as used in \cite{Hambrock:2013zya}) 
for $H^{T_t}_\la$ if all of the operators present in the equation were used in 
the effective Hamiltonian. Since the higher derivative operators are not present 
in the effective Hamiltonian used in this paper, such a simplification does not occur.

\section{Specific Results for $\bar B \to \bar K \lone \ltwobar$}
\label{app:BtoK}
The angular distribution for this decay is
\begin{equation}
\frac{d^2 \Gamma}{dq^2 \; d\textrm{cos}\thetal} =   \GK{0} \WignerD{0}{0}{0}{\Omega_\ell} + \GK{1} \WignerD{1}{0}{0}{\Omega_\ell} + \GK{2}  \WignerD{2}{0}{0}{\Omega_\ell}  \;,\nonumber \\[0.1cm]
\end{equation}
where, using the general leptonic HAs in appendix~\ref{app:leptonHAs} and taking 
$m_{\lone} \neq m_{\ltwo}$, the functions $\GK{l_\ell} =  \NN \GKt{l_\ell} $  (with $\NN$ defined in \eqref{eq:NN}),
are given in terms of 
$\bar B \to \bar K$ HAs by
\begin{alignat}{2}
\label{eq:gidiff}
&\GKt{0} &&=  \left(4 \left( E_1 E_2 + \mlone \mltwo \right) + \frac{\Kallengas}{3 q^2}  \right) \left| \HVK \right|^2 +  \left(4 \left( E_1 E_2 - \mlone \mltwo \right) + \frac{\Kallengas}{3 q^2}  \right) \left| \HAK \right|^2  \nonumber \\
& && \quad {} + \left(4 \left( E_1 E_2 - \mlone \mltwo \right) + \frac{\Kallengas}{ q^2}  \right) \left| \HSK \right|^2 + \left(4 \left( E_1 E_2 + \mlone \mltwo \right) + \frac{\Kallengas}{ q^2}  \right) \left| \HPK \right|^2  \nonumber \\
& && \quad {} + 16 \left( E_1 E_2 + \mlone \mltwo - \frac{\Kallengas}{12 q^2}  \right) \left| \HTenstK \right|^2 + 8 \left( E_1 E_2 - \mlone \mltwo- \frac{\Kallengas}{12 q^2}  \right) \left| \HTensK \right|^2  \nonumber \\ 
& && \quad {} + 16 \left( \mlone  E_2 +  \mltwo E_1 \right) \Ima \left[ \HVK \HTenstKbar \right] + 8 \sqrt{2} \left( \mlone E_2 - \mltwo E_1 \right) \Ima \left[ \HAK \HTensKbar \right]  \; , \nonumber \\[0.2cm]
&\GKt{1} &&=  -4\sqrt{\Kallengas}\left( \Rea \left[\frac{\mlone + \mltwo}{\sqrt{q^2}} \HVK \HSKbar +\frac{\mlone - \mltwo}{\sqrt{q^2}} \HAK \HPKbar \right] \right. \nonumber \\
& && \left. {} \phantom{=-4\sqrt{\Kallengas}\Bigg(} - \Ima \left[2  \HTenstK \HSKbar  + \sqrt{2} \HTensK \HPKbar\right]  \right) \; , \displaybreak[0]\nonumber \\[0.2cm]
&\GKt{2} &&= - \frac{4\Kallengas}{3q^2} \left( \left| \HVK \right|^2 + \left| \HAK \right|^2 - 2 \left| \HTensK \right|^2 - 4 \left|\HTenstK \right|^2\right)\;.
\displaybreak[0]
\end{alignat}
 The equivalent expressions for equal lepton masses are, using the notation $\GK{l_\ell} =  \NN q^2 \GKo{l_\ell} $ 
\begin{alignat}{2}
\label{eq:gi}
&\GKo{0} &&= \frac{4}{3} \left(1 + 2\hat{\ml}^2 \right) \left| \HVK \right|^2 +  \frac{4}{3}  \beta_\ell^2 \left| \HAK \right|^2  \nonumber + 2  \beta_\ell^2 \left| \HSK \right|^2   +2   \left| \HPK \right|^2  \nonumber \\
& &&  {} + \frac{8}{3} \left( 1 + 8 \hat{\ml}^2  \right) \left| \HTenstK \right|^2 + \frac{4}{3} \beta_\ell^2\left| \HTensK \right|^2   + 16 \hat{\ml} \,\Ima \left[ \HVK \HTenstKbar \right]   \; , \nonumber \\[0.2cm]
&\GKo{1} &&=  - 4  \beta_\ell \left(2  \hat{\ml}   \, \Rea \left[ \HVK \HSKbar  \right] - \Ima \left[2  \HTenstK \HSKbar  + \sqrt{2} \HTensK \HPKbar\right]  \right) \; , \displaybreak[0]\nonumber \\[0.2cm]
&\GKo{2} &&= - \frac{4  \beta_\ell^2 }{3} \left( \left| \HVK \right|^2 + \left| \HAK \right|^2 - 2 \left| \HTensK \right|^2 - 4 \left|\HTenstK \right|^2\right)\;, \displaybreak[0]
\end{alignat}
where we have used the shorthand $\hat{\ml} \equiv  m_\ell /\sqrt{q^2}$.

\subsection{Explicit $ \bar B \to \bar K $ Helicity Amplitudes in terms of Form Factors}
\label{app:BtoKHAs}

As for $\bar B \to \bar K^* \lone \ltwobar$ we quote the HAs for form factor 
contributions only, which allows for comparison with the literature. 
Form factor computations are available for low $q^2$  and  high $q^2$ from  LCSR \cite{Ball:2004ye,Khodjamirian:2012rm}
and lattice QCD \cite{Bouchard:2013mia} respectively.  
Contributions to long-distance processes can be found in the same references 
as for the $K^*$-meson final state (quoted in appendix \ref{app:HadronHAs}).
The form factor matrix elements relevant to $\bar B \to \bar K$ transition, in standard parametrisation, are 
\begin{align}
\langle \bar K(p) |\bar{s} \gamma_\mu  b| \bar B(p_B)\rangle & = \left ( p_B + p \right )_\mu f_{+} (q^2) + \frac{m_B^2 - m_K^2}{q^2}q_\mu \left( f_0 (q^2) - f_{+} ( q^2 ) \right) \; , \nonumber \\
\langle \bar K(p) |\bar{s}  \sigma_{\mu \nu}  b| \bar B(p_B)\rangle & = i \left[ \left ( p_B + p \right )_\mu q_\nu -  \left ( p_B + p \right )_\nu q_\mu \right] \frac{f_T (q^2)}{m_B + m_K}  \; , \nonumber \\
\langle \bar K(p) |\bar{s}    b| \bar B(p_B)\rangle & = \frac{m_B^2 - m_K^2 }{m_b - m_s} f_0 (q^2) \; ,
\end{align}
with $\langle \bar K(p) |\bar{s} \gamma_\mu  \ga_5     b| \bar B(p_B)\rangle = 
\langle \bar K(p) |\bar{s}  \ga_5   b| \bar B(p_B)\rangle = 0$ in QCD.
The hadronic HA is defined by 
\begin{equation}
\label{eq:HXBK}
h^X = \matel{\bar K}{\bar s \overline\Gamma^X b}{\bar B}  \;,
\end{equation}
where $\overline \Gamma^X|_{\la_X \to 0}$ as in Tab.~\ref{tab:Gammas} with $\gpol \to \bar{\gpol}$, containing the full set of dimension-six operators in the effective Hamiltonian \eqref{eq:Heffexplicit}. We find
\begin{align}
\HVK &= \frac{\sqrt{\KallenBK} }{2 \sqrt{q^2}} \left(\frac{2 m_b}{ \mB + \mK }  ( \WC{7} + \WC{7}') f_T + ( \WC{V} + \WC{V}')   f_+  \right) \; , \displaybreak[0] \nonumber \\
\HAK &= \frac{\sqrt{\KallenBK}  }{2 \sqrt{q^2}} ( \WC{A} + \WC{A}')  f_+ \; , \nonumber \\
\HSK &=  \frac{\mB^2-\mK^2}{2}\left(\frac{ ( \WC{S} + \WC{S}')}{m_b - m_s} + \frac{ \mlone - \mltwo }{ q^2  }   ( \WC{V} + \WC{V}') \right)f_0 \; ,  \displaybreak[0]\nonumber \\
\HPK &=  \frac{\mB^2-\mK^2}{2}\left(\frac{ ( \WC{P} + \WC{P}')}{m_b - m_s} + \frac{ \mlone + \mltwo }{ q^2  }   ( \WC{A} + \WC{A}') \right)f_0 \; , \displaybreak[0] \nonumber \\
\HTensK &=  - i \frac{\sqrt{\KallenBK}  }{\sqrt{2} \left( \mB + \mK \right)} \left(\WC{\mathcal{T}} - \WC{\mathcal{T}}' \right) f_T \; , \displaybreak[0] \nonumber \\
\HTenstK &= - i \frac{\sqrt{\KallenBK}  }{2 \left( \mB + \mK \right)} \left(\WC{\mathcal{T}} + \WC{\mathcal{T}}' \right) f_T \; , \displaybreak[0]
\end{align}
where the K\"all\'en function (cf. equation~\eqref{eq:Kallen}) $\KallenBK \equiv \la(m_B^2,m_K^2,q^2)$ replaces $\KallenB \equiv \la(m_B^2,m_{K^*}^2,q^2)$ 
and $\WC{V(A)} = \WC{9(10)}$ in the standard notation used in the literature.

\subsection{Comparison with the literature}

The results for equal lepton masses \eqref{eq:gi} do agree with the results of 
reference \cite{BHP2007} when $\thetal \to \pi - \thetal$ is taken into account. 
This is consistent with the angular conventions. In this paper we use 
the same conventions as LHCb \cite{angularK14}, which differ from the ones 
of \cite{BHP2007} by the transformation stated above.

\section{$\Lambda_b \to \Lambda \left( \to (p,n) \pi \right) \lone \ltwobar $ Angular Distribution}
\label{app:baryons}
The decay $\Lambda_b \to \Lambda \left( \to (p,n) \pi \right)  \lone \ltwobar $ with a final-state proton or neutron, 
 recently measured by the LHCb Collaboration \cite{LHCbbaryon2015},
 can also be considered within the generalised helicity formalism, and is particularly relevant because this decay can also be described using the effective Hamiltonian defined in \ref{app:Hamiltonian}. In this case \eqref{eq:helamp} becomes, in the rest frame of the $\Lambda_b$,
\begin{equation}
\begin{split}
& \amp(\Omega_{\Lambda_b},\Omega_\ell ,\Omega_{\Lambda}| \la_{\Lambda_b},\la_{N},\la_1,\la_2 )  \sim  \\
& \quad  \sum_{\la_{\ga}, \la_{\Lambda} , J_\ga} 
\delta_{\la_{\Lambda_b}, \la_{\ga}-\la_{ \Lambda } } 
 {\cal \Had}_{\la_{\ga} \la_{ \Lambda } }
 \WignerD {\frac{1}{2}}{\la_{\Lambda} }{\la_{N}}{\Omega_{\Lambda}} 
{\cal N}_{\la_{N}}
 \WignerD {J_\ga}{\la_{\ga}}{\la_\ell}{\Omega_\ell}{\cal \ell}_{\la_1
\la_2} \\
&  =  \sum_{\la_{\ga}, J_\ga} 
 {\cal \Had}_{\la_{\ga} ,\la_{\ga} - \la_{\Lambda_b} }
 \WignerD {\frac{1}{2}}{\la_{\ga} - \la_{\Lambda_b} }{\la_{N}}{\Omega_{\Lambda}} 
{\cal N}_{\la_{N}}
 \WignerD {J_\ga}{\la_{\ga}}{\la_\ell}{\Omega_\ell}{\cal \ell}_{\la_1
\la_2} \;,
\end{split}
\end{equation} 
where the leptonic HAs are the same as before and ${\cal N}_{\la_{N}}$ is the HA for the decay $ \Lambda \to N \pi$ analogous to the $g_{K^{*} K \pi}$ factor in the $B \to K^{*}$ decay, this time carrying non-trivial dependence on helicities owing to the final state particle $N$ having spin-$\frac{1}{2}$. The terms $ {\cal \Had}_{\la_{\ga} \la_{ \Lambda } }$ are the HAs for the $\Lambda_b \to \Lambda$ decay and can be again expressed in the form
\begin{equation}
{\cal \Had}_{\la_{\ga} \la_{ \Lambda } } = \matel{\Lambda (\la_\Lambda)}{\bar{s} \overline\Gamma^X b}{\Lambda_b (\la_{\Lambda_b})} \; , 
\end{equation}
with the $\overline\Gamma^X$ the same as defined in Tab.~\ref{tab:Gammas}. The resulting angular distribution can then be expressed as 
\begin{align}
K(q^2, \Omega_\Lambda , \Omega_\ell ) \sim \Rea  \left[\vphantom{ \Gbaryon{1}{0}{0} } \right. & \Gbaryon{0}{0}{0} \DD{0}{0}{0} \left(\Omega_\Lambda ,\Omega_\ell \right)  +\Gbaryon{0}{1}{0} \DD{0}{1}{0} \left(\Omega_\Lambda ,\Omega_\ell \right)  + \Gbaryon{0}{2}{0} \DD{0}{2}{0} \left(\Omega_\Lambda ,\Omega_\ell \right)   \nonumber \\
 {} + {} &  \Gbaryon{1}{0}{0} \DD{1}{0}{0} \left(\Omega_\Lambda ,\Omega_\ell \right) + \Gbaryon{1}{1}{0} \DD{1}{1}{0} \left(\Omega_\Lambda ,\Omega_\ell \right)  +   \Gbaryon{1}{2}{0} \DD{1}{2}{0} \left(\Omega_\Lambda ,\Omega_\ell \right) \nonumber \\
 {} + {} &\Gbaryon{1}{1}{1} \DD{1}{1}{1} \left(\Omega_\Lambda ,\Omega_\ell \right)  +  \left.  \Gbaryon{1}{2}{1} \DD{1}{2}{1} \left(\Omega_\Lambda ,\Omega_\ell \right) \right] \; ,
\end{align}
where $\Omega_\Lambda = (0,\theta_\Lambda,0)$ and $\Omega_\ell = (\hel,\thetal,- \hel)$. A theoretical angular analysis of this decay has been performed in \cite{GIKLS13,BDF2015}; in terms of the functions defined in \cite{BDF2015}, the $\Gbaryon{l_\Lambda}{l_\ell}{m}$ above are

\begin{alignat}{3}
& \Gbaryon{0}{0}{0} = \frac{1}{3} \left( K_{1cc} + 2 K_{1ss} \right) \; , \quad  && \Gbaryon{0}{1}{0} = K_{1c} \; ,  && \Gbaryon{0}{2}{0} = \frac{2}{3} \left( K_{1cc} - K_{1ss} \right)  \;,\nonumber \\
& \Gbaryon{1}{0}{0} = \frac{1}{3} \left( K_{2cc} + 2 K_{2ss} \right) \; , \quad  && \Gbaryon{1}{1}{0} = K_{2c} \; , \quad  && \Gbaryon{1}{2}{0} = \frac{2}{3} \left( K_{2cc} - K_{2ss} \right) 
\;, \nonumber \\
&   \Gbaryon{1}{1}{1} = K_{3s} + i K_{4s} \; , &&  && \Gbaryon{1}{2}{1} = \frac{1}{\sqrt{3}} \left( K_{3sc} + i K_{4sc} \right) \; .
\end{alignat}
These results can also be compared with those found in \cite{BCDS2015}; it follows that the MoM will be equally useful in future angular analyses of this decay.

\bibliographystyle{utphys}
\bibliography{referencesangdist}

\end{document}